 \makeatletter
 \@ifundefined{@parse@version@dash}{%
 \def\@parse@version#1{\@parse@version@0#1}
 \def\@parse@version@#1/#2/#3#4#5\@nil{%
 \@parse@version@dash#1-#2-#3#4\@nil}
 \def\@parse@version@dash#1-#2-#3#4#5\@nil{%
 \if\relax#2\relax\else#1\fi#2#3#4 }
 }{}
 \makeatother

\documentclass[aps,prl,a4paper,twocolumn,superscriptaddress]{revtex4-2}
\usepackage{dcolumn}

\usepackage{amsmath,bbm,amsthm}
\usepackage{amssymb}
\usepackage{graphicx}
\usepackage{epstopdf}
\usepackage{mathrsfs}
\usepackage{xcolor}
\usepackage{blkarray}

\definecolor{myurlcolor}{rgb}{0,0,0.7}
\definecolor{myrefcolor}{rgb}{0.8,0,0}
\usepackage[unicode=true,pdfusetitle, bookmarks=false,bookmarksnumbered=false,
bookmarksopen=false, breaklinks=false,pdfborder={0 0 0},backref=false,
colorlinks=true, linkcolor=myrefcolor,citecolor=myurlcolor,urlcolor=myurlcolor]{hyperref}

\newcommand{\Tr}{\mathrm{Tr}}
\newcommand{\upe}{\mathrm{e}}
\newcommand{\upi}{\mathrm{i}}

\newcommand{\ket}[1]{\left|#1\right\rangle}
\newcommand{\bra}[1]{\left\langle #1\right|}

\newcommand{\ketbra}[2]{|#1\rangle\langle#2|}
\newcommand{\proj}[1]{\ket{#1}\!\bra{#1}}

\newtheorem{theorem}{Theorem}
\newtheorem{lemma}{Lemma}
\newtheorem{corollary}{Corollary}
\newtheorem{observation}{Observation}

\newcommand{\rvec}{\boldsymbol{r}}
\newcommand{\mvec}{\boldsymbol{m}}
\newcommand{\bvec}{\boldsymbol{b}}
\newcommand{\evec}{\boldsymbol{e}}
\newcommand{\id}{\mathbbm{1}}
\newcommand{\pauliz}{\sigma_\mathrm{z}}
\newcommand{\paulix}{\sigma_\mathrm{x}}

\newcommand{\Jx}{\hat{J}_\mathrm{x}}
\newcommand{\Jy}{\hat{J}_\mathrm{y}}
\newcommand{\Jz}{\hat{J}_\mathrm{z}}

\newcommand{\phivec}{\vec{\phi}}

\newcommand{\Phivec}{\vec{\Phi}}
\newcommand{\Vphi}{V_{\phivec}}
\newcommand{\VPhi}{V_{\Phivec}}
\newcommand{\Vcalphi}{\cV_{\phivec}}
\newcommand{\VcalPhi}{\cV_{\Phivec}}
\newcommand{\Piphi}[1]{\Pi_{#1,\phivec}}
\newcommand{\Mphi}[1]{M_{#1,\phivec}}

\newcommand{\pphitheta}{p_{\theta,\phivec}}
\newcommand{\dotpphitheta}{\dot{p}_{\theta,\phivec}}
\newcommand{\qphitheta}{q_{\theta,\phivec}}

\newcommand{\piphitheta}{p_{\theta,\phivec}(i)}
\newcommand{\dotpiphitheta}{\dot{p}_{\theta,\phivec}(i)}
\newcommand{\qxphitheta}{q_{\theta,\phivec}(x)}
\newcommand{\dotqxphitheta}{\dot{q}_{\phivec,\theta}(x)}

\newcommand{\opt}[1]{#1_\mathrm{opt}}
\newcommand{\varphiopt}{\opt{\varphi}}
\newcommand{\phiopt}{\opt{\phi}}

\newcommand{\lambdazero}{\lambda_{\ket{0}}}
\newcommand{\lambdaone}{\lambda_{\ket{1}}}

\newcommand{\FoneSingle}{\bar{F}}
\newcommand{\Ftwobin}{F_{\text{2-bin}}^*}
\newcommand{\Ftwobinmax}{\bar{F}_{\text{2-bin}}}
\newcommand{\Ftwobinmaxstar}{\bar{F}_{\text{2-bin}}^*}
\newcommand{\Fthreebinmax}{\bar{F}_{\text{3-bin}}}

\newcommand{\Fdown}[1]{F^{(#1)}}
\newcommand{\FQ}{\mathcal{F}}
\newcommand{\FQim}{\FQ^{\mathrm{\scriptstyle{(im)}}}}
\newcommand{\FQimbar}{ \bar{\FQ}^{\mathrm{\scriptstyle{(im)}}}}
\newcommand{\FQimbarloc}{ \bar{\FQ}^{\mathrm{\scriptstyle{(im,l)}}}}

\newcommand{\FNCE}{F_N^{\mathrm{(CE)}}}
\newcommand{\FNCEas}{F_N^{\mathrm{(CE,as)}}}
\newcommand{\FNdown}{F_{N}^{\downarrow}}

\newcommand{\krausvec}{\boldsymbol{K}}
\newcommand{\rhocl}{\rho_\mathrm{cl}}

\newcommand{\cP}{\mathcal{P}}
\newcommand{\cM}{\mathcal{M}}
\newcommand{\cH}{\mathcal{H}}
\newcommand{\cB}{\mathcal{B}}
\newcommand{\cV}{\mathcal{V}}
\newcommand{\cU}{\mathcal{U}}
\newcommand{\cE}{\mathcal{E}}
\newcommand{\cW}{\mathcal{W}}

\newcommand{\cPloss}{\cP_{\mathrm{loss}}}
\newcommand{\cPdc}{\cP_{\mathrm{dc}}}

\newcommand{\uu}{\mathsf{u}}
\newcommand{\hh}{\mathsf{h}}
\newcommand{\pp}{\mathsf{p}}
\newcommand{\qq}{\mathsf{q}}

\newcommand{\dimM}{{|X|}}

\DeclareGraphicsExtensions{.png,.pdf,.eps,.jpg}
\graphicspath{ {./figs/} }
\newcommand{\eref}[1]{(\ref{#1})}
\newcommand{\eqnref}[1]{Eq.~(\ref{#1})}
\newcommand{\eqnsref}[2]{Eqs.~(\ref{#1}) and (\ref{#2})}
\newcommand{\figref}[1]{Fig.~\ref{#1}}

\newcommand{\secref}[1]{Sec.~\ref{#1}}

\newcommand{\lemref}[1]{Lemma~\ref{#1}}
\newcommand{\thmref}[1]{Thm.~\ref{#1}}

\begin{document}

\title{Quantum metrology with imperfect measurements}

\author{Yink Loong Len}
\email{y.len@cent.uw.edu.pl}
\affiliation{Centre for Quantum Optical Technologies, Centre of New Technologies, University of Warsaw, Banacha 2c, 02-097 Warszawa, Poland}

\author{Tuvia Gefen}
\email{tgefen@caltech.edu}
\affiliation{Institute for Quantum Information and Matter, Caltech, Pasadena, CA, USA}

\author{Alex Retzker}
\affiliation{Racah Institute of Physics, The Hebrew University of Jerusalem, Jerusalem 91904, Givat Ram, Israel}
\affiliation{AWS Center for Quantum Computing, Pasadena, CA 91125, USA}

\author{Jan Ko\l{}ody\'{n}ski}
\email{jan.kolodynski@cent.uw.edu.pl}
\affiliation{Centre for Quantum Optical Technologies, Centre of New Technologies, University of Warsaw, Banacha 2c, 02-097 Warszawa, Poland}

\date{\today}

\begin{abstract}
The impact of measurement imperfections on quantum metrology protocols 
has not been approached in a systematic manner so far. In this work, we tackle this issue 
by generalising firstly the notion of quantum Fisher information to account for noisy detection, and propose tractable methods allowing for its approximate evaluation. We then show that in canonical scenarios involving $N$ probes with local measurements undergoing readout noise, the optimal sensitivity depends crucially on the control operations allowed to counterbalance the measurement imperfections---with global control operations, the ideal sensitivity (e.g.~the Heisenberg scaling) can always be recovered in the asymptotic $N$ limit, while with local control operations the quantum-enhancement of sensitivity is constrained to a constant factor. We illustrate our findings with an example of NV-centre magnetometry, as well as schemes involving spin-$1/2$ probes with bit-flip errors affecting their two-outcome measurements, for which we find the input states and control unitary operations sufficient to attain the ultimate asymptotic precision.
\end{abstract}

\keywords{ }

\maketitle

\section{Introduction}
One of the most promising quantum-enhanced technologies are the \emph{quantum sensors}~\cite{Degen2017} that by utilising quantum features of platforms such as solid-state spin systems~\cite{awschalom_quantum_2018,Barry2020}, atomic ensembles~\cite{Pezze2018} and interferometers~\cite{bongs_taking_2019}, or even gravitational-wave detectors~\cite{Tse2019} are capable of operating at unprecedented sensitivities. They all rely on the architecture in which the parameter to be sensed (e.g.~a magnetic or gravitational field) perturbs a well-isolated quantum system, which after being measured allows to precisely infer the perturbation and, hence, estimate well the parameter from the measurement data. In case the sensor consists of multiple probes (atoms, photons) their inter-entanglement opens doors to beating classical limits imposed on the estimation error~\cite{Giovannetti2004}---a fact that ignited a series of breakthrough experiments~\cite{Leibfried2004,Mitchell2004,Esteve2008,Appel2009,Sewell2012,Hosten2016}, being responsible also for the quantum-enhancement in gravitational-wave detection~\cite{Tse2019}. 

These demonstrations are built upon various seminal theoretical works, in particular Refs.~\cite{Helstrom1976, Holevo1982,Braunstein1994} that adopted parameter-inference problems to the quantum setting, and generalised the \emph{Fisher information} (FI)~\cite{Kay1993} to quantum systems. This general formalism provides tools to identify optimal probe states and measurements for any given quantum metrology task~\cite{Giovannetti2006}.  Interestingly it was shown that in many multi-probe scenarios, even those that involve entangled probes, optimal readout schemes turn out to be local---each of the probes can in principle be measured independently~\cite{Giovannetti2006}. 

In practice, however, engineering a measurement of a quantum system is a challenge \emph{per se}---it relies on a scheme in which a meter component, typically light, interacts with the quantum sensor before being subsequently detected~\cite{Hammerer2010,Clerk2010}. This allows the state of the probes to be separately controlled, at the price of the meter component carrying intrinsic noise that cannot be completely eradicated. As a result, the implemented measurement becomes \emph{imperfect} with the measured data being noisy due to, e.g., finite resolution of the readout signal. Such an issue naturally arises across different sensing platforms:~in nitrogen-vacancy (NV) centres in diamond~\cite{PhysRevLett.100.077401,jelezko2006single,schirhagl2014nitrogen}, superconducting based quantum information processors~\cite{sete2015quantum,heinsoo2018rapid,krantz2019quantum,bergquist1986observation}, trapped ions~\cite{nagourney1986shelved,PhysRevLett.57.1696,PhysRevLett.100.200502,marciniak_optimal_2021}, and interferometers involving photodetection~\cite{Harris2017,Xu2020}. Although for special detection-noise models (e.g.~Gaussian blurring) the impact on quantum metrological performance and its compensation via the so-called interaction-based readout schemes has been studied~\cite{Davis2016,Frowis2016,Nolan2017,Haine2018} and demonstrated~\cite{Linnemann2016}, a general analysis has been missing thus far.

Crucially, such a detection noise affecting the measurement cannot be generally put on the same grounds as the ``standard'' decoherence disturbing the (quantum) dynamics of the sensor before being measured~\cite{Maccone2011}. In the latter case, the impact on quantum metrological performance has been thoroughly investigated~\cite{Fujiwara2008,Escher2011,Demkowicz2012} and, moreover, shown under special conditions to be fully compensable by implementing methods of \emph{quantum error correction}~\cite{Duer2014,Arrad2014,Sekatski2017,Demkowicz2017,zhou_achieving_2018}. This contrasts the setting of readout noise that affects the classical output (outcomes) of a measurement, whose impact cannot be inverted by employing, e.g., the methods of \emph{error mitigation}~\cite{maciejewski_mitigation_2020,Bravyi2021} designed to recover statistical properties of the ideal readout data at the price of overhead, which cannot be simply ignored in the context of parameter estimation by increasing the sample size.

In our work, we formalize the problem of imperfect measurements in quantum metrology by firstly generalising the concept of 
\emph{quantum Fisher information}~(QFI)~\cite{Braunstein1994} to the case of noisy readout. For pure probe-states, we explicitly relate the form of the resulting \emph{imperfect QFI} to the perfect QFI, i.e.~to the one applicable in presence of ideal detection. However, as we find the imperfect QFI not always to be directly computable, we discuss two general methods allowing one to tightly bound its value, as illustrated by a specific example of precision magnetometry performed with help of a NV centre~\cite{Maze2008,Taylor2008}, for which the measurement imperfection is naturally inbuilt in the readout procedure~\cite{Jiang2009,Neumann2010}. Using the \emph{conjugate-map decomposition} formalism, we also study when the measurement imperfections can be effectively interpreted as an extra source of ``standard'' decoherence, in order to show that this may occur only under very strict conditions.

Secondly, we focus on the canonical metrology schemes involving multiple probes~\cite{Giovannetti2006}, in order investigate how do the measurement imperfections affect then the attainable sensitivity as a function of the probe number $N$, which in the ideal setting may scale at best quadratically with $N$---following the so-called ultimate \emph{Heisenberg scaling} (HS)~\cite{Giovannetti2004}. Considering general local measurements undergoing detection noise, we demonstrate that the achievable precision strongly depends on the type, i.e.~global vs local, of control operations one is allowed to apply on the probes before the readout is performed.

In the former case, we prove a \emph{go-theorem} which states that there always exists a \emph{global} control unitary such that for pure states the imperfect QFI converges to the perfect QFI with $N$, and the detection noise can then be effectively ignored in the $N\to\infty$ limit. We provide a recipe how to construct the required global unitary operation, and conjecture the general form of the optimal unitary from our numerical evidence. On the contrary, when restricted to \emph{local} control unitaries, we resort to the concept of \emph{quantum-classical channels}~\cite{Holevo1998} that describe then not only the evolution of each probe, but also the noisy measurement each probe is eventually subject to. For this complementary scenario, we establish a \emph{no-go theorem} which states that whenever measurements exhibit any non-trivial local detection noise, attaining the HS becomes ``elusive''~\cite{Demkowicz2012}---the maximal quantum-enhancement becomes restricted to a constant factor with the estimation error asymptotically following at best a classical behaviour ($\sim1/N$), which we refer to as the \emph{standard scaling} (SS). 

In order to illustrate the applicability of both theorems, 
we consider the 
phase-estimation example involving $N$ spin-$1/2$ probes, whose binary measurements undergo bit-flip errors. On one hand, we explicitly construct the global unitary control operation, thanks to which the sensitivity quickly attains the HS with $N$, using for example the \emph{GHZ state}~\cite{GHZ}. On the other, when only local control operations are allowed, we evaluate the asymptotic SS-like bound on precision analytically, and prove its saturability with $N\to\infty$ by considering the probes to be prepared in a \emph{spin-squeezed state}~\cite{Wineland1992,Kitagawa1993} and measuring effectively the mean value of their total angular momentum by adequately interpreting the noisy readout data. Furthermore, we apply the above analysis in Methods to the setting of optical interferometry involving $N$-photon states and imperfect detection, which suffers from both photonic losses and dark counts.

\section{Results}

\subsection{Metrology with imperfect measurements}	                
\begin{figure*}[t!]
	\centering
	\includegraphics[width=0.90\textwidth]{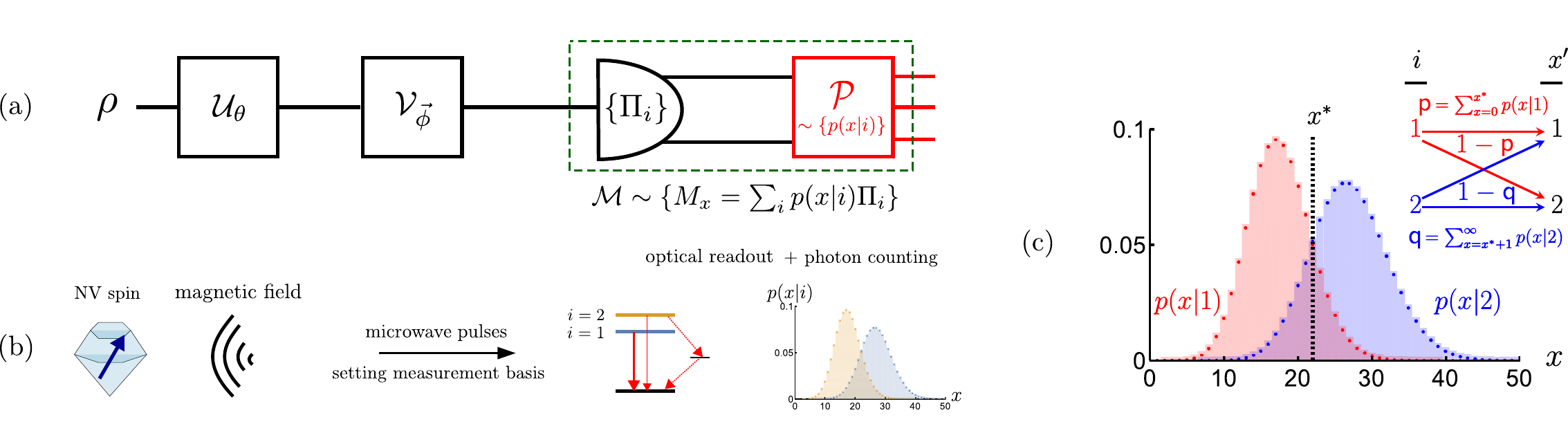}
	\caption{(a) \textbf{Scheme of quantum metrology with an imperfect measurement}. A quantum state $\rho$ is fed into a unitary channel $\cU_\theta$ which encodes the parameter of interest, $\theta$. The probe is then rotated by a unitary $\cV_{\phivec}$, so that a given projective measurement $\{\Pi_i\}$ can be performed in the preferred basis. The measurement $\cM$ is, however, \emph{imperfect}, i.e.:~different $\{i\}$ outcomes are `inaccessible'
    being mapped onto another set of `observable' outcomes $\{x\}$, as specified by the \emph{noisy detection channel} (stochastic map) $\cP\sim\{p(x|i)\}$. (b) \textbf{Phase sensing with the nitrogen-vacancy (NV) centre used as a spin probe}. The spin of the NV is initialised in an equal superposition between the $m_s=0,1$ ($i=1,2$) energy-level states and evolves in presence of an external magnetic field, which induces a relative phase proportional to its strength. Microwave pulse is then applied to transform the relative phase into the population difference of the energy levels, which is then readout optically. The measurement procedure is inherently imperfect:~the two populations indicating either $i=1\vee2$ each yield a (photon-number) signal that is stochastic and distributed according to a Poissonian profile, whose overlap renders the observed outcome $x$ ambiguous. (c) \textbf{Binary binning strategy or the ``threshold method'':} Infinite outcomes from Poissonian imperfections are categorised into two ``\emph{bins}'' containing $x\le x^*$ and $x> x^*$, respectively. As a result, the effective post-processing map $\cP$ simplifies to an \emph{asymmetric bit-flip channel} of the (projective) measurement outcomes summarised in the inset. 
	}
	\label{fig:QMwIM-fig1}
\end{figure*}

Let us consider a 
quantum metrology scenario depicted in \figref{fig:QMwIM-fig1}(a), in which a $d$-dimensional qudit probe is prepared in a quantum state $\rho$, before it undergoes the dynamics encoding the parameter of interest $\theta$ that is represented by a unitary channel $\cU_\theta\sim\{U_\theta\}$~\cite{footnote1}.
The probe state thus transforms onto $\rho(\theta)=\cU_\theta[\rho]=U_\theta\rho U_\theta^\dagger$, and is subsequently rotated by a control unitary transformation 
$\cV_{\phivec}\sim\{\Vphi\}$ specified by the vector of parameters $\phivec$. It is then subjected to a fixed projective (von Neumann) measurement 
formally represented by a set of projection operators $\{\Pi_i\}_{i=1}^{d}$, i.e.~$\Pi_{i} \Pi_{j} =\delta_{i,j}\Pi_{i}$ and $\sum_{i=1}^{d}\Pi_i=\id_{d}$. As a consequence, any projective measurement $\Piphi{i}:=\Vphi^\dagger \Pi_{i}\Vphi$ with $d$ outcomes can be implemented, where the purpose of the unitary operation $\cV_{\phivec}$ is to select a particular measurement basis. In an \emph{ideal} setting, every outcome $i$ can be directly observed with its probability being given by the Born's rule $\piphitheta=\Tr\{\rho(\theta)\Piphi{i}\}$. Repeating the procedure over many rounds, an estimate $\tilde{\theta}$ can then be constructed based on all the collected data, which most accurately reproduces the true parameter value $\theta$.

In particular, it is then natural to seek $\tilde{\theta}$ that minimises the \emph{mean squared error} (MSE), $\Delta^2\tilde{\theta}$, while also minimising it over different measurement bases and initial states of the probe. For unbiased estimators, considering $\nu$ repetitions, the MSE is generally lower limited by the \emph{quantum Cram\'{e}r-Rao bound} (QCRB)~\cite{Kay1993,Braunstein1994}:
\begin{equation}
\label{eq:QCRB}
\nu \Delta^2\tilde{\theta}
\;\geq\;
\frac{1}{\FQ}
\;\geq\;  
\frac{1}{\bar{\FQ}},
\end{equation}
where $\FQ$ is the \emph{quantum Fisher information} (QFI) that corresponds to the maximal (classical) \emph{Fisher information} (FI), $F$, defined for a given distribution $\pphitheta$ and its derivative w.r.t.~the estimated parameter, $\{\dotpiphitheta\equiv\partial_\theta \piphitheta\}$, i.e.~\cite{Kay1993,Braunstein1994}:
\begin{equation}
\label{eq:FI}
\FQ:=\max_{\phivec} F 
\quad\text{with}\quad
F[\pphitheta]:=\sum_i \frac{\dotpiphitheta^2}{\piphitheta},
\end{equation}
that is optimised over all possible measurement bases $\phivec$. ${\bar{\FQ}}$ in \eqnref{eq:QCRB} is the \emph{channel QFI} which includes a further optimisation over all possible input probe states $\rho$, i.e.~$\bar{\FQ}:=\max_{\rho}\FQ=\max_{\rho,\phivec}F.$

For perfect projective measurements this theory is well established---close analytical expressions for the QFI and the channel QFI exist. The QFI 
for any $\rho(\theta)$ reads~\cite{Braunstein1994}:
\begin{equation}
\label{eq:FQ}
\FQ[\rho(\theta)]=\Tr\{\rho(\theta) L^2\},
\end{equation}
where $L$ is the symmetric-logarithmic derivative operator defined implicitly as $\partial_\theta \rho(\theta)=\frac{1}{2}(L\rho(\theta)+\rho(\theta)L)$, whose eigenbasis provides then the optimal measurement basis $\phivec$ that yields the QFI. Moreover, as the QFI is convex over quantum states~\cite{Alipour2015}, its maximum is always achieved by pure input states $\psi=\proj{\psi}$. Hence, for the unitary encoding $\psi(\theta)=\cU_\theta[\psi]$ the channel QFI in \eqnref{eq:QCRB} just reads~\cite{Pang2014}:
\begin{align}
\label{eq:channelQFI}
\bar\FQ[\cU_\theta]=(\lambda_\mathrm{max}(h_\theta)-\lambda_\mathrm{min}(h_\theta))^2,
\end{align}
where $h_{\theta}=-\upi\left(\partial_{\theta}U_\theta\right)U_\theta^{\dagger}$, $\lambda_\mathrm{max}(h_\theta)$ and $\lambda_\mathrm{min}(h_\theta)$ are the maximum and minimum eigenvalues of $h_\theta$, respectively, and $\bar{\FQ}$ is attained by $\psi(\theta)$ being an equal superposition of the corresponding two eigenvectors~\cite{Giovannetti2006,Pang2014}.

In practical settings, however, perfect measurements are often beyond reach. Instead, one must deal with an \emph{imperfect measurement} $\cM$ that is formally described by a positive operator-valued measure (POVM)---a set consisting of $\dimM$  
positive operators $\cM\sim \left\{ M_{x}\right\}_x$ that satisfy $\sum_x M_{x}=\id_{d}$ and are now no longer projective. In \figref{fig:QMwIM-fig1}(a) we present an important scenario common to many quantum-sensing platforms---e.g.~NV-centre-based sensing depicted in \figref{fig:QMwIM-fig1}(b). In particular, it includes a \emph{noisy detection channel} $\cP$ which distorts the ideal projective measurement $\{\Piphi{i}\}_{i=1}^d$, so that its $d$ outcomes become `inaccessible', as they get randomised by some stochastic post-processing map $\cP\sim \{p(x|i)\}$ into another set $X\sim\{x\}$ of 
$\dimM$ outcomes. The noise of the detection channel is then specified by the transition probability $p(x|i)$, which describes the probability of observing an outcome $x$, given that the projective measurement $i$ was actually performed. In such a scenario any `observable' outcome $x$ occurs with probability $\qxphitheta=\sum_{i=1}^d  p(x|i)\,\piphitheta=\Tr\{\rho(\theta)\Mphi{x}\}$, where the corresponding imperfect measurement is then described by $\Mphi{x}=\sum_{i=1}^d p(x|i)\, \Piphi{i}$. 

In presence of measurement imperfections, the QCRB \eref{eq:QCRB} must be modified, so that it now contains instead the \emph{imperfect QFI} and the \emph{imperfect channel QFI}, which are then respectively defined as:
\begin{equation}\label{eq:FQim+FQimbar}
\FQim:=\underset{\phivec}{\text{max}}\,F[\qphitheta]
\quad \text{and} \quad
\FQimbar:=\max_{\rho,\phivec}\,F[\qphitheta].
\end{equation}
Once the assumption of perfect measurements is lifted, very little is known.
In particular, although $\FQimbar$ can still be attained with some pure encoded state $\psi(\theta)$ by the convexity argument, there are no established general expression for $\FQim$ and $\FQimbar$, as in \eqnsref{eq:FQ}{eq:channelQFI}. 

Firstly, we establish a formal relation between $\FQim$ and $\FQ$ for  
all quantum metrology protocols involving pure states with arbitrary $\theta$-encoding and imperfect measurements, which can be summarised as follows:
\begin{lemma}[Quantum Fisher information with imperfect measurements]
	\label{lem:FIwIM}
	For any pure encoded probe state, $\psi\!\left(\theta \right)$, and imperfect measurement, $\cM$, the imperfect QFI reads
	\begin{align} \label{eq:FIwIM}
		\FQim =\gamma_\cM\;\FQ[\psi(\theta)],
	\end{align}
	where
	\begin{equation}
		\label{eq:gamma_M}
		\gamma_\cM
		=
		\max_{\ket{\xi},\ket{\xi_{\perp}}} \sum_x \frac{\mathrm{Re}\!\left\{\langle\xi_{\perp}|M_{x}|\xi\rangle\right\}^{2}}{\langle\xi|M_{x}|\xi\rangle}
	\end{equation}
	is a constant $0\leq\gamma_\cM\leq1$ depending solely on the imperfect measurement, with the maximisation being performed over all pairs of orthogonal pure states $\ket{\xi}$ and $\ket{\xi_\perp}$.
\end{lemma}
We leave the explicit proof of \lemref{lem:FIwIM} to the Supplement, but let us note that when assuming a unitary encoding, $\psi\!\left(\theta \right)=\cU_\theta[\psi]$ and maximising \eqnref{eq:FIwIM} over all pure input states, $\psi$, it immediately follows that:
	\begin{align} \label{eq:FIwIMbar}
	\FQimbar = \gamma_\cM\;\bar{\FQ}[\cU_\theta]. 
\end{align}
The constant $\gamma_\cM$ specified in \eqnref{eq:gamma_M} has an intuitive meaning:~it quantifies how well the imperfect measurement $\cM$ can distinguish at best a pair of orthogonal states. In fact, we prove explicitly in the Supplement that if there exist two orthogonal states that can be distinguished perfectly using $\cM$, then $\gamma_{\cM}=1$ and $\FQim=\FQ$. 

Unfortunately, $\gamma_\cM$ need not be easily computable, even numerically---consider, for instance, 
noisy detection channels $\cP$ (e.g.~the NV-centre example of \figref{fig:QMwIM-fig1} discussed below) that yield imperfect measurements with infinitely many outcomes $X$ and, hence, the sum in \eqnref{eq:gamma_M} not even tractable. For this, we introduce in Methods two techniques that allow us to approximate well both $\FQim$ and $\FQimbar$ in \eqnsref{eq:FIwIM}{eq:FIwIMbar}, respectively, by considering tight lower bounds on the corresponding FIs.

\subsubsection{Example: Phase sensing with an NV centre}
The utilisation of NV centres as quantum spin probes allows for precise magnetic-field sensing with unprecedented resolution~\cite{Degen2017}. For detailed account on sensors based on NV centres we refer the reader to Refs.~\cite{Doherty2013,Rondin2014,Barry2020}; here, we focus on the very essence and briefly outline the canonical NV-centre-based sensing protocol based on a Ramsey-type sequence of pulses, schematically depicted in \figref{fig:QMwIM-fig1}(b), and as described in the Methods section.

In short, the sensing of a magnetic field with an NV centre fits into the general formalism introduced above, whereby now the encoding channel is $\cU_\theta\sim \{U_\theta=\upe^{\upi h\theta}\}$, $h=\pauliz/2=(\ketbra{0}{0}-\ketbra{1}{1})/2$, with $\Pi_1=\proj{0}$ and $\Pi_2=\proj{1}$, which can be rotated into another measurement basis by a Ramsey pulse. These projective measurements are, however, not ideally implemented, as the fluoresence readout technique is inherently noisy. The final `observed' outcomes are the number of collected photons $X=\{0,1,2,\cdots\}$, distributed according to the two Poissonian distributions $p(x|1)=\upe^{-\lambdazero}(\lambdazero)^x/x!$ and $p(x|2)=\upe^{-\lambdaone}(\lambdaone)^x/x!$, whose means, $\lambdazero$ and $\lambdaone$, differ depending on which energy state the NV spin was previously projected onto by $\Piphi{1}$ or $\Piphi{2}$.

\begin{figure}[t!] 	
	\includegraphics[width=0.45\textwidth]{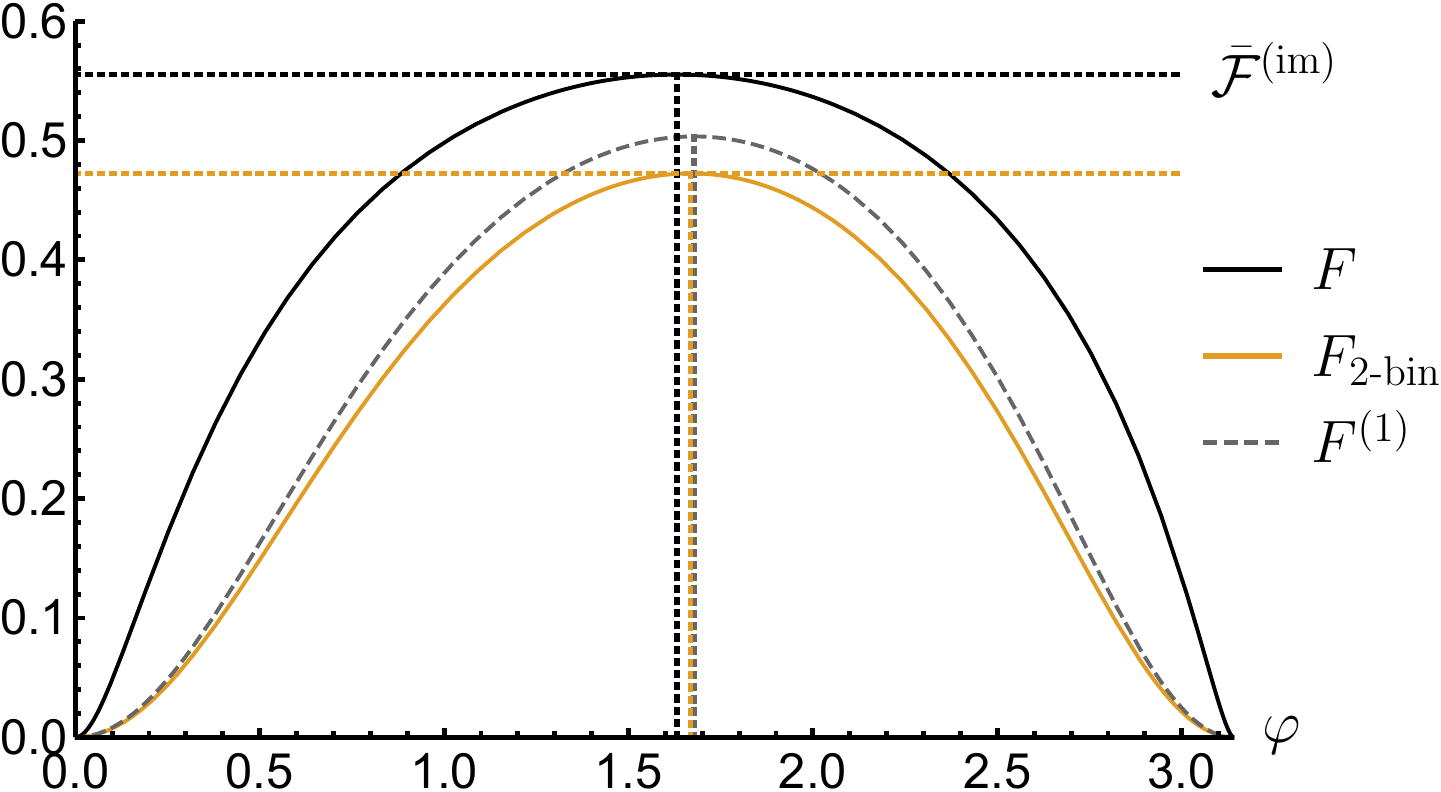}
	\caption{\textbf{Computing FI for sensing phase $\theta$ with measurements experiencing Poissonian noise.} The FIs are presented as a function of the relative measurement-basis angle, $\varphi=\theta+\phi$, whose changes are equivalent to varying the input-state angle, $\phi$, and/or the estimated phase, $\theta$ (we set $\lambdaone/\lambdazero=0.65$ and $\lambdazero=27$~\cite{Boss2017}). The exact $F$ (\emph{solid black}) is numerically approximated by summing over $x\le100$ in \eqnref{eq:FNV}, while $F_{2\text{-bin}}$ (\emph{solid orange}) accounts for the binning method in \eqnref{eq:F2bin} with the choice of the binning boundary $x^*$ further optimised over. We also plot $\Fdown{1}$ (\emph{dashed grey}), the FI approximated with using just the first two moments of the observed probability distribution. The vertical dotted lines indicate the (optimal) $\varphi$ at each of the respective quantities is maximised. Note that when the measurement is perfect, $\bar{\FQ}[\cU_{\theta}]$ is unity, and is for all choices of the angle $\varphi$ (not shown). The horizontal dotted lines depict the (numerically approximated) values of \eqnref{eq:FIwIMbar}, $\FQimbar=\gamma_\cM$, for respective cases of the exact $F$ and its two-binned version.} 
	\label{fig:QMwIM-fig2}
\end{figure}

In order to determine $\FQimbar$ in this case, we first note that only pure input states and projective measurements, whose elements lie in the equatorial plane in the Bloch-ball representation need to be considered (see Supplement for the proof). Hence, after fixing the measurement to $\Pi_{1(2),\phivec}=\proj{\pm}, \ket{\pm}=(\ket{0}\pm\ket{1})/\sqrt{2}$, the maximisation in \eqnref{eq:FI} simplifies to optimising over a single parameter $\phi$ of the input state $\ket{\psi}=(\ket{0}+\upi\upe^{-\upi\phi}\ket{1})/\sqrt{2}$, so that $\FQimbar=\max_{\varphi} F$ with
\begin{align}\label{eq:FNV}
	F=\sum_x \frac{\frac{1}{2}\big(p(x|1)-p(x|2)\big)^2 \cos^2\varphi}{p(x|1)+p(x|2)+\big(p(x|1)-p(x|2)\big)\sin\varphi}, 
\end{align}
and $\varphi:=\theta+\phi$. As neither $F$ nor $\FQimbar$ can be evaluated analytically due to the infinite summation in \eqnref{eq:FNV}, their values may only be approximated numerically by considering a sufficient cut-off---as done in \figref{fig:QMwIM-fig2} (see the solid and dashed black lines).

A systematic and practically motivated approach allowing to lower-bound well $F$ and $\FQimbar$ corresponds to grouping the infinite outcomes $X$ into a finite number of categories:~``bins''. Although complex ``binning'' strategies are possible (see Methods), the crudest one considers just two bins (2-bin)---an approach known as the ``threshold method" in the context of NV-readout~\cite{Jiang2009,Neumann2010}. The binary outcome $X'$ is then formed by interpreting all the photon-counts from $x=0$ up to a certain $x^*$ as $x'=1$, while the rest as $x'=2$. This results in an effective \emph{asymmetric bit-flip channel}~\cite{Cover1991}, $\cP$, mapping the ideal outcomes $I$ onto $X'$, which we depict in \figref{fig:QMwIM-fig1}(c) for the case of photon-counts following Poissonian distributions, upon defining
$\pp:=p(x'=1|1)=\sum_{x=x^*+1}^{\infty} p(x|1)$ and $\qq:=p(x'=2|2)=\sum_{x=0}^{x^*} p(x|2)$, as well as $\eta:= \pp+\qq-1$ and $\delta:= \pp-\qq$.

As a result, we can analytically compute for the 2-bin strategy both the corresponding imperfect QFI and the imperfect channel QFI as, respectively:
\begin{align}
\Ftwobin &= \frac{\eta^2\cos^2\varphi}{1-(\delta+\eta\sin\varphi)^2},\label{eq:F2bin0}\\
\Ftwobinmaxstar&=\max_\varphi \Ftwobin = \eta(\eta+\delta\sin \varphiopt) \nonumber\\
&= 1-\left(\sqrt{\pp\left(1-\qq\right)}+\sqrt{\qq\left(1-\pp\right)}\right)^{2},\label{eq:F2bin}
\end{align}
where the optimal angle parametrising the input state reads $\phiopt=\varphiopt-\theta$, with $\varphiopt=\sin^{-1} (\Theta)$ and
\begin{equation}
\Theta= \frac{1-\delta^2-\eta^2-\sqrt{(1-\delta^2-\eta^2)^2-4\delta^2\eta^2}}{2\delta\eta}. 
\label{eq:CosDeltaopt}
\end{equation}
In \figref{fig:QMwIM-fig2} we plot $F_{\text{2-bin}}$ that corresponds to $\Ftwobin$ being further maximised over the binning boundary $x^*$---it allows us to verify that $\phiopt$ provides indeed a very good approximation of the optimal input state. 

We close the analysis of imperfect measurements in the single-probe scenario by briefly discussing another general method to approximate $F$ and $\FQimbar$. It relies on a construction (see Methods for the full methodology) of a convergent hierarchy of lower bounds on the FI, $F^{(k)}\le F$, which are obtained by considering subsequent $2k$ moments of the probability distribution $\qphitheta$ describing the set of `observed' outcomes $X$, even if infinite \cite{Jarrett1984}.
In \figref{fig:QMwIM-fig2}, we present $F^{(1)}$ based on only first two moments of $\qphitheta$, which, however, contain most information about the estimated phase $\theta$, so that the method also predicts the optimal input state very well.

\subsection{Relations to quantum metrology with noisy encoding}
\begin{figure}[t!]  
    \includegraphics[width=0.325\textwidth]{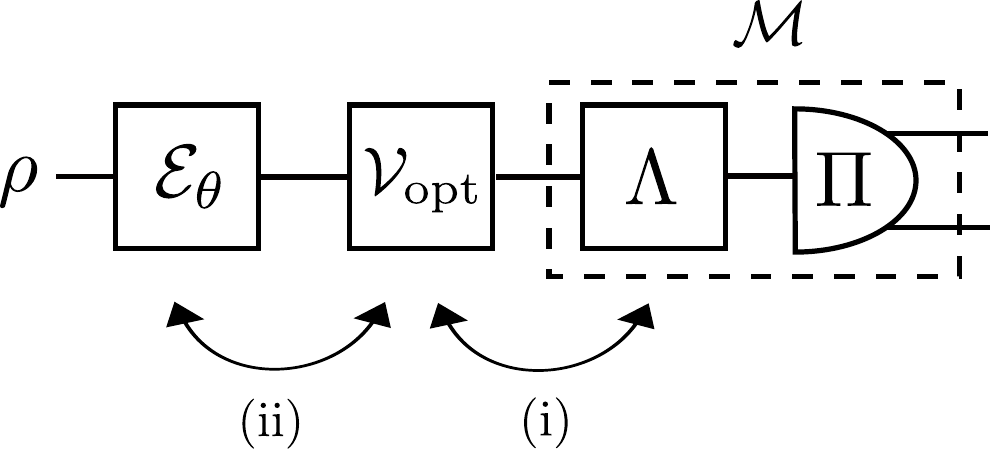}
    \caption{
    \textbf{The scheme of \figref{fig:QMwIM-fig1}(a) with the imperfect measurement $\cM$ decomposed according to its conjugate-map decomposition}, i.e.~$\cM=\Lambda^\dagger[\Pi]$ where $\Lambda$ is a quantum channel such that $\{\Pi_x\}$ forms a projective measurement in the output space of $\dimM$ `observable' outcomes. By $\opt{\cV}$ we denote the optimal unitary control required by the imperfect (channel) QFI $\FQim$ ($\FQimbar$), which in principle depends on the particular form of the input state $\rho$, encoding $\cE_\theta$ and the measurement $\cM$. Still, if (i):~$\opt{\cV}$ commutes with the map $\Lambda$, then $\FQim\le\FQ\!\left[\Lambda\circ\cE_\theta[\rho]\right]$ and, hence, $\FQimbar \le \bar{\FQ}[\Lambda\circ\cE_\theta]$. However, the latter is also true if (ii):~$\opt{\cV}$ commutes with the encoding $\cE_\theta$.
    } 
    \label{fig:QMwIM-fig3}
\end{figure}

Any imperfect measurement $\cM$ admits a \emph{conjugate-map decomposition}, $\cM=\Lambda^\dagger[\Pi]$, i.e.~all its elements can be expressed as $M_x=\Lambda^\dagger[\Pi_x]$, where $\Pi\sim\{\Pi_x\}_{x=1}^{\dimM}$ form a projective measurement in $\cH_{\dimM}$ and $\Lambda\!:\,\cB(\cH_d)\to\cB(\cH_{\dimM})$ is a quantum channel that may always be constructed (see Supplement), where  $\cB(\cH_{\ell})$ denotes the set of bounded linear operators on the Hilbert space $\cH_{\ell}$ of dimension $\ell$. Hence, given the channel $\Lambda$, for any two operators $A\in\cB(\cH_{d})$, $B\in\cB(\cH_{|X|})$, $\Lambda^{\dagger}$ is defined by $\Tr\{\Lambda^{\dagger}[B]A\}\equiv\Tr\{B\Lambda[A]\}$. This implies that any imperfect measurement $\cM$ can always be represented by the action of a fictitious channel $\Lambda$, followed by a projective (``ideal'') measurement $\{\Pi_x\}$ that acts in the space of the `observed' outcomes---compare \figref{fig:QMwIM-fig3} with \figref{fig:QMwIM-fig1}(a).

\begin{figure*}[t!]
    \centering
    \includegraphics[width=0.93\textwidth]{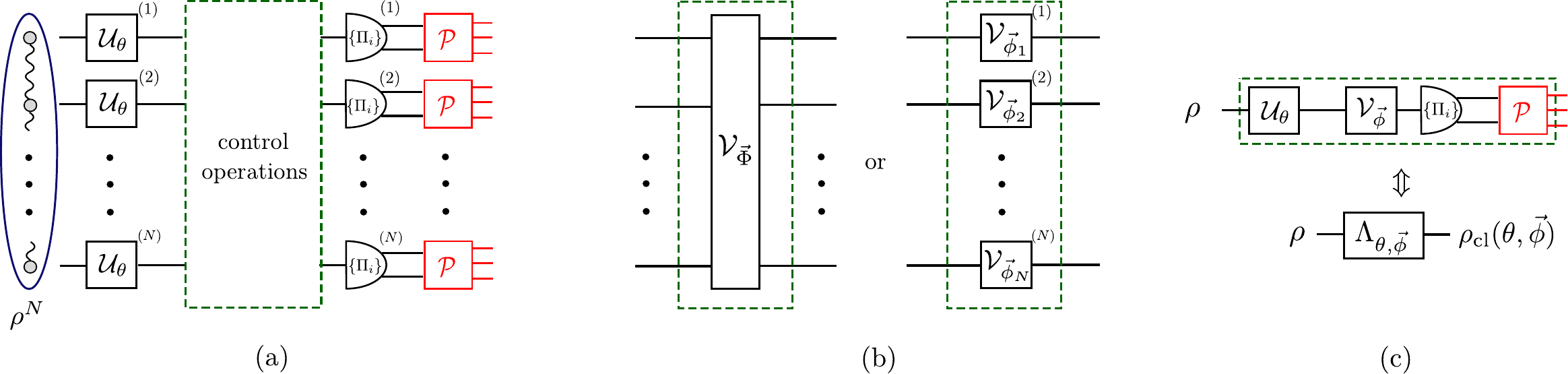}
    \caption{\textbf{(a) Canonical multi-probe scenario of quantum metrology with imperfect measurements}. $N$ probes, generally prepared in a entangled state, $\rho^N$, undergo identical parameter encoding, $\cU_\theta$, and are subject to identical local projective measurements $\{\Pi_i\}$, whose outcomes are affected by the noisy detection channel (stochastic map) $\cP$. In between the encoding and measurement, control operations are applied and optimised in order to compensate for local measurement imperfections, so that the minimal error in estimating $\theta$ can be attained. \textbf{(b)} \textbf{Control operations} in (a) may always be represented by a \emph{global} unitary transformation, $\VcalPhi$; or be rather constrained to a product of general \emph{local} unitaries, $\bigotimes_{j=1}^{N}\cV_{\phivec_j}^{(j)}$. \textbf{(c) Single-probe evolution as a \emph{quantum-classical channel}}, denoted as $\Lambda_{\theta,\phivec}$ that transforms the $d$-dimensional state $\rho$ of the probe into a classical state $\rhocl(\theta,\phivec)$ defined in a fictitious Hilbert space, whose dimension is specified by the number of outcomes of the noisy detection channel $\cP$.}
    \label{fig:QMwIM-fig4}
\end{figure*}        

However, the channel $\Lambda$ acts \emph{after} both the (here, arbitrary) parameter encoding $\cE_\theta$ and the optimal unitary control denoted by $\opt{\cV}$ in \figref{fig:QMwIM-fig3}, of which the latter generally depends on a particular form of all:~the input state $\rho$, the encoding $\cE_\theta$, and the imperfect measurement $\cM$. Only in the very special case when one can find $\Lambda$ such that it commutes with $\opt{\cV}$---case (i) in \figref{fig:QMwIM-fig3}---the problem can be interpreted as an instance of the ``standard'' noisy metrology scenario~\cite{Maccone2011}. It is so, as the corresponding imperfect QFI must then obey $\FQim\le\FQ\!\left[\Lambda\circ\cE_\theta[\rho]\right]$, which upon maximisation over input states implies also $\FQimbar \le \bar{\FQ}[\Lambda\circ\cE_\theta]$ for the imperfect channel QFI. The latter inequality may be independently assured if the optimal control $\opt{\cV}$ commutes with the parameter encoding $\cE_\theta$ instead---case (ii) in \figref{fig:QMwIM-fig3}---as this allows $\opt{\cV}$ to be incorporated into the maximisation over $\rho$.

Although we demonstrate that both above bounds can be computed via a semi-definite programme via a `seesaw' method (see Supplement), 
which may also incorporate optimisation over all valid conjugate-map decompositions of $\cM$, their applicability is very limited. In particular, their validity can only be \emph{a priori} verified if the problem exhibits some symmetry---the $G$-covariance that we discuss in Methods---that must ensure the commutativity (i) or (ii) of the optimal control $\opt{\cV}$ in \figref{fig:QMwIM-fig3}, without knowing its actual form.

Even in the simple qubit case with unitary encoding, $U_\theta=\upe^{\upi \frac{\pauliz}{2}\theta}$, and the binary outcome of a projective measurement being randomly flipped---equivalent to the NV-motivated scenario with 2-binning that yields the imperfect channel QFI \eref{eq:F2bin}---$\Lambda$ must be \emph{phase-covariant}~\cite{Smirne2016} for $\FQim\le\FQ\!\left[\Lambda\circ\cU_\theta[\rho]\right]$ to hold (see Supplement). While this may be satisfied \emph{only if} there exists some $\phi\in[0,2\pi)$ such that $\frac{4\eta^{2}}{\sin^{2}\phi}+\frac{\delta^{2}}{\cos^{2}\phi}\leq1$, the resulting bound is tight \emph{only} for symmetric bit-flips ($\delta=0$). Furthermore, considering already a two-qubit system with local imperfect measurements of this type, $\opt{\cV}$ ceases to commute with the encoding, $U_\theta^{\otimes2}$, so that even $\FQimbar \le \bar{\FQ}[(\Lambda\circ\cU_\theta)^{\otimes 2}]$ cannot be assured (see Supplement). This opens doors to circumvent the no-go theorems of quantum metrology with uncorrelated noise~\cite{Escher2011,Demkowicz2012}, as exploited in the multi-probe schemes discussed below.

%


\subsection{Multi-probe scenarios}	           
We turn now our focus to \emph{multi-probe} scenarios of quantum metrology, in particular, the canonical one in which the parameter is encoded locally onto each probe, so that the inter-probe entanglement can prove its crucial usefulness, e.g.~to reach the HS of precision, whereas the ideal projective measurement can be considered to be \emph{local} without loss of generality~\cite{Giovannetti2006}. While including imperfect measurements into the picture, we depict such a scheme in \figref{fig:QMwIM-fig4}(a), in which $N$ qudits are prepared in a (possibly entangled) state $\rho^N$ before undergoing a unitary transformation $\cU_\theta^{N}\sim\{U_\theta^{N}\}$, so that $\rho^N(\theta)=U_\theta^{N}\rho^N U_\theta^{N\dagger}$, where $U_\theta^{N}=U_\theta^{\otimes N}$ in the canonical scenario~\cite{Giovannetti2006}. Each probe is still measured independently but in an imperfect manner, so that the overall POVM corresponds now to $\cM^{\otimes N}\!\sim \{M_{\boldsymbol{x}}\}=\{M_{x_1}\otimes M_{x_2}\otimes\dots\otimes M_{x_N}\}$. For instance, as shown in \figref{fig:QMwIM-fig4}(a), each $\{M_{x}\}$ may be obtained by randomising outcomes of projectors $\{\Pi_{i}\}$ according to some stochastic map $\cP\sim \{p(x|i)\}$ representing the noisy detection channel.

In order to compensate for measurement imperfections, we allow for control operations to be performed on all the probes before being measured. However, we differentiate between the two extreme situations, see \figref{fig:QMwIM-fig4}(b), in which the control operations can act collectively on all the probes---being represented by a \emph{global} unitary channel $\VcalPhi\sim\{\VPhi\}$ specified by the vector of parameters $\Phivec$;~or can affect them only locally---corresponding to a product of (possibly non-identical) \emph{local} unitary channels $\bigotimes_{\ell=1}^N\cV_{\phivec_\ell}^{(\ell)}$ with $\cV_{\phivec_\ell}\sim\{V_{\phivec_\ell}\}$, each of which is specified by a separate vector of parameters $\phivec_\ell$.

As in the general case, the QCRB \eqref{eq:QCRB} determines then the ultimate attainable sensitivity. In particular, given a large number $\nu$ of protocol repetitions, the MSE $\Delta^2\tilde{\theta}_N$, which now depends on the number of probes $N$ employed in each protocol round, is ultimately dictated by the lower bounds:
\begin{align}\label{eq:CRBN}
	\nu \Delta^2\tilde{\theta}_N \geq \frac{1}{\FQim_{N}}  \geq \frac{1}{\FQimbar_N},
\end{align}
where
\begin{align} \label{eq:FIN}
\FQim_N:=\max_{\Phivec\,\mathrm{or}\,\{\phivec_\ell\}_\ell} F_N
\quad\text{and}\quad
\FQimbar_N:=\max_{\rho^N}\, \FQim_N[\rho^N(\theta)], 
\end{align}
are again the imperfect QFI and the imperfect channel QFI, respectively, see \eqnref{eq:FQim+FQimbar}, but evaluated now for the case of $N$ probes. Similarly, $F_N$ is the $N$-probe version of \eqnref{eq:FI}, whose maximisation over all local measurement settings becomes now incorporated into the optimisation over control operations, either global $\Phivec$ or local $\{\phivec_\ell\}_\ell$.

\subsubsection{Global control operations}
We first consider multi-probe scenarios in which one is allowed to perform global unitary control operations, $\VcalPhi$ in \figref{fig:QMwIM-fig4}(b), to compensate for measurement imperfections. In such a case, let us term an imperfect measurement $\cM$ \emph{information-erasing}, if all its elements $M_x$
are proportional to identity, so that no information can be extracted. Then,
it follows from \lemref{lem:FIwIM} that:
\begin{theorem}[Multi-probe metrology scheme with global control]
	\label{thm:HLwIM-Global}
	For any pure encoded $N$-probe state $\psi^N(\theta)=\proj{\psi^N \left( \theta \right) }$, and any imperfect measurement ${\cM^{\otimes N}}$ that is not information-erasing and operates independently on each of the probes, the imperfect QFI converges to the perfect QFI for large enough $N$:
\begin{equation}
\FQim_{N} \underset{N\to\infty}{=}\FQ[\psi^N(\theta)].
\label{eq:Fim_conv_F_perf}
\end{equation}
\end{theorem}%
We differ the 
proof to the Supplement, where we explicitly show that for any non--information-erasing imperfect measurement $\cM$, the resulting constant factor 
$\gamma_{\cM^{\otimes N}}$ appearing in \lemref{lem:FIwIM}---which now depends and must monotonically grow~%
\footnote{As generally $\gamma_{\cM}\leq\gamma_{\cM\otimes I}\leq\gamma_{\cM\otimes\cM}$, see the Supplement for the proof.} 
with the probe number $N$---satisfies $\gamma_{\cM^{\otimes N}}\rightarrow1$ as $N\rightarrow\infty$. Intuitively, recall that $\gamma_{\cM^{\otimes N}}$ quantifies how well one can distinguish at best some two orthogonal states $\ket{\xi^N}$ and $\ket{\xi^N_{\perp}}$. Hence, we can always consider $\ket{\xi^N}=\ket{\xi}^{\otimes N}$ and $\ket{\xi_\perp^N}=\ket{\xi_\perp}^{\otimes N}$, whose effective ``overlap'' for the resulting imperfect measurement ${\cM^{\otimes N}}\!\sim \{M_{\boldsymbol{x}}\}$ reads
\begin{equation}
\sum_{\boldsymbol{x}}\sqrt{\langle\xi^{N}|M_{\boldsymbol{x}}|\xi^{N}\rangle\langle\xi_{\perp}^{N}|
 M_{\boldsymbol{x}}|\xi_{\perp}^{N}\rangle} 
 =
 c^{N}
\end{equation}
with $c=\sum_x\sqrt{\langle\xi|M_{x}|\xi\rangle\langle\xi_{\perp}|M_{x}|\xi_{\perp}\rangle}<1$, and is thus assured to be exponentially decaying to zero with $N$. This implies perfect distinguishability and, hence, attaining perfect QFI as $N\to\infty$, with the convergence rate depending solely on the single-probe POVM, $\cM\sim\{M_x\}$.

More formally, we establish the existence of a global unitary $\VPhi$, such that the following lower bound holds:
\begin{equation} 
\label{eq:FNbound-1}
		F_N(\VPhi)
		\geq\;
		 (1-c^N)\, \FQ[\psi^N(\theta)] :=\FNdown(\VPhi),
\end{equation}
where $0\leq c\leq1$ depends only on $\cM$ and the unitary $\VPhi$ used. As $\gamma_{\cM^{\otimes N}}\ge 1-c^N$ should be interpreted as a distinguishability measure similar to the ones of quantum hypothesis testing~\cite{audenaert_discriminating_2007,Calsamiglia2008,audenaert_asymptotic_2008}, it is rather its \emph{asymptotic rate exponent}, $\Gamma_\cM:=\lim_{N\to\infty}-\frac{1}{N}\ln(1-\gamma_{\cM^{\otimes N}})$, that quantifies metrological capabilities of $\cM^{\otimes N}$ in the asymptotic $N$ limit. Hence, we formally determine the lower bound $\chi\le\Gamma_\cM$, where $\chi:=-\ln c$. Nonetheless, the form of $\VPhi$ we use, and the discussion on its optimality we leave to the Supplement. Crucially, in case of the canonical multi-probe scenario of \figref{fig:QMwIM-fig4}(a), we may directly conclude from \eqnref{eq:FNbound-1} that:
\begin{corollary}[Go-theorem for the HS with imperfect measurements and global control]
	\label{corr:HLwIM-Global}
	For any non--information-erasing detection channel, the HS ($\Delta^2\tilde{\theta}_N\sim1/N^2$) can always be asymptotically attained, by choosing any global unitary $\VPhi$ such that \eqnref{eq:FNbound-1} holds, and any pure input state with QFI $\mathcal{F}[\psi^N(\theta)]\sim N^2$ for $N\to\infty$.
\end{corollary}
Note that in the view of relations to ``standard'' noisy quantum metrology protocols~\cite{Maccone2011}, $\VPhi$ required by \eqnref{eq:FNbound-1} must not allow for its commutation as in (i) or (ii) of \figref{fig:QMwIM-fig3}---as shown explicitly in the Supplement already for two qubits ($N=2$), each measured projectively with bit-flip errors---so that the corresponding no-go theorems~\cite{Escher2011,Demkowicz2012} forbidding the HS no longer apply.

\begin{figure}[t!] 	
	\includegraphics[width=0.45\textwidth]{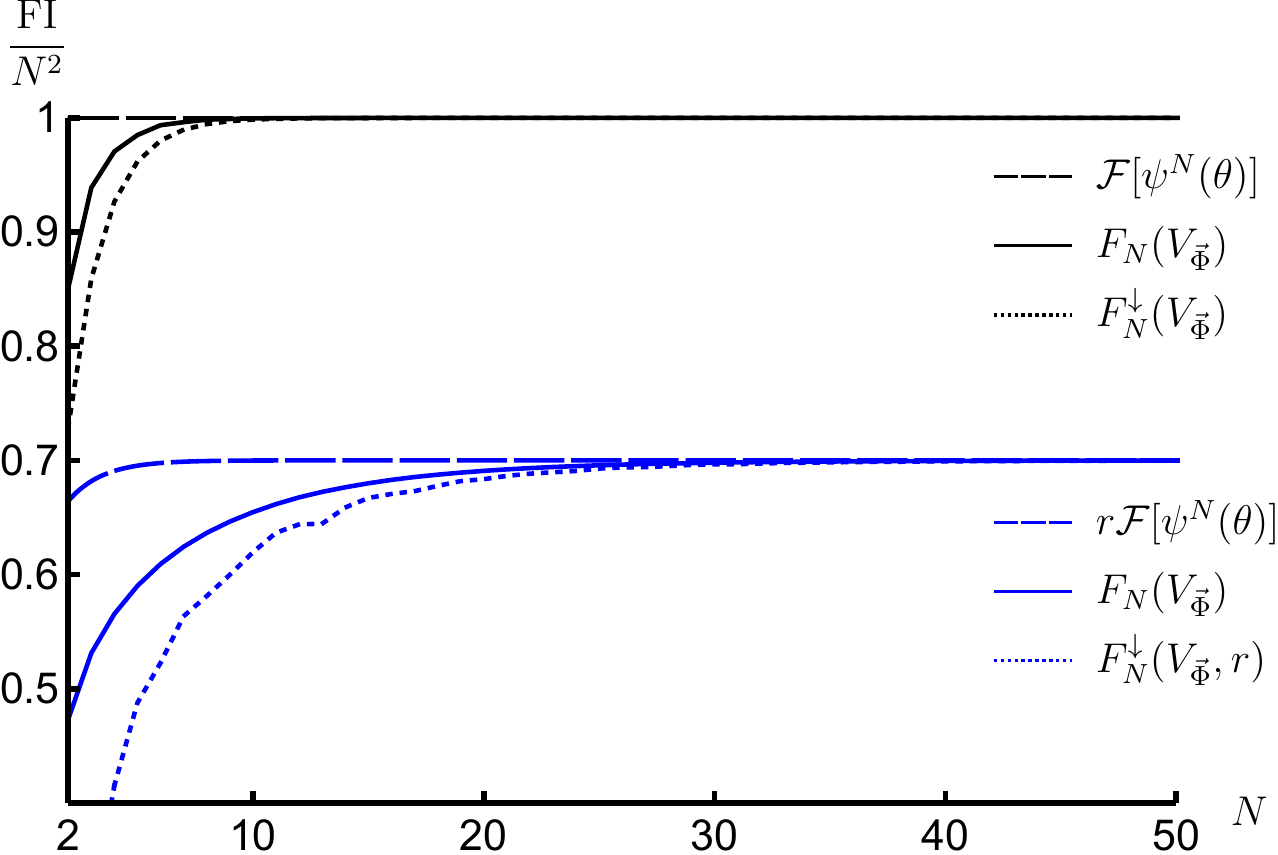}
	\caption{
		\textbf{
        Attaining the ultimate HS of precision in presence of measurement imperfections and \emph{global} unitary control}. The case of phase estimation with $N$ qubit probes is considered, which are initialised in a GHZ state, whereas the outcomes of ideal local measurements undergo an asymmetric bit-flip channel with $\pp=0.95$ and $\qq=0.9$.
		The \emph{black solid line} is the exact (numerical) FI for a specific choice of the control global unitary $\VPhi$, while the \emph{black dotted line} is its lower bound $\FNdown(\VPhi)$ defined in \eqnref{eq:FNbound-2}---both converge to the optimal achievable $\mathcal{F}[\psi^N(\theta)]=N^2$ (\emph{black dashed line}). The family of lines in \emph{blue} are the corresponding FIs for the case of a distorted GHZ state, with an admixture of white noise (with $r=0.7$ in \eqnref{eq:werner_state}) being added.
		}
	\label{fig:QMwIM-fig5}
\end{figure}

However, one should also verify whether the above corollary, relying on convergence \eref{eq:Fim_conv_F_perf}, is not a ``measure-zero'' phenomenon. In particular, whether, if the assumption of state purity in \eqnref{eq:Fim_conv_F_perf} is dropped, the preservation of different scalings in $N$ is still maintained. That is why, we prove the robustness of \thmref{thm:HLwIM-Global} by generalising it to the case of noisy (mixed) input states, which after $\theta$-encoding take the form: 
\begin{equation}
\rho_r^N(\theta) \;:=\; r\psi^N(\theta)+\frac{1-r}{d^N}\id_{d^N},	
\label{eq:werner_state}
\end{equation}
and can be interpreted in the canonical multi-probe scenario of \figref{fig:QMwIM-fig4}(a) as \emph{white noise} (or \emph{global depolarisation}~%
\cite{footnote5}) 
of fixed strength $0<r<1$ being admixed to a pure input state $\psi^N$. 
Nonetheless, all our claims hold if one replaces $\id_{d^N}/d^N$ in \eqnref{eq:werner_state} by any product state. In particular, we prove (see Supplement) the following lemma:
\begin{lemma}[Robustness of \thmref{thm:HLwIM-Global}]
	\label{lem:robustness_HLwIM-Global}
	For any mixed encoded state $\rho_{r}^N(\theta)$ of the form \eqref{eq:werner_state}, and any detection channel that is non--information-erasing, the imperfect QFI about $\theta$ converges to the perfect QFI as $N \rightarrow \infty$: 
\begin{equation} \label{eq:FI_ineq_sand}
\FQim_N\underset{N\rightarrow\infty}{=}\FQ\!\left[\rho_{r}^N(\theta)\right]\underset{N\rightarrow\infty}{=}r\,\FQ\!\left[\psi^{N}(\theta)\right].
\end{equation}
\end{lemma}
\noindent The proof is very similar to that of \thmref{thm:HLwIM-Global}, while it relies also (see \eqnref{eq:FNbound-1}) on existence of lower bounds $\FQim_N\geq F_N(\VPhi) \geq\FNdown(\VPhi,r)$, where 
$\FNdown(\VPhi,r)\to r\mathcal{F}[\psi^N(\theta)]$ as $N\to\infty$. 
Focussing on the asymptotic scaling of precision in the canonical multi-probe scenario, it directly follows that despite the white noise, if $\FQ[\psi^N(\theta)]\sim N^2$, then $\FQim_N\sim r N^2$ and the HS is still attained.

As an example, let us explicitly discuss how the \thmref{thm:HLwIM-Global} and \lemref{lem:robustness_HLwIM-Global} apply in the canonical multi-qubit scenario of \figref{fig:QMwIM-fig4}(a), in which $U_\theta^{N}=U_\theta^{\otimes N}$ with $U_\theta=\upe^{\upi h\theta}$ and $h=\pauliz/2$
, while measurement imperfections arise due to a noisy detection channel $\cP$ that flips the binary outcome for each qubit with probabilities $\pp$ and $\qq$, respectively---as depicted within the inset of \figref{fig:QMwIM-fig1}(c). Then, by initialising the probes in the GHZ state, $\psi^N = \ketbra{\psi^N}{\psi^N}$ with $\ket{\psi^N}=(\ket{0}^{\otimes N}+\ket{1}^{\otimes N})/\sqrt{2}$, we find (see Supplement) a global control unitary $\VPhi$ for which
 the lower bound in \eqnref{eq:FNbound-1} reads:
\begin{align} \label{eq:FNbound-2}
\FNdown(\VPhi)=&\, N^2\;[1-(\sqrt{\pp(1-\qq)}+\sqrt{\qq(1-\pp)})^N]\nonumber\\ 
=  &\, N^2\;[1-\upe^{-\chi N}]
\end{align}
with $\chi\approx\frac{1}{4}\frac{\left(\pp+\qq-1\right)^{2}}{\pp\left(1-\pp\right)+\qq\left(1-\qq\right)}$. The ultimate precision with $N$ is attained and, hence, the HS---as illustrated in \figref{fig:QMwIM-fig5} for $\pp=0.95$ and $\qq=0.9$. Furthermore, we repeat the above procedure of finding $\VPhi$ to attain the ultimate asymptotic precision for an input GHZ state subjected to white noise according to \eqnref{eq:werner_state}. In such a setting, we determine analytically the required lower bound $\FNdown(\VPhi,r)$ (see Supplement), which we similarly depict in \figref{fig:QMwIM-fig5} for $r=0.7$, together with the exact behaviour of $F_N(\VPhi)$ determined numerically. Note that an expression similar to \eqnref{eq:FNbound-2} has been established for the noisy detection channel corresponding to Gaussian coarse-graining~\cite{Frowis2016, Haine2018}, while in Methods we derive its form for lossy photonic interferometry with dark counts.

\subsubsection{Local control operations}
We next turn our attention to canonical multi-probe scenarios with unitary encoding, $\rho^N(\theta)=\cU_\theta^{\otimes N}[\rho^N]$, in which only local control operations are allowed, $\otimes_{\ell=1}^N\cV_{\phivec_\ell}^{(\ell)}$ with every $\cV_{\phivec_\ell}\sim\{V_{\phivec_\ell}\}$ in \figref{fig:QMwIM-fig4}(b), in order to verify whether these are already sufficient to compensate for measurement imperfections. We denote the corresponding imperfect channel QFI as $\FQimbarloc_N$.
Crucially, in such a case the quantum metrology protocol of \figref{fig:QMwIM-fig4}(a) can be recast using the formalism of \emph{quantum-classical channels}~\cite{Holevo1998}. For each probe we introduce a fictitious $|X|$-dimensional Hilbert space spanned by orthogonal states $\ket{x}$ that should be interpreted as flags marking different outcomes $x$ being observed. As a result, focusing first on the evolution of a single probe illustrated in \figref{fig:QMwIM-fig4}(c), the observed outcome of the imperfect measurement may be represented by a classical state $\rhocl(\theta, \phivec)=\sum_x \qxphitheta\ket{x}\bra{x}$, with the transformation $\rho\rightarrow\rhocl(\theta, \phivec)=\Lambda_{\theta,\phivec}[\rho]$ governed by the quantum-classical channel $\Lambda_{\theta,\phivec}$. Then, in the canonical multi-probe scenario of \figref{fig:QMwIM-fig4}(a), each of the $N$ probes is independently transformed by the quantum-classical channel 
$\Lambda_{\theta,\phivec_\ell}=\Lambda_\cM \circ \cV_{\phivec_\ell} \circ \cU_\theta$ (see Supplement for the explicit form of $\Lambda_\cM$), 
and the overall input state undergoes $\rho^N\rightarrow\bigotimes_{\ell=1}^N\Lambda_{\theta,\phivec_\ell}^{(\ell)}[\rho^N]=\rhocl^N(\theta, \{\phivec_\ell\})$, where the output classical state $\rhocl^N(\theta, \{\phivec_\ell\})$ is now diagonal in the total $N\times|X|$-dimensional fictitious Hilbert space---describing the probability distribution of all the $N$ measurement outcomes. By treating quantum-classical channels as a special class of quantum maps that output diagonal states in a fixed basis, we apply the \emph{channel extension} (CE) method introduced in Refs.~\cite{Fujiwara2008,Demkowicz2012,Kolodynski2013} in order to construct the so-called \emph{CE-bound}, i.e.~
\begin{align}\label{eq:CEbound}
F_N\leq \FNCE(\{\phivec_\ell\}).
\end{align}
While leaving the technical derivation and expression of $\FNCE(\{\phivec_\ell\})$ to Methods,
let us emphasise that the CE-bound \eqref{eq:CEbound} is independent of the probes' state $\rho^N$ and allows even for extending---hence, the name---them to include extra $N$ ancillae, which do not undergo the parameter encoding but can be prepared in a state entangled with the probes before being (ideally) measured to further enhance the precision. Still, the bound \eqref{eq:CEbound} depends, in principle, on the setting of each (local) measurement $\phivec_\ell$, as well as the parameter $\theta$ itself. Nonetheless, we prove (see Methods for the prescription and Supplement for further details) the following lemma:
\begin{lemma}[Linear scaling of the asymptotic CE bound]
	\label{lem:no-go-QMwIM}
For unitary encoding $U_\theta^{N}=U_\theta^{\otimes N}$ with $U_\theta=e^{\upi h\theta}$, we may further define the asymptotic CE bound $\FNCEas$, which satisfies $F_N\leq \FNCE(\{\phivec_\ell\})\leq \FNCEas(\{\phivec_\ell\})$ and  $\lim_{N\to\infty}\FNCE=\FNCEas$, whenever there exists a set of Hermitian operators $\{A_{x}(\phivec_{\ell})\}_{x}$ such that for each $\phivec_{\ell}$: 
\begin{align}
\label{eq:no-go-conditions}
h=\sum_{x=1}^{|X|} V_{\phivec_{\ell}}^{\dagger}\sqrt{M_{x}}A_{x}(\phivec_{\ell})\sqrt{M_{x}}V_{\phivec_{\ell}}.
\end{align}
Moreover, upon optimising $\FNCEas$  
over all local control unitaries, we obtain
\begin{align}
\FQimbarloc_N \;\leq\; \bar{F}_N^{\mathrm{(CE,as)}}
&:=\;
\max_{\{\phivec_\ell\}}\FNCEas(\{\phivec_\ell\}) =4Nc, 
\label{eq:CEasboundmax}
\end{align}
where $0\le c < \infty$ is an $N$-independent constant factor that is fully determined by a single copy of the channel $\Lambda_{\theta,\phivec}$, and \eqnref{eq:CEasboundmax} applies for any local control $\{\phivec_\ell\}$~\cite{footnote2}.
\end{lemma}

Although the condition \eqref{eq:no-go-conditions} may look abstract, it actually has an intuitive meaning, when considering imperfect measurement that arises due to some noisy detection channel $\cP$, such that $M_{x}=\sum_{i}p(x|i)\Pi_{i}$. Let us call a detection channel $\cP$ \emph{non-trivial} if its transition probabilities $p(x|i)$ are such that for all pairs of `inaccessible' outcomes $i,i'$, 
there is at least one `observable' outcome $x$ such that $p(x|i)p(x|i')>0$. Then, we have (see Supplement for an explicit proof):

\begin{corollary}[No-go theorem for HS with imperfect measurements and local control]
	\label{HLwIM-5}
 	Consider the canonical multi-probe scenario depicted in \figref{fig:QMwIM-fig4}(a) that incorporates a non-trivial noisy detection channel $\cP$, whose impact one may only compensate for by means of local control unitaries, see \figref{fig:QMwIM-fig4}(b). Then, the condition \eqref{eq:no-go-conditions} can always be satisfied and, as \eqnref{eq:CEasboundmax} implies that $\Delta^2\tilde{\theta}_N\ge \varepsilon/N$ for some $\varepsilon>0$, the HS cannot be attained with the MSE following at best the SS.
\end{corollary}

\begin{figure}[t!]
	\centering
	\includegraphics[width=0.45\textwidth]{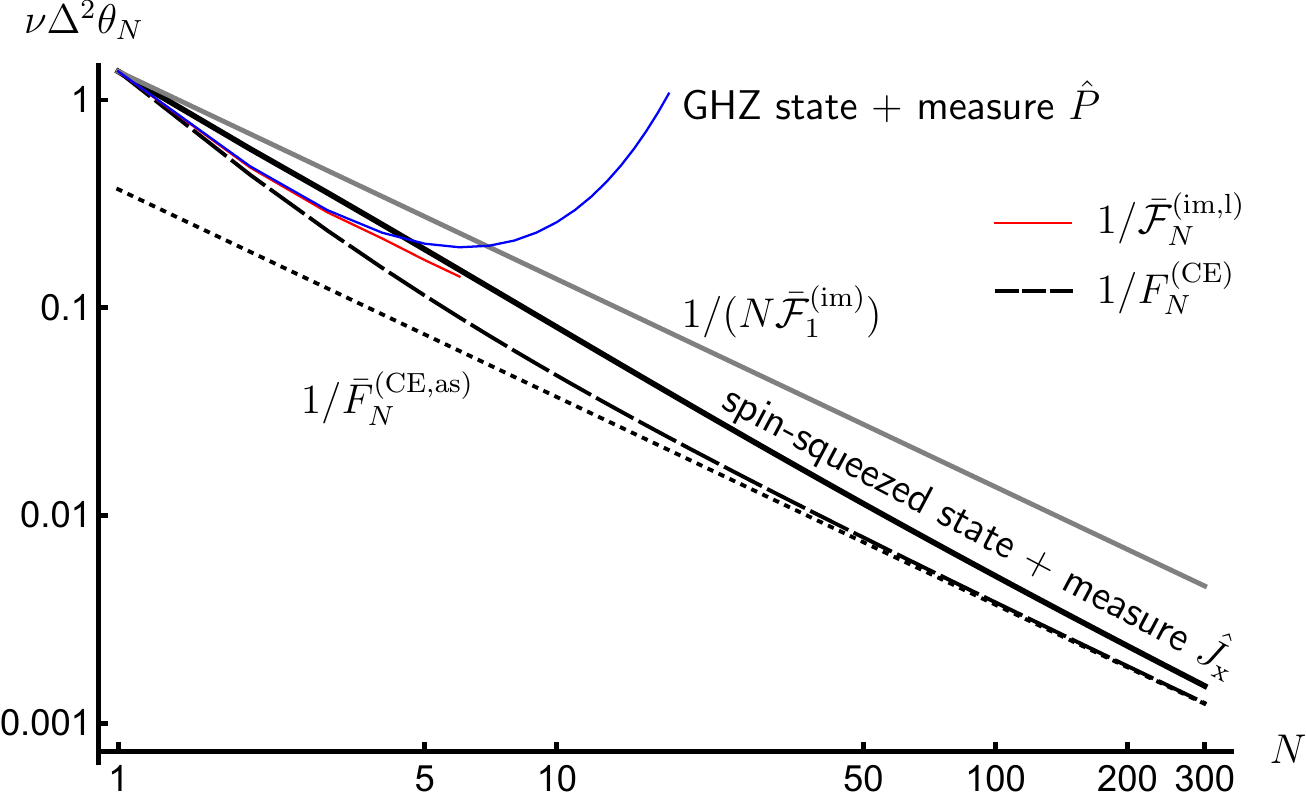}
	\caption{\textbf{
Attaining the optimal SS of precision in presence of measurement imperfections and \emph{local} unitary control.} The \emph{thick solid black line} depicts the MSE in estimating phase $\theta$ from an imperfect measurement of the angular momentum operator $\Jx$, while the $N$ qubit probes are prepared in a one-axis spin-squeezed state~\cite{Kitagawa1993}, optimised by local control (see Methods). The noisy detection channel corresponds to an asymmetric bit-flip map with probabilities $\pp=0.95$ and $\qq=0.9$. The \emph{dotted black line} denotes the asymptotic CE bound with $\bar{F}_N^{\mathrm{(CE,as)}}$ given by \eqnref{eq:FNCEas}, while the \emph{thin red solid line} is the exact achievable precision $1/\FQimbarloc_N$, which we compute numerically up to $N=6$ by brute-force heuristic methods. The \emph{dashed black line} corresponds to $1/F_N^{\mathrm{(CE)}}$ in \eqnref{eq:CEbound} applicable in absence of control ($\forall_\ell:\;V_{\phivec_{\ell}}^{(\ell)}=\id$). At small $N$ ($\lesssim 4$), the ultimate precision can be attained by performing (imperfect) parity measurements with input GHZ states (\emph{thin blue line}). For comparison, we also include the optimal precision attained by uncorrelated probe states, $1/(N\FQimbar_1)$, (\emph{solid gray}).}
	\label{fig:QMwIM-fig6}
\end{figure}

In order to illustrate our result, let us consider again the canonical multi-qubit scenario with every qubit being subject to a projective measurement, whose outcome suffers an asymmetric bit-flip noise parametrised by $\pp$ and $\qq$, see \figref{fig:QMwIM-fig4}. We evaluate the corresponding asymptotic CE bound (see Supplement):
\begin{align}\label{eq:FNCEas}
\bar{F}_N^{\mathrm{(CE,as)}}=N\Big(\frac{\sqrt{\pp(1-\pp)}-\sqrt{\qq(1-\qq)}}{\pp-\qq}\Big)^2,
\end{align}
which, however, must be further verified to be asymptotically attainable. Indeed, in Methods we show this to be true even for a simple inference strategy, in which an (imperfect) measurement of the total angular momentum $\Jx$ is performed with the $N$ probes prepared in an \emph{one-axis spin-squeezed state}~\cite{Kitagawa1993} with the correct amount of squeezing and rotation, as illustrated graphically in \figref{fig:QMwIM-fig6}. Moreover, we demonstrate that up to $N\lesssim 4$, the ultimate precision determined numerically can also be attained by considering the parity observable incorporating the imperfect measurement, with probes being prepared in a GHZ state rotated at an optimal angle (see Supplement for details). 

Note that our recipe to construct the bound \eref{eq:CEasboundmax} applies generally, not relying on any properties of the imperfect measurement $\cM$, e.g.~see Methods for its application to the photonic setting in which $\dimM>d$. Still, for the above multi-qubit case with detection bit-flip noise, for which $d=\dimM=2$, we observe (see Supplement) that the corresponding bound \eref{eq:FNCEas} can be postulated based on a conjecture of the optimal local controls corresponding to \emph{phase-covariant} rotations~\cite{Smirne2016}, what allows then to invoke the results of ``standard" noisy metrology~\cite{Maccone2011}.

\section{Discussions}
We have analysed 
the impact of measurement imperfections on quantum metrology protocols and, in particular, the prospects of recovering the ideal quantum enhancement of sensitivity, e.g.~the Heisenberg scaling (HS) of precision, despite the readout noise. The contrasting results obtained with global or local control operations can be understood by the following simple intuition. 

With global control operations available, one may effectively construct a global measurement tailored to the two-dimensional subspace containing the information about any tiny changes of the parameter. Importantly, thanks to the exponential increase of the overall dimension
with the number of probes, one may then distinguish (exponentially) better and better the two states lying in this two-dimensional subspace, 
within which the effective amount of readout noise diminishes, and the perfect optimal scaling prevails.
Our work, thus, motivates explicitly the use of variational approaches in identification of such global unitary control not only at the level of state preparation~\cite{kaubruegger_variational_2019,koczor_variational-state_2020}, but also crucially in the optimisation of local measurements~\cite{kaubruegger_quantum_2021,marciniak_optimal_2021}. On the other hand, it demonstrates that control operations form the key building-block in fighting measurement imperfections in quantum metrology. Although we have provided control strategies that allow to maintain the HS both in the qubit and photonic settings, these employ $N$-body interactions, while it is known that $2$-body interactions suffice in presence of Gaussian blurring arising in cold-atom experiments~\cite{Davis2016,Frowis2016,Nolan2017,Haine2018}. Thus, we believe that our results open an important route of investigating the complexity of such global control required, depending on the form of the readout noise encountered.

On the contrary, there is no exponential advantage gained when only local control operations are available. Hence, as the overall amount of noise also rises limitlessly as we increase the number of probes, the asymptotic scaling of sensitivity is constrained to be classical. Note that this conclusion is valid also in the Bayesian scenario, as by the virtue of the Bayesian CRB~\cite{Trees1968} also the \emph{average} MSE is then lower-bounded by $\langle\Delta^2\tilde{\theta}\rangle \gtrapprox 1/\langle \FQimbar_N \rangle \gtrsim 1/N$, where $\langle\dots \rangle$ denotes now the averaging over some prior distribution of the parameter.

Finally, although we have primarily focussed here on phase-estimation protocols, let us emphasise once more that \thmref{thm:HLwIM-Global} applies to any quantum metrology scheme involving pure states and imperfect measurements. Hence, it holds also when sensing, e.g., `critical' parameters at phase transitions with noisy detection~\cite{Mirkhalaf2021}. Still, generalisation to the case with mixed states (beyond product-state admixtures) remains open. This would allow us, for instance, to approach quantum thermometry protocols utilising thermalised (Gibbs) probe states with the temperature being then estimated despite coarse-graining of measurements~\cite{Hovhannisyan2021}. In such cases, one should then also characterise the (mixed) states for which the imperfect QFI is actually guaranteed to converge to the perfect QFI in the asymptotic $N$ limit.

\section{Methods}
\begin{figure*}[t!]
	\centering
	\includegraphics[width=0.75\textwidth]{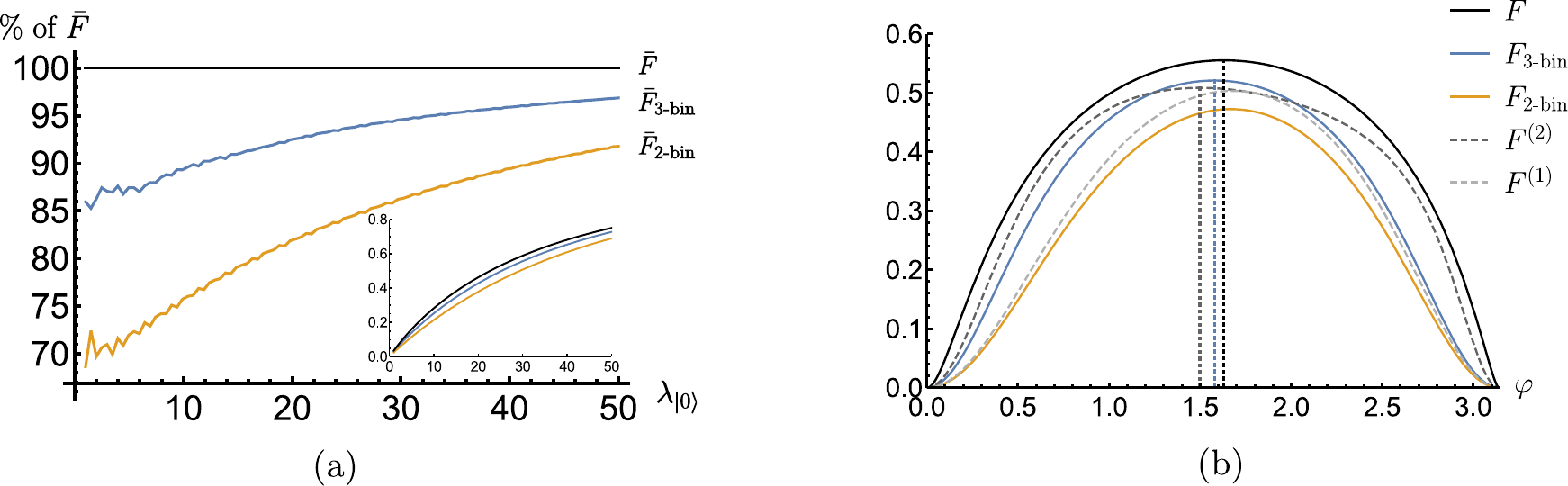}
	\caption{\textbf{FI for phase $\theta$ with noisy measurement with Poissonian noise.} (a) \textbf{With binning strategies}: the corresponding FI---$\Ftwobinmax$ (orange) and $\Fthreebinmax$ (blue) with optimal binning into two and three categories, respectively---compared against the exact $\FQimbar$ (ratio in \%) computed by performing large enough cut-off ($x\le100$) in \eqnref{eq:FNV}. The ratio of means for the Poissonnian distributions is set to $\lambdaone/\lambdazero=0.65$~\cite{Boss2017}, while $\lambdazero$ is varied. The inset shows the absolute values of FIs. (b) \textbf{With binning strategies and the moment method}: The FIs ($F$--black, $F_{2\text{-bin}}$--orange, $F_{3\text{-bin}}$--blue) presented now as a function of the input state angle $\phi=\varphi-\theta$ (for $\lambdaone/\lambdazero=0.65$ and $\lambdazero=27$~\cite{Boss2017}) in comparison to the lower bounds on $F$ constructed by taking into account up to the second ($\Fdown{1}$, light gray dash) and fourth moment ($\Fdown{2}$, dark gray dash) of the distribution describing the observed outcomes, $\qphitheta$. The vertical dotted lines indicate the  (optimal) state angle at each of the respective quantities is maximised. Note that when the measurement is perfect, the FI is unity for all choices of the angle $\varphi$ (not shown).}
	\label{fig:QmwIM-fig7}
\end{figure*}


\subsection{Phase sensing with an NV centre}    

Within this protocol, the NV centre is firstly initialised into some superposition 
state $\rho=\ket{\psi}\bra{\psi}$ of the $m_s=0$ (corresponding to $\ket{0}$) and $m_s=1$ (corresponding to $\ket{1}$) ground-state energy levels with help of a Ramsey pulse.
The NV spin is then used to sense a magnetic field of strength $B$ in the $\mathrm{z}$-direction for time $t$ (usually chosen to be as long as the decoherence allows for, i.e.~$T_2^*$ or $T_2$ for either static or alternating fields), gaining the relative phase $\theta=-t\gamma B$, where $\gamma$ is the gyromagnetic ratio characteristic to the NV centre~\cite{Taylor2008,santagati_magnetic-field_2019}. 
For our purpose we assume the evolution time to be perfectly known (and so the gyromagnetic ratio), so that the problem of estimating the field strength $B$ is effectively equivalent to estimating the relative phase $\theta$. Effectively then, the encoding channel is $\cU_\theta\sim \{U_\theta=\upe^{\upi h\theta}\}$, with $h=\pauliz/2$, where $\sigma_{\ell}$ is the usual Pauli-$\ell$ operators with $\ell=\mathrm{x,y,z}$. 

In order to read out $\theta$, a measurement is performed on the NV spin. Since the energy levels are fixed and not directly accessible, a microwave pulse is again applied to rotate the qubit basis, such that the phase is now carried in state populations instead. Afterwards, the NV-spin is optically excited, so that $\ket{0}\rightarrow\ket{0'}$ and $\ket{1}\rightarrow\ket{1'}$, where $\ket{0'}$ and $\ket{1'}$ correspond respectively to the $m_s=0$ and $m_s=1$ excited energy levels. While the optical transitions between the two $m_s=0$ energy levels are essentially exclusive, there is a metastable singlet state to which the excited $m_s=1$ energy state can decay non-radiatively. As a consequence, when performing now the measurement of photon emissions in such a spin-dependent fluorescence process over a designated time window, a dark signal indicates the original NV spin to be projected onto $\ket{1}$, while a bright signal corresponds to the projection onto $\ket{0}$. That is, within our general formalism, $\Pi_1=\proj{0}$ and $\Pi_2=\proj{1}$, so that after fixing the second Ramsey pulse to e.g.~$\Vphi=\upe^{\upi\pi\paulix/4}$, we have $\Pi_{1(2),\phivec}=\proj{\pm}$, with $\paulix\ket{\pm}=\pm\ket{\pm}$.

The bright versus dark distinction is however not perfect:~the $m_s=1$ excited state could still decay radiactively into the ground state, with the dark signal typically reducible to about 65\% of the bright signal. Moreover, as the photon emissions are spontaneous and random, the same photon-number being recorded can actually come from both the dark and bright signals, albeit with different 
probabilities. These for the readout of an NV-centre are modelled as two Poissonian distributions of distinct means, depending also on the number of QND repetitions \cite{Boss2017, schmitt2021optimal}, often approximated by Gaussians~\cite{Neumann2010}---see \figref{fig:QMwIM-fig1}(c). As a result, the `observed' outcomes correspond to the number of collected photons, $X=\{0,1,2,\cdots\}$, which are distributed according to the two Poissonian distributions $p(x|1)=\upe^{-\lambdazero}(\lambdazero)^x/x!$ and $p(x|2)=\upe^{-\lambdaone}(\lambdaone)^x/x!$, whose means, $\lambdazero$ and $\lambdaone$, differ depending on which energy state the NV spin was previously projected onto by $\Piphi{1}$ or $\Piphi{2}$.

\subsection{Estimating the FI with binning strategies}	                
In this section we discuss in more depth the binning method for estimating the FI for the single-probe scenario. Firstly, let us remark that for the strategy with two bins and $\delta=0$ in \eqnref{eq:F2bin}, we deal with a symmetric bit-flip channel mixing the two outcomes regardless of the choice of measurement basis. In this special case, the noisy detection affecting the measurement has exactly the same effect as if a dephasing noise acted \emph{before} an ideal measurement. Indeed, taking the limit $\delta\rightarrow0$ in \eqnsref{eq:F2bin}{eq:CosDeltaopt}, the optimal state angle becomes $\phiopt=-\theta$ and $\Ftwobinmaxstar=\eta^2$, agreeing with the well known result for the dephasing noise~\cite{Demkowicz2012,Escher2011}. Still, for any asymmetric bit-flip detection with $\pp\neq \qq$, the imperfect measurement model can no longer be interpreted as decoherence affecting rather the parameter encoding.

Secondly, let us note that when adopting a binning strategy one can freely choose the boundaries that define the bins. For binary binning $\Ftwobin$ and $\Ftwobinmaxstar$ depend on a single boundary (``threshold''~\cite{Jiang2009,Neumann2010}) $x^*$ via the parameters $\delta$ and $\eta$, so that upon maximising the choice of $x^*$ we can also define:
	\begin{align}\label{eq:F2binbarreal}
	F_{2\text{-bin}}=\max_{x^*}\Ftwobin,\quad \bar{F}_{2\text{-bin}} = \max_{x^*}\bar{F}_{2\text{-bin}}^*.
	\end{align}
Intuitively, one should choose $x^*$ such that the distributions $p(x|i)$ have the smallest overlap with the bins that yield errors in inferring the outcome $i$. Indeed, for the NV-sensing problem, the optimal choice of $x^*$ is located around the point where the two Poissonians cross in \figref{fig:QMwIM-fig4}(a), $p(x|1)=p(x|2)$, so that the probability of $x< x^*$ occurring when $i=2$ is minimised (and similarly for $x> x^*$ when $i=1$). More generally, in case of $k$-binning strategy with the corresponding FIs:~$F_{k\text{-bin}}^*, \bar{F}_{k\text{-bin}}^*, F_{k\text{-bin}}, \bar{F}_{k\text{-bin}}$; constituting natural generalisations of \eqnsref{eq:F2bin}{eq:F2binbarreal} and $\boldsymbol{x}^*$ being now a $(k-1)$-entry vector specifying boundaries between all the bins. Consistently, the more bins are considered the closer the corresponding FIs are to the exact $F$ (and $\FQimbar$) defined in \eqnref{eq:FNV}. 
	
For illustration, we revisit the case of sensing the relative phase with a NV spin, with the measurement suffering a Poissonian noise. In \figref{fig:QmwIM-fig7}(a), the performances of the optimal FIs for two- and three-binning strategies, $\Ftwobinmax$ and $\Fthreebinmax$, are investigated and compared against the exact maximal FI, $\FoneSingle$, which we numerically approximate by maximising $F$ with $x$ summed in \eqnref{eq:FNV} up a cut-off large enough ($x\leq100$) to be effectively ignorable. Within the plot the optical contrast is fixed to the typical experimental value of $0.35$, i.e. $\lambdaone/\lambdazero=0.65$~\cite{Boss2017}, while the FIs are plotted as a fraction of $\FoneSingle$ for different values of $\lambdazero$, which can be varied experimentally by having different repetitions of the QND measurement~\cite{Jiang2009,Neumann2010,Boss2017}. From the figure, we see that despite their simplicity, the strategy of binning into just two (orange) or three outcomes (blue) is pretty effective, as they are able to account for at least $70\%$ of $\FoneSingle$, and reach $90\%$ with increasing $\lambdazero$ already at $\lambdazero\approx50$. Then, similar to \figref{fig:QMwIM-fig2}, in \figref{fig:QmwIM-fig7}(b) we further include $F_{3\text{-bin}}$ in the plot of FI for different choices of input state angles $\phi=\varphi-\theta$, for the specific value of $\lambdazero=27$ which has been experimentally used in Ref.~\cite{Boss2017}.

\subsection{Lower-bounding the FI via the moments of a probability distribution}
It can be shown (see Supplement for derivation) that by including up to the first $2K$ moments of the distribution $\qphitheta\sim\{\qxphitheta\}$ a lower bound on the corresponding FI, $\Fdown{K}\le F[\qphitheta]$ in \eqnref{eq:FI}, can be constructed that corresponds to an inner product of two $K\times K$ matrices:
\begin{align} \label{eq:FdownK}
\Fdown{K}:=\Tr\{A^{-1}B\},
\end{align}
where $B=\boldsymbol{b}\boldsymbol{b}^T$ with $\boldsymbol{b}=\begin{pmatrix}0,\,\dot{\mathbb{E}}[x],\,\cdots\,,\dot{\mathbb{E}}[x^K] \end{pmatrix}^T$,
and
\begin{align}\label{eq:FdownK2}
A&=\begin{pmatrix}
1 & \mathbb{E}[x]&\cdots & \mathbb{E}[x^K] \\ 
\mathbb{E}[x] & \mathbb{E}[x^2]&\cdots & \mathbb{E}[x^{K+1}] \\ 
\vdots & \, & \ddots & \, \\ 
\mathbb{E}[x^K] & \mathbb{E}[x^{K+1}]&\cdots & \mathbb{E}[x^{2K}] 
\end{pmatrix} 
\end{align}
with $\mathbb{E}[x^j]=\sum_x \qxphitheta x^j$, $\dot{\mathbb{E}}[x^j]=\sum_x\dotqxphitheta x^j$ and $\dotqxphitheta=\partial_\theta \qxphitheta$. Note that for the simplest case of $K=1$, one obtains $\Fdown{1}=\dot{\mathbb{E}}[x]^2/(\mathbb{E}[x^2]-\mathbb{E}[x]^2)=|\partial_\theta\left<X\right>|^2/\mathrm{Var[X]}$ that constitutes the standard lower-bound on $F$ formed by considering the error-propagation formula applied to the distribution of the outcomes $X$~\cite{Wineland1992}. Evidently, we have the hierarchy $\Fdown{K}\leq\Fdown{K+1}$, whereby the more we know about its moments, the more we recover the underlying probability distribution, and $\Fdown{K}$ converges to $F[\qphitheta]$. For demonstration on the improvement of FI lower bound with higher moments considered, in \figref{fig:QmwIM-fig7}(b), we reproduce \figref{fig:QMwIM-fig2}, with now $\Fdown{2}$ included as well.

\subsection{Upper-bounding the imperfect QFI given the $G$-covariance of a conjugate-map decomposition}
We formalise the condition when the results of ``standard'' noisy metrology~\cite{Maccone2011}, in which the decoherence affects the parameter encoding, can be applied to the setting of imperfect measurements by resorting to the notion of symmetry, in particular, the $G$-covariance~\cite{Holevo1993,holevo_covariant_1996}. 

Given a compact group $G$, we say that a quantum channel $\cE$ is \emph{$G$-covariant} if~\cite{Holevo1993,holevo_covariant_1996}
\begin{equation}
    \forall_{g\in G}:\;\cE\circ\cV_{g}=\cW_g\circ\cE,
    \label{eq:G_cov_def}
\end{equation}
where $\cV_{g}$, $\cW_g$ form some unitary representations of $G$. 

Now, by denoting the FI in \eqnref{eq:FQim+FQimbar} as $F[\qphitheta]\equiv F[\cV_{\phivec}[\rho(\theta)], \cM]$ to separate its dependence on the state and the POVM, we formulate the following observations.
\begin{observation}[Imperfect measurement with a $G$-covariant conjugate-map decomposition]
\label{lem:UB_impQFI_conjmap}
Given an imperfect measurement with a conjugate-map decomposition $\cM=\Lambda^\dagger[\Pi]$, and a parametrised state $\rho(\theta)$, if the following conditions are satisfied:\\
(a)~$\Lambda$ is covariant with respect to a compact group $G$. \\
(b)~the optimal unitary that yields the imperfect QFI ($\opt{\cV}$ in \figref{fig:QMwIM-fig3}) is guaranteed to be in $G$, so that
\begin{align} 
    \FQim =\max_{g\in G}\,F\!\left[\cV_{g}[\rho(\theta)], \Lambda^\dagger[\Pi] \right],
\end{align}
then
\begin{equation} 
    \FQim \le \FQ\!\left[\Lambda[\rho(\theta)]\right]
    \label{eq:UB_impQFI_conjmap}
\end{equation}
If further $G=SU(d)$, then equality in \eqnref{eq:UB_impQFI_conjmap} is assured.
\end{observation}
Moreover, if the parameter encoding is provided in a form of a quantum channel, $\rho(\theta)=\cE_\theta[\rho]$, and the optimal unitary $\opt{\cV}$ remains within $G$ for the optimal input state, the upper bound \eref{eq:UB_impQFI_conjmap} applies also to the corresponding imperfect channel QFI, i.e.~$\FQimbar\le\bar\FQ[\Lambda\circ\cE_\theta]$. 

However, in case the parameter encoding satisfies the $G$-covariance property itself, we independently have that:
\begin{observation}[Imperfect channel QFI for $G$-covariant parameter encodings]
\label{lem:UB_impchQFI_conjmap}
Given an imperfect measurement with some conjugate-map decomposition $\cM=\Lambda^\dagger[\Pi]$ and the parameter encoding $\rho(\theta)=\cE_\theta[\rho]$, if the following conditions are satisfied:\\
(a)~both $\cE_\theta$ and $\dot{\cE}_\theta\equiv\partial_\theta\cE_\theta$ are $G$-covariant.\\ 
(b)~the optimal unitary that yields the optimal channel QFI ($\opt{\cV}$ in \figref{fig:QMwIM-fig3}) is guaranteed to be in $G$, so that
\begin{align} 
    \FQimbar  
    &=\max_{\rho} \max_{g\in G}\,F\!\left[\cV_g[\cE_\theta[\rho]], \Lambda^\dagger[\Pi] \right];
\end{align}
then
\begin{align} 
    \FQimbar  
    &\le \bar{\FQ}[\Lambda\circ\cE_\theta]. 
    \label{eq:UB_impchQFI_conjmap}
\end{align}
\end{observation}
We refer the reader to the Supplement for explicit proofs and further discussions of the above conditions.

\subsection{Upper-bounding the FI with the CE method for quantum-classical channel}
A thorough account on the CE method is available at Refs.~(\cite{Fujiwara2008,Demkowicz2012,Kolodynski2013}); here we simply highlight the general idea. In the CE method, when the probe state $\rho$ undergoes an effective encoding described by a given channel $\cE_{\theta}$, such that $\rho(\theta)=\cE_{\theta}[\rho]$, the corresponding FI for $\theta$ is bounded by considering an enlarged space with a corresponding input state $\rho_\mathrm{ext}$, such that  $\max_\rho\FQ\big[\rho(\theta)\big]\leq \max_{\rho_\mathrm{ext}}\FQ\big[(\cE_{\theta}\otimes\id)[\rho_\mathrm{ext}]\big]$,
where the r.h.s. can be shown to be equal to $4\min_{\tilde{\kappa}} \|\alpha_{\tilde{\kappa}}\|$, with $\tilde{\kappa}=\{\tilde{\kappa}_{i}\}$ denoting all the equivalent sets of Kraus operators for $\cE_{\theta}$, and $\alpha_{\tilde{\kappa}} :=\sum_{i}\dot{\tilde{\kappa}}_{i}^\dagger\dot{\tilde{\kappa}}_{i}$.

In order to apply the CE method to the canonical multi-probe metrology scheme with local control unitaries and local imperfect measurements, which has the corresponding product quantum-classical channel $\cE(\theta,\{\phivec_\ell\}):=\bigotimes_{\ell=1}^N\Lambda_{\theta,\phivec_\ell}^{(\ell)}$ with 
$\Lambda_{\theta,\phivec}=\Lambda_\cM \circ \cV_{\phivec} \circ \cU_\theta$
as depicted in \figref{fig:QMwIM-fig4}(b), we first specify the `canonical' set of Kraus operators for $\Lambda_{\theta,\phivec}$, $K(\theta,\phivec)=\{K_{x,j}(\theta, \phivec)=\ket{x}\bra{j}\sqrt{U_\theta^\dagger \Mphi{x} U_\theta}\}$, given some orthonormal basis of states $\{\ket{j}\}_{j=1}^d$ spanning the qudit ($d$-dimensional) probe space. Importantly also, as the output classical state is diagonal in the flag basis, its QFI corresponds just to the (classical) FI of the eigenvalue distribution~\cite{Braunstein1994} which we denote simply as $F_N=\FQ[\rhocl^N(\theta, \phivec)]$, with the corresponding (quantum-classical) channel QFI reads $\FQimbarloc_{N}:=\max_{\rho^N,\phivec}F_N$. Hence, upon further restricting the domain of minimisation over $\tilde{\kappa_{i}}$, where we only consider Kraus operators of $\cE(\theta,\{\phivec_\ell\})$ with the product structure $\tilde{\kappa}_{i}(\theta,\{\phivec_\ell\}) = \bigotimes_{\ell=1}^N \tilde{K}_{x_\ell,j_\ell}^{(\ell)}(\theta,\phivec_\ell)$, where $\tilde{K}(\theta,\phivec_\ell)=\{\tilde{K}_{x_\ell,j_\ell}^{(\ell)}(\theta,\phivec_\ell)\}$ is the set of Kraus operators for $\Lambda_{\theta,\phivec_\ell}^{(\ell)}$, it is then straightforward to arrive at
\begin{align}\label{eq:FNext}
&F_N\leq 4\min_{\{\tilde{K}(\theta,\phivec_\ell)\}} \Big\|\bigoplus_{\ell=1}^N\alpha_{\tilde{K}(\theta,\phivec_\ell)}^{(\ell)}+\bigoplus_{\ell\neq m}^N\beta_{\tilde{K}(\theta,\phivec_\ell)}^{(\ell)} \beta_{\tilde{K}(\theta,\phivec_m)}^{(m)}\Big\|, 
\end{align}
where 
\begin{align}
\alpha_{\tilde{K}(\theta,\phivec)} &:=\sum_{x,j}\dot{\tilde{K}}_{x,j}^\dagger(\theta,\phivec)\dot{\tilde{K}}_{x,j}(\theta,\phivec), \label{eq:alpha} \\
\beta_{\tilde{K}(\theta,\phivec)} &:=\upi\sum_{x,j}\dot{\tilde{K}}_{x,j}^\dagger(\theta,\phivec)\tilde{K}_{x,j}(\theta,\phivec), \label{eq:beta}
\end{align} 
with $\dot{\tilde{K}}_{x,j}:=\partial_\theta \tilde{K}_{x,j}$, and $||\cdots||$ is the operator norm. 
Finally then, we obtain our CE-bound in \eqnref{eq:CEbound} directly from applying the triangle inequality of the operator norm to the r.h.s. of \eqnref{eq:FNext}, which gives
\begin{align}\label{eq:CEbound2}
&\FNCE(\{\phivec_\ell\}):=\nonumber\\
&4\min_{\{\tilde{K}(\theta,\phivec_\ell)\}} \Big\{\sum_{\ell=1}^N||\alpha_{\tilde{K}(\theta,\phivec_\ell)}||+\sum_{\ell\neq m}^N||\beta_{\tilde{K}(\theta,\phivec_\ell)}||\,||\beta_{\tilde{K}(\theta,\phivec_m)}||\Big\}, 
\end{align} 
and the minimisation in both \eqnsref{eq:FNext}{eq:CEbound2} is performed independently for each $\phivec_\ell$ over all possible single-probe Kraus representations $\tilde{K}(\theta,\phivec_\ell)$.

As we prove in the Supplement, whenever 
the noisy detection channel, 
$\cP\sim\{p(x|i)\}$ such that $\forall_x:\;M_x=\sum_x p(x|i)\Pi_i$, 
is non-trivial, 
we can always find a Kraus representation such that $\beta_{\tilde{K}(\theta,\phivec_\ell)}=0$ for all $\ell$. Then, we define the asymptotic CE bound by
\begin{align} \label{eq:CEasbound}
\FNCEas(\{\phivec_\ell\}):= 4\,\sum_{\ell=1}^N\min_{\substack{\tilde{K}(\theta,\phivec_\ell)\\ \beta_{\tilde{K}(\theta,\phivec_\ell)}=0}} ||\alpha_{\tilde{K}(\theta,\phivec_\ell)}||, 
\end{align}
which evidently satisfies  $\FNCE(\{\phivec_\ell\})\leq \FNCEas(\{\phivec_\ell\})$ and $\FNCE\underset{n\rightarrow\infty}{\rightarrow}\FNCEas$. Finally, upon optimising $\FNCEas$ further over all local control unitaries gives us
\begin{align}
\bar{F}_N^{\mathrm{(CE,as)}}&:=\max_{\{\phivec_\ell\}}\FNCEas(\{\phivec_\ell\})
= 4N\,c,
\label{eq:CEasboundmax2}
\end{align}
such that $\FQimbarloc_N\le\bar{F}_N^{\mathrm{(CE,as)}}$, where
\begin{equation}
	c:= \max_{\phivec} \min_{\substack{\tilde{K}(\theta,\phivec)\\ \beta_{\tilde{K}(\theta,\phivec)}=0}} ||\alpha_{\tilde{K}(\theta,\phivec)}||, 
\end{equation}
is a constant factor that requires maximisation over $\phivec$ describing \emph{only} a single local unitary, and can be proven to be bounded, given the condition $\beta_{\tilde{K}(\theta,\phivec)}=0$ is fulfilled.

\subsection{Saturating $\bar{F}_N^{\mathrm{(CE,as)}}$ with an angular momentum measurement and spin-squeezed states}
To obtain the optimal asymptotic CE bound \eqref{eq:FNCEas}, we consider the measurement operators $\Piphi{1(2)}=\proj{\pm}$, followed by an asymmetric bit-flip channel $\cP$ with $p(1|1)=\pp, p(2|2)=\qq$ for all qubits. As a result, the measurements whose outcomes are actually observed read:~$\Mphi{1}=\pp\Piphi{1}+(1-\qq)\Piphi{2}=(1+\delta)\id/2+\eta\paulix/2$, $\Mphi{2}=(1-\pp)\Piphi{1}+\qq\Piphi{2}=(1-\delta)\id/2-\eta\paulix/2$, where $\eta=\pp+\qq-1$ and $\delta=\pp-\qq$. Constructing a qubit observable taking values $\pm1/2$ depending on the outcomes $x=1$ or $x=2$, it is not hard observe that when measured in parallel on each of the $N$ probes and summed, one effectively conducts a measurement of the operator $\hat{O}=N\delta\id/2+\eta\Jx$ that constitutes a modification of the total angular momentum $\Jx=\sum_{\ell=1}^N\frac{\sigma_\mathrm{x}^{(\ell)}}{2}$, being tailored to the (binary bit-flip) noisy detection channel. A simple estimator of $\theta$ may then be directly formed by inverting the expectation-value relation $O(\theta)=\Tr\{\rho(\theta)\hat{O}\}$ from the outcomes (repeating the protocol $\nu\gg1$ times).

As derived in the Supplement, the MSE of such an estimator, given sufficiently large number $\nu$ of measurement repetitions, is well approximated by the (generalised) error-propagation formula,
\begin{align}\label{eq:errorpropJx}
\nu\Delta^2\tilde{\theta}_N = \frac{\Delta^2 \Jx}{|\partial_\theta\langle \Jx\rangle|^2}-\frac{\delta \langle \Jx\rangle}{\eta|\partial_\theta\langle \Jx\rangle|^2}+\frac{N}{4\eta^2}\frac{1-\eta^2-\delta^2}{|\partial_\theta\langle \Jx\rangle|^2}, 
\end{align}
where $\langle \dots\rangle=\Tr\{\rho^N(\theta)\dots\}$ and $\Delta^2 \Jx=\langle \Jx^2\rangle-\langle \Jx\rangle^2$.

Consider now $\rho^N=\ket{\phi,\mu}\bra{\phi,\mu}$ with $\ket{\phi,\mu}=\upe^{\upi\phi\Jz}\ket{\mu}$ and 
\begin{equation}
\label{eq:spin_squeezed_state}
	\ket{\mu}=W_\mu\upe^{-\upi\Theta_\mu W_\mu^\dagger\Jy W_\mu}\ket{j,m_y=j}_\mathrm{y}
\end{equation}
being the one-axis spin-squeezed state~\cite{Kitagawa1993} expressed in the angular momentum eigenbasis defined by the $\hat{J}^2$ and $\Jy$ operators, where $W_\mu=\upe^{-\upi\mu\Jz^2/2}$ is the unitary squeezing operation of strength $\mu$, while $\Theta_\mu=\pi/2-\epsilon$ with $\epsilon=\arctan(b/a)$, $a=1-\cos^{2j-2}\mu$ and $b=4\sin(\mu/2)\cos^{2j-2}(\mu/2)$. For our purpose we will consider states \eqref{eq:spin_squeezed_state} obtained by squeezing a completely polarised ensemble spins along the $y$-axis, i.e.~prepared in a state $\ket{j,m_y=j}_\mathrm{y}$ with $j=m_j=N/2$. Substituting such choice into the error-propagation expression \eqref{eq:errorpropJx}, we arrive after lengthy but straightforward algebra at 
\begin{align}\label{eq:errorpropJx2}
\nu\Delta^2\tilde{\theta}_N
=
& \frac{\cos^2\!\varphi\,(\Delta^2 \Jx)_\mu+\sin^2\!\varphi\,(\Delta^2 \Jy)_\mu}{\cos^2\!\varphi\,\langle \Jy\rangle_\mu^2}\nonumber\\
&-\frac{\delta \sin\varphi\,\langle \Jy\rangle_\mu}{\eta \cos^2\!\varphi\,\langle \Jy\rangle_\mu^2}+\frac{N}{4\eta^2}\frac{1-\eta^2-\delta^2}{\cos^2\!\varphi\,\langle \Jy\rangle_\mu^2},
\end{align}
where the subscripts $\mu$ indicate expectations to be evaluated w.r.t.~the state $\ket{\mu}$ in \eqnref{eq:spin_squeezed_state}, having defined $\varphi:=\phi+\theta$ as in \eqnref{eq:FNV}. For large $N$, we find that after choosing the squeezing strength to scale as $\mu\sim N^{-8/9}$, one has $(\Delta^2 \Jx)_\mu\sim N^{7/9}, (\Delta^2 \Jy)_\mu\sim N^{4/9}/128, \langle \Jy\rangle_\mu\sim N/2$, and therefore:
\begin{align}\label{eq:errorpropJx3}
\nu\Delta^2\tilde{\theta}_N\sim \frac{1}{N}\frac{1}{\cos^2\varphi}\Big( \frac{1-\delta^2-\eta^2}{\eta^2}-\frac{2\delta}{\eta}\sin\varphi \Big).
\end{align}
Finally, by choosing now $\varphi=\varphiopt$ as the angle derived in single-probe (qubit) setting in \eqnref{eq:CosDeltaopt}, the r.h.s.~of \eqnref{eq:errorpropJx3} converges exactly to $1/\bar{F}_N^{\mathrm{(CE,as)}}$ with $\bar{F}_N^{\mathrm{(CE,as)}}$ stated in \eqnref{eq:FNCEas}. In other words, the asymptotic ultimate precision is achieved by (imperfectly) measuring the total angular momentum in the x-direction, while preparing the $N$ probes (spin-$1/2$s) in a spin-squeezed state rotated by the same optimal angle \eqref{eq:CosDeltaopt} as in the single-probe scenario, with squeezing parameter scaling as $\mu\sim N^{-8/9}$ with $N$.

\subsection{Application to lossy photonic interferometry with dark counts}
\begin{figure}[t!]
	\centering
	\includegraphics[width=0.45\textwidth]{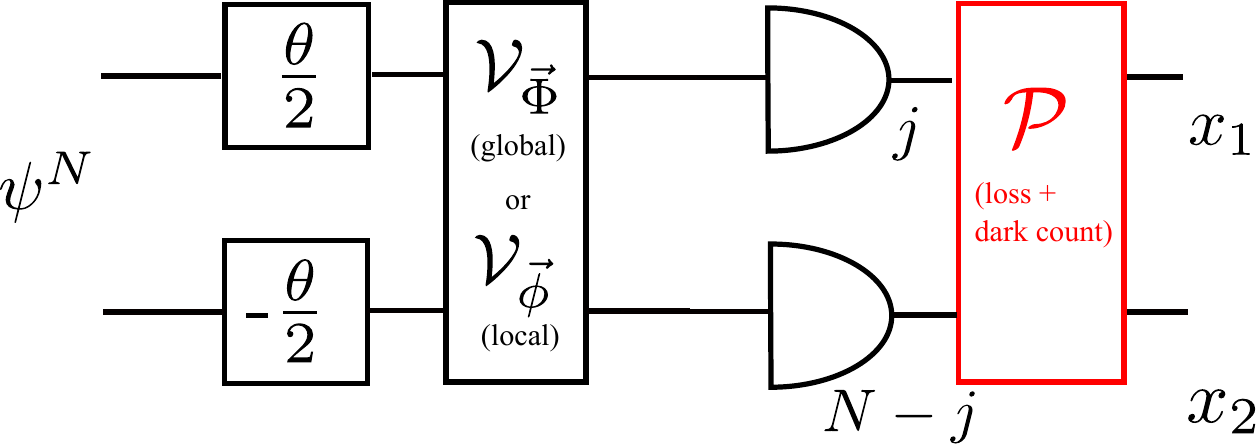}
	\caption{\textbf{Sensing the relative phase $\theta$ with a pure $N$-photon state, $\psi^{N}$, in a lossy interferometer with dark counts.} Either a global (affects all the photons) or local (affects each photon separately) control unitary is allowed. The imperfect measurement is described by an ideal photon-number detection $(j,N-j)$ at the output modes, followed by a noisy detection channel $\cP$ that incorporates losses and dark counts, resulting in $(x_{1},x_{2})$ registered detection counts.
}
	\label{fig:QMwIM-fig8}
\end{figure}

In this section, we consider another standard problem in quantum metrology, namely, two-mode interferometry involving $N$-photon quantum states of light. The effective full Hilbert space is thus spanned by the Fock basis $\left\{ \ket{j,N-j}\right\}_{j=0}^{N}$. The detection channel is two identical photodetectors measuring each mode, which independently suffer from losses and dark counts. Specifically, we quantify the losses by efficiency $\eta$, i.e.~a photon is detected (or lost) with probability $\eta$ (or $1-\eta$)~\cite{Datta2011}. Moreover, we assume that each photodetector may experience a single dark count with probability $\pp$ per each photon that enters the interferometer~\cite{footnote3}.
Note that this results in extending the effective measured Hilbert space to $\left\{ \ket{x_1,x_2}\right\}$ with $0\le x_1,x_2\le 2N$ and $0\le x_1+x_2\le 3N$.
A schematic of the problem is depicted in \figref{fig:QMwIM-fig8}. 

Given the above imperfect measurement model, let us find the corresponding $\gamma_{\cM^{\otimes N}}$ in \eqnref{eq:gamma_M} that effectively determines the imperfect (channel) QFI \eref{eq:FIwIM}, while requiring a global control unitary to be performed, $\VcalPhi$ in \figref{fig:QMwIM-fig8}~\footnote{Note that the photon-counting measurement is local w.r.t.~the Hilbert spaces associated with each photon, while the global control may now in principle require $N$-photon interactions , i.e.~may be highly non-linear within the second quantisation \cite{Rafal2015}}. Firstly, we ignore the dark counts, so that $0\le x_1+x_2\le N$ and
\begin{align} 
&\gamma_{\cM^{\otimes N}}=
\max_{\ket{\xi},\ket{\xi_{\perp}}} \sum_{x_{1},x_{2}} \frac{\mathrm{Re}\!\left\{\langle\xi_{\perp}|M_{x_{1},x_{2}}|\xi\rangle\right\}^{2}}{\langle\xi|M_{x_{1},x_{2}}|\xi\rangle} \label{eq:gammaM_photon}\\
&=\text{\ensuremath{\underset{{\bf a\cdot b=0,\;}|{\bf a}|^{2}=|{\bf b}|^{2}=1}{\text{max}}}} \sum_{x_{1},x_{2}} \frac{\left(\sum_j p\!\left(x_{1},x_{2}|j,N-j\right)a_{j}b_{j}\right)^{2}}{\sum_j p\!\left(x_{1},x_{2}|j,N-j\right)a_{j}^{2}},
\nonumber
\end{align}
where $\ket{\xi}=\sum_{j=0}^{N} a_{j}\ket{j,N-j}$, $|\xi_{\perp}\rangle=\sum_{j=0}^{N} b_{j}\ket{j,N-j}$ are some orthogonal vectors with real coefficients. The imperfect photon-count measurement corresponds then to the POVM:~$\cM^{\otimes N}\sim \{M_{x_{1},x_{2}}=\sum_{j=0}^{N}p(x_{1},x_{2}|j,N-j)\proj{j}\otimes\proj{N-j}\}$, whose mixing coefficients are defined by the noisy detection channel responsible for \emph{photon loss}, $\cPloss\sim\{p\!\left(x_{1},x_{2}|j,N-j\right)\}$ with
\begin{eqnarray}
\begin{split}
&p\!\left(x_{1},x_{2}|j,N-j\right)=p_{\eta}\!\left(x_{1}|j\right)p_{\eta}\!\left(x_{2}|N-j\right)\\
&={j \choose x_{1}}{N-j \choose x_{2}}\eta^{x_{1}+x_{2}}\left(1-\eta\right)^{N-x_{1}-x_{2}},
\end{split}
\end{eqnarray} 
where $p_{\eta}(x|k):={k \choose x}\eta^x(1-\eta)^{k-x}$ is a binomial distribution arising due to the finite detection efficiency $\eta$.

We observe from numerical simulations that
\begin{align}
\label{eq:opt_xi}
     \ket{\xi} &=  \sin\varphi\ket{N,0} + \cos\varphi\ket{0,N}
     \\
     \ket{\xi_\perp} &=\cos\varphi|N,0\rangle -\sin\varphi\ket{0,N}
     \nonumber
\end{align} 
are optimal for any $\varphi\in\mathbb{R}$. Assuming the form of $\ket{\xi}$ and $\ket{\xi_\perp}$ as above, we can calculate $\gamma_{\cM^{\otimes N}}$ explicitly in \eqnref{eq:gammaM_photon}: 
\begin{align}
 \gamma_{\cM^{\otimes N}}=&\sum_{x_{1}\neq0}\frac{\left(p_{\eta}\!\left(x_{1}|N\right)\sin\varphi\cos\varphi\right)^{2}}{p_{\eta}\!\left(x_{1}|N\right)\sin^2\!\varphi}+ \nonumber\\
& \sum_{x_{2}\neq0}\frac{\left(p_{\eta}\!\left(x_{2}|N\right) \sin\varphi\cos\varphi\right)^{2}}{p_{\eta}\!\left(x_{2}|N\right)\cos^2\!\varphi} \nonumber\\
=&\sum_{x\neq0}p_{\eta}\!\left(x|N\right)=1-\left( 1-\eta \right)^{N}. \label{eq:GammaMlossy}
\end{align}
The above expression has a simple interpretation:~for $\varphi=0$ the states $\ket{\xi}=\ket{0,N}$, $\ket{\xi_\perp}=\ket{N,0}$ remain orthogonal unless all photons are lost in both arms, what may happen only with probability $\left( 1-\eta \right)^{N}$. Moreover, this expression coincides with the lower bound on $\gamma_{\cM^{\otimes N}}$ used in \eqnref{eq:FNbound-1} (i.e.~$1-\sum_{\boldsymbol{x}}\sqrt{p_{+}(\boldsymbol{x})p_{-}(\boldsymbol{x})}$ in Eq.~(S.58) of the Supplement~\cite{footnote4}%
). 
Finally, we observe that by lifting the assumption of photodetection efficiency being equal in both arms, no longer $\varphi$ may be arbitrary chosen in \eqnref{eq:opt_xi}, but rather must also be optimised.

As a result, considering e.g.~the N00N state as the the input probe $\psi^{N}$ for which $\FQ[\psi^N(\theta)]=N^2$, we obtain the equivalent of \eqnref{eq:FNbound-1} in the form
\begin{equation}
  F_N(\VPhi) =\;
     N^2 \,[1-(1-\eta)^N] = N^2 \,[1-\upe^{-\chi N}]
     \label{eq:N00N_convergence}  
\end{equation}
with $\chi=-\ln(1-\eta)$ to be compared with the one in \eqnref{eq:FNbound-2}. As anticipated, the HS is maintained despite the imperfect measurement, however, the necessary global control unitary, $\VcalPhi$ in \figref{fig:QMwIM-fig8}, must rotate the encoded state $\ket{\psi^N(\theta)}=(\upe^{\upi N\theta/2}\ket{N,0}+\upe^{-\upi N\theta/2}\ket{0,N})/\sqrt{2}$ and its orthogonal $\ket{\psi_{\perp}^N(\theta)}=\upi(\upe^{\upi N\theta/2}\ket{N,0}-\upe^{-\upi N\theta/2}\ket{0,N})/\sqrt{2}$ onto the optimal $\ket{\xi}$ and $\ket{\xi_{\perp}}$ in \eqnref{eq:opt_xi}. Note that it is a highly non-linear operation allowing to ``disentangle'' N00N states. For instance, for $\varphi,\theta=0$ it rotates the input N00N state and its perpendicular component onto the desired $\ket{0,N}$ and $\ket{N,0}$, respectively, which are product w.r.t.~the Hilbert spaces associated with each photon, $\mathbb{C}_2^{\otimes N}$.

Consider now also the presence of dark counts parametrised by the rate $\pp$~\cite{footnote3}. Similarly to $\cPloss$, the detection channel responsible for \emph{dark counts} is characterised by a binomial distribution $p_{\pp}(y|N):=\binom{N}{y}\pp^{y}(1-\pp)^{N-y}$, so that $\cPdc\sim \{p(y_{1}, y_{2}|N)\}$ with
\begin{eqnarray}
p\!\left(y_{1},y_{2}|N\right)=p_{\pp}\!\left(y_{1}|N\right)p_{\pp}\!\left(y_{2}|N\right),
\end{eqnarray} 
where $y_{1}$ and $y_{2}$ are respectively the number of dark counts in each detector. The resultant \emph{overall noisy detection channel} corresponds to the composition of $\cPloss$ and $\cPdc$, i.e.~$\cP \sim \{p\!\left(x_{1},x_{2}|j,N-j\right)\}$ with elements 
\begin{equation}
    p\!\left(x_{1},x_{2}|j,N-j\right)=p_{\eta,\pp}\!\left(x_{1}|j\right)p_{\eta,\pp}\!\left(x_{2}|N-j\right),
    \label{eq:cPlossdc}
\end{equation}
where $p_{\eta,\pp}\!\left(x|k\right)=\sum_{m=0}^{x} p_{\eta}\!\left(m|k\right)p_{\pp}\!\left(x-m|N\right)$, while $x_{1}$ and $x_{2}$  are respectively the final number of photons registered in each of the photodetectors.

\begin{figure}[t!]
\includegraphics[width=0.45\textwidth]{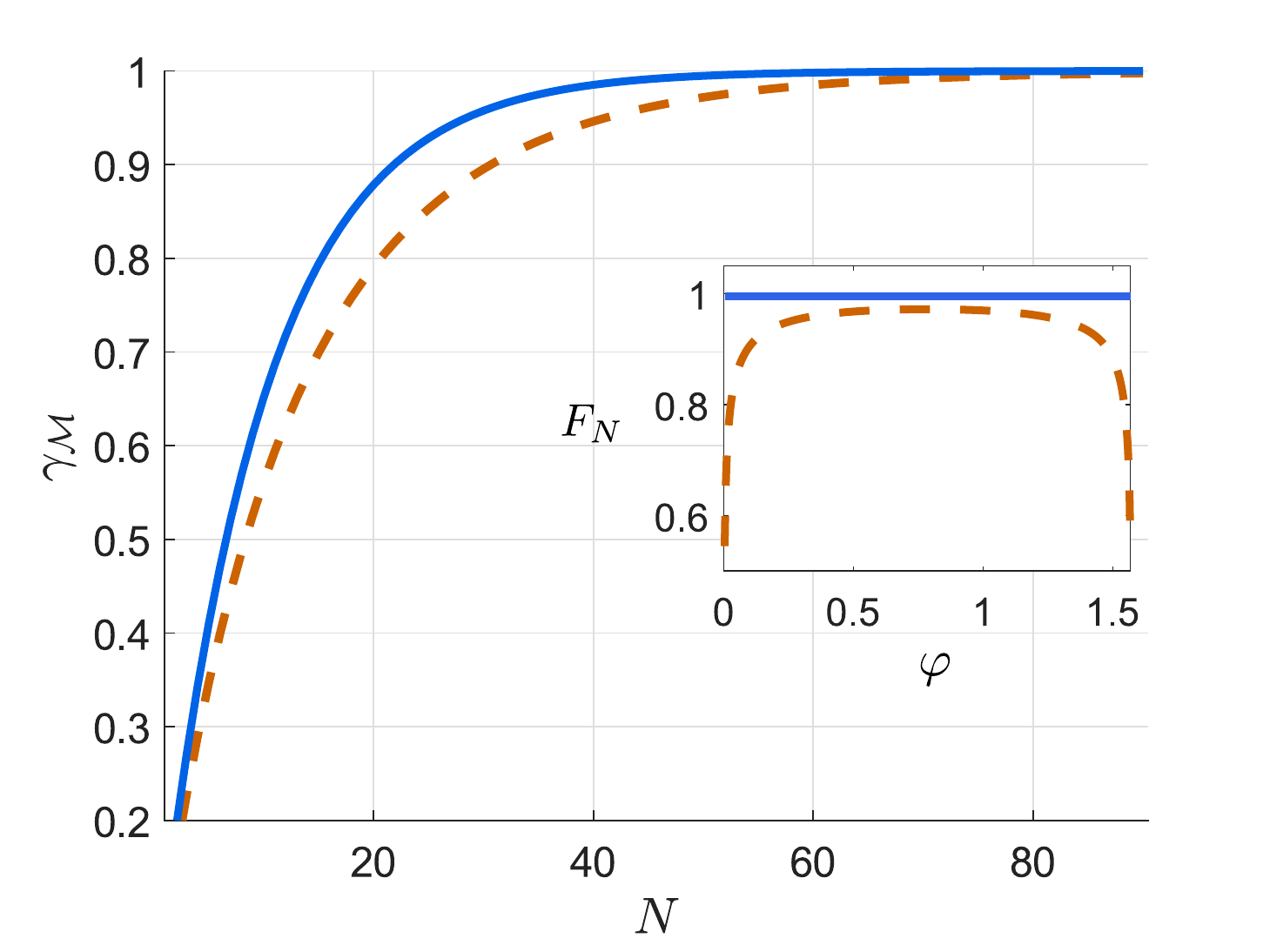}
\caption{
\textbf{Convergence to the perfect QFI with the photon-number $N$ for different noisy photodetection channels:}
$\gamma_{\cM^{\otimes N}}$ for different dark-count rates, $\pp$, given a finite detection efficiency (losses) $\eta=0.1$. The blue (solid) line corresponds to losses without dark counts $\pp=0$, while the orange dashed line also includes a dark-count rate of $\pp=0.01$. \emph{Inset:}~The FI as a function of the angle $\varphi$ between $\ket{N,0}$ and $\ket{0,N}$ in \eqnref{eq:opt_xi} that parametrises the global control unitary $\VPhi$.  While with only losses (solid blue line) any angle is optimal, dark counts (dashed orange line) remove this symmetry. In this illustration, $N=50.$}
\label{fig:QMwIM-fig9}
\end{figure}

Verifying numerically again that optimal $|\xi \rangle$, $| \xi_{\perp} \rangle$ take the form \eref{eq:opt_xi}, we find  $\gamma_{\cM^{\otimes N}}$ in \eqnref{eq:gammaM_photon} to read 
\begin{eqnarray}
& \gamma_{\cM^{\otimes N}} = \underset{x_{1},x_{2}}{\sum}\left(\sin\varphi\cos\varphi\right)^{2}\times & \\ 
& \frac{\left(p_{\eta,p}\left(x_{1}|N\right)p_{\eta,p}\left(x_{2}|0\right)-p_{\eta,p}\left(x_{1}|0\right)p_{\eta,p}\left(x_{2}|N\right)\right)^{2}}{\sin^2\!\varphi\,p_{\eta,p}\left(x_{1}|N\right)p_{\eta,p}\left(x_{2}|0\right)+\cos^2\!\varphi\,p_{\eta,p}\left(x_{1}|0\right)p_{\eta,p}\left(x_{2}|N\right)}. &
\nonumber
\end{eqnarray}
The dark counts lift the degeneracy in $\varphi$ and reduce $\gamma_{\cM^{\otimes N}}$. A comparison between $\gamma_{\cM^{\otimes N}}$ evaluated numerically for different levels of dark-counts rate, $\pp$, is presented in \figref{fig:QMwIM-fig9}. Importantly, $\gamma_{\cM^{\otimes N}}$ still approaches unity as $N$ increases, however, the optimal control unitary $\VcalPhi$ is more involved and the convergence is slower. Finding an analytical expression for the convergence rate with dark counts, $\chi$ as in \eqnref{eq:N00N_convergence}, we leave as an open challenge.

We now turn to the scenario when only local control unitary operations are allowed, i.e.~ones that may affect only a single (two-mode) photon, denoted by $\Vcalphi$ in \figref{fig:QMwIM-fig8}. For this, let us imagine a more advanced interferometry scheme in which the input $N$ photons can be resolved into different time-bins, despite all of them occupying a bosonic (permutation invariant) state~\cite{Berry2000}. Within such a picture, each time-bin is represented by a qubit with basis states $\ket{0}\equiv\ket{1,0}$, $\ket{1}\equiv\ket{0,1}$ corresponding to a single photon occupying either of the two optical modes. Moreover, the ideal measurement in each time-bin is then described by projectors ($\{\Pi_{1(2)}\}$) onto the above basis states, each yielding a ``click'' in either of the detectors.

We include the loss and dark-count noise within the detection process, as described by \eqnref{eq:cPlossdc}, but the rate of the latter, $\pp$, to be small enough, so that at most one false detection event may occur per  time-bin (photon). Consequently, the noise leads to six possible `observable' outcomes (i.e.~$d=2\to\dimM=6$), namely, $\binom{2}{0}, \binom{0}{2}, \binom{1}{1}, \binom{0}{0}, \binom{1}{0}$, and $\binom{0}{1}$, which we will respectively label as outcomes $x=1$ to $6$---by $\binom{x_1}{x_2}$ we denote that $x_1$ ($x_2$) ``clicks'' were recorded in the upper (lower) detector. The resulting imperfect measurement $\cM$ performed in each time-bin is then specified by $M_x=\sum_{i=1}^2 p(x|i)\Pi_i$, where $p(x|i)$ is the $(x,i)$-th entry of the stochastic matrix:
\begin{align}
\label{eq:P_1}
	\cP^{\mathrm{(1)}}=\begin{pmatrix}
		\pp\eta & 0 \\
		0 & \pp\eta \\
		\pp\eta & \pp\eta \\
		1-2\pp-\eta+2\pp\eta & 1-2\pp-\eta+2\pp\eta \\
		\pp+\eta-3\pp\eta & \pp-\pp\eta \\
		\pp-\pp\eta & \pp+\eta-3\pp\eta  
	\end{pmatrix},
\end{align}
which can be obtained equivalently by evaluating the form of detection channel $\cP$ defined in Eq.~\eqref{eq:cPlossdc} for $N=1$, and truncating the quadratic terms in $\pp$. 

Possessing the form of the local (single-photon) detection noise \eref{eq:P_1}, we follow our technique based on the CE-method~\cite{Fujiwara2008,Demkowicz2012,Kolodynski2013} to compute upper bounds on the precision in estimating $\theta$, where thanks to employing the quantum-classical channel formalism we are able explicitly determine the asymptotic CE-bound, as defined in \eqnsref{eq:CEasboundmax}{eq:CEasboundmax2}, despite $\dimM\neq d$, i.e.~the number of 'observable' outcomes differing from the `inaccessible' ones, i.e.:
\begin{align}\label{eq:FNCEasLoss+DC}
	\FNCEas=N\frac{\eta(\eta-3\pp\eta+2\pp^2)}{2\pp+\eta^2(3\pp-1)+\eta(1-4\pp-2\pp^2)},
\end{align}
whereas the finite CE-bound, $\FNCE$ in \eqnsref{eq:CEbound}{eq:CEbound2}, can be computed efficiently via a semi-definite programme (SDP). We leave an explicit proof open, however, the derivation of \eqnref{eq:CEasboundmax2} suggests the asymptotic CE-bound \eref{eq:FNCEasLoss+DC} to apply also to protocols involving (local) adaptive measurements~\cite{Berry2000}.

\begin{figure}[t!]
    \centering
    \includegraphics[width=0.45\textwidth]{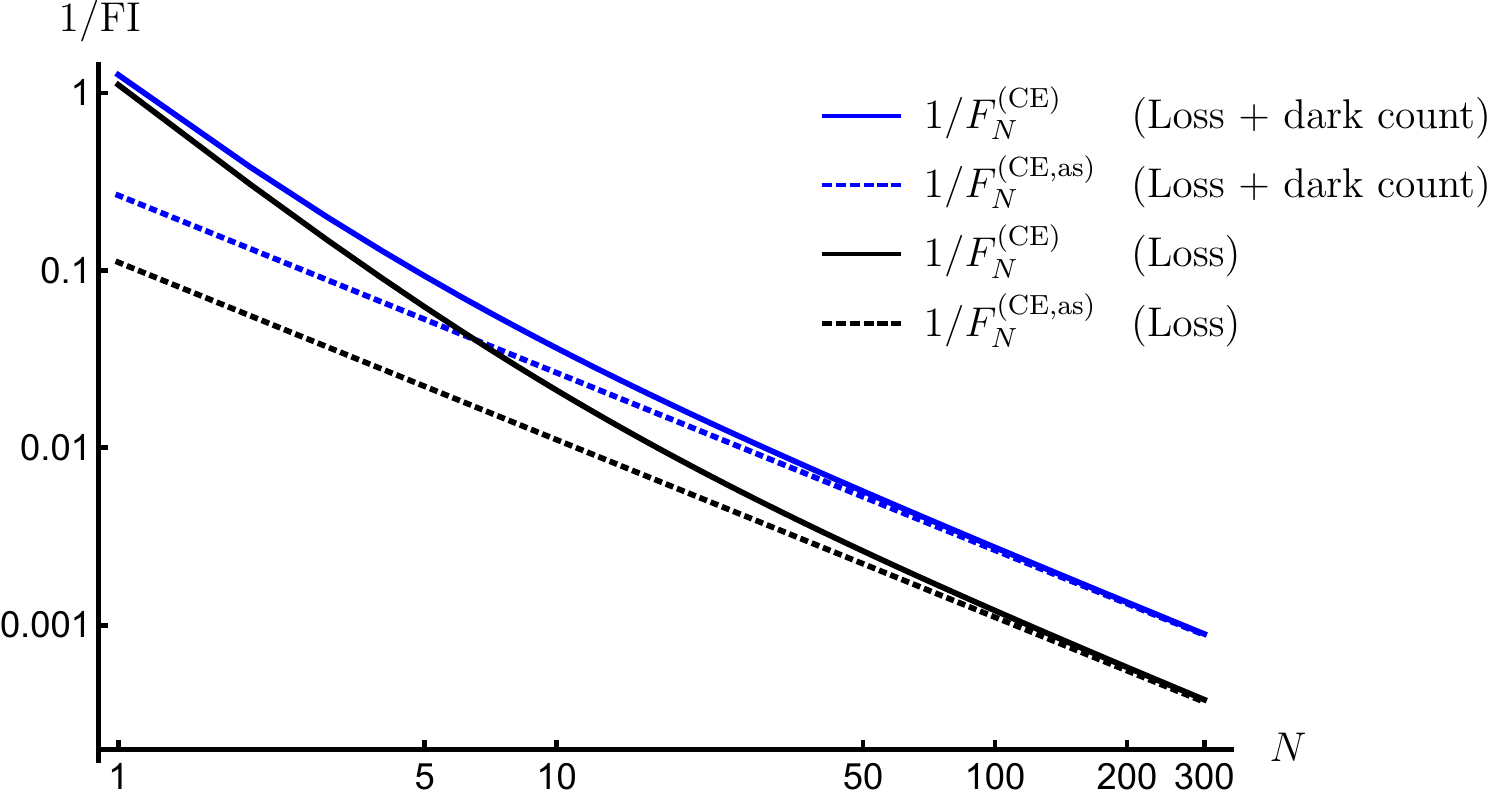}
    \caption{
    \textbf{Ultimate bounds on precision in lossy photonic interferometry with dark counts and local unitary control}. The \emph{solid and dotted blue} lines are the inverse of the finite-$N$ (evaluated via an SDP) and asymptotic (see \eqnref{eq:FNCEasLoss+DC}) CE bounds, respectively, for detection efficiency $\eta=0.9$ and dark-count rate $\pp=0.1$. Control gates may act on each individual dual-rail photon. For comparison, we also plot, \emph{solid and dotted black} lines, the corresponding quantities when only photon loss is present ($\eta=0.9$, $\pp=0$).
    }
    \label{fig:QMwIM-fig10}
\end{figure}

For illustration, in \figref{fig:QMwIM-fig10} we plot in blue the respective inverses of $\FNCE$ and $\FNCEas$ for $\pp=0.1$ and $\eta=0.9$. Note that, as it should, the presence of dark counts worsen the estimation precision as compared to the having just lossy detectors, which can be obtained by taking $\pp\rightarrow0$, and are plotted in black. As a side note, in the latter case, $\FNCE$ and $\FNCEas=\eta/(1-\eta)$ have also been obtained by expanding the space to a qutrit state, where an auxiliary mode 3 is introduced to keep track of the $\binom{0}{0}$ outcome \cite{Kolodynski2013}. Meanwhile, on the other extreme, when the detector has unity efficiency but only just dark count, we have $\FNCEas=N(\pp^{-1}-1)$. Interestingly, we observe that the CE-bound for the effect of pure loss with rate $1-\eta$ is the same as that of a pure dark count with rate $\pp=1-\eta$, true both for finite-$N$ and asymptotically.


\section*{Acknowledgments}
The authors are thankful to Spyridon Michalakis and Micha\l{} Oszmaniec for fruitful discussions.
YLL \& JK acknowledge financial support from the Foundation for Polish Science within the ``Quantum Optical Technologies'' project carried out within the International Research Agendas programme co-financed by the European Union under the European Regional Development Fund. TG acknowledges the support of the Israel Council for Higher Education Quantum Science and Technology Scholarship.

\section{Supplementary Information}
\setcounter{equation}{0}
\renewcommand{\theequation}{S.\arabic{equation}}
\setcounter{figure}{0}
\renewcommand{\figurename}{Fig.}
\renewcommand{\thefigure}{S\arabic{figure}}
\newcommand{\cc}{\mathrm{c.c.}}
\newcommand{\upd}{\mathrm{d}}
\newcommand{\refmain}[1]{{\color{black} #1}}
\newcommand{\Vopt}{\opt{V}}
\newcommand{\rhoopt}{\opt{\rho}}
\newcommand{\FQbar}{\bar{\FQ}}
\newcommand{\xvec}{\boldsymbol{x}}
\renewcommand{\gg}{\mathsf{g}}
\setcounter{secnumdepth}{5}


\subsection{Proof of Lemma 1}
Consider an imperfect measurement $\cM\sim\{M_x\}_x$. Then, for a pure encoded state $\psi(\theta)=\proj{\psi(\theta)}$, and a control unitary $\cV_{\phivec}\sim V_{\phivec}$ allowing a change of measurement basis, the Fisher information (FI) is given by 
\begin{align} \label{eq:AF}
F&=\sum_{x} \frac{[\partial_\theta(\bra{\psi(\theta)}\Vphi^\dagger M_x\Vphi\ket{\psi(\theta)})]^2}{\bra{\psi(\theta)}\Vphi^\dagger M_x\Vphi\ket{\psi(\theta)}}\nonumber\\
&=\sum_{x} \frac{[\bra{\partial_\theta\psi(\theta)}\Vphi^\dagger M_x\Vphi\ket{\psi(\theta)}+\cc]^2}{\bra{\psi(\theta)}\Vphi^\dagger M_x\Vphi\ket{\psi(\theta)})},
\end{align}
where $\cc$ stands for complex conjugation, and $\ket{\partial_\theta\psi(\theta)}$ is the shorthand for $\partial_\theta\ket{\psi(\theta)}$. We decompose $\ket{\partial_\theta\psi(\theta)}$ into the orthogonal and parallel parts to $\ket{\psi(\theta)}$, i.e.: 
\begin{align} \label{eq:Adpsi}
\ket{\partial_\theta\psi(\theta)}&=\ket{\partial_\theta\psi_{\perp}(\theta)}+\ket{\partial_\theta\psi_{\parallel}(\theta)},\\
\mathrm{with}\quad \ket{\partial_\theta\psi_{\perp}(\theta)}&:=(\id-\psi(\theta))\ket{\partial_\theta\psi(\theta)},\nonumber\\
\ket{\partial_\theta\psi_{\parallel}(\theta)}&:=\psi(\theta)\ket{\partial_\theta\psi(\theta)}\nonumber.
\end{align}
It is straightforward to show that $\bra{\partial_\theta\psi_{\parallel}(\theta)}\Vphi^\dagger M_x\Vphi\ket{\psi(\theta)}+\cc=0$, and upon defining $\ket{\psi_\perp(\theta)}:=\ket{\partial_\theta\psi_\perp(\theta)}/\sqrt{\langle \partial_\theta\psi_\perp(\theta)\ket{\partial_\theta\psi_\perp(\theta)}}$, \eqnref{eq:AF} is equal to
\begin{align}\label{eq:AF2}
F=&4\;\gamma(\phivec,\psi(\theta))\;\langle \partial_\theta\psi_\perp(\theta)\ket{\partial_\theta\psi_\perp(\theta)}, 
\end{align}
with
\begin{align}\label{eq:Agamma}
&\gamma(\phivec,\psi(\theta))\nonumber\\
&:=\frac{1}{4}\sum_{x}  \frac{[\bra{\psi_\perp(\theta)}\Vphi^\dagger M_x\Vphi\ket{\psi(\theta)}+\cc]^2}{\bra{\psi(\theta)}\Vphi^\dagger M_x\Vphi\ket{\psi(\theta)}}.
\end{align}
Note that $4\langle \partial_\theta\psi_\perp(\theta)\ket{\partial_\theta\psi_\perp(\theta)}=\FQ[\psi(\theta)]$ is nothing but the (perfect) quantum Fisher information (QFI) of $\psi(\theta)$.

The imperfect QFI is thus: 
\begin{equation}
\FQim=\left[ \max_{\phivec} \gamma(\phivec,\psi(\theta)) \right] \FQ[\psi(\theta)].
\end{equation}
Let us denote $\gamma_\cM:=\max_{\phivec} \gamma(\phivec,\psi(\theta))$. Clearly by an appropriate choice of $\vec{\phi}$ we can map $\ket{\psi(\theta)}, \ket{\psi_\perp(\theta)}$ to any two arbitrary orthogonal states $\ket{ \xi }, \ket{ \xi_{\perp} }$. Therefore, the optimization over $\vec{\phi}$ is basically an optimization over any two orthogonal states $\ket{ \xi }, \ket{ \xi_{\perp} }$, namely:
\begin{equation}
\label{eq:gamma_M}
\gamma_\cM=\underset{|\xi\rangle,|\xi_{\perp}\rangle}{\max} \sum_{x}\frac{\text{Re}\{\langle\xi_{\perp}|M_x|\xi\rangle\}^{2}}{\langle\xi|M_x|\xi\rangle}.
\end{equation}
Evidently, $\gamma_\cM$ is completely independent of the encoding of the parameter, $\psi(\theta)$,
and depends only on the imperfect measurement $\cM\sim\{ M_x\}_{x}$. 
Clearly $\gamma_\cM \geq 0$, because the FI is non-negative, and $\gamma_\cM \leq 1$, because the imperfect QFI cannot be bigger than the perfect QFI. The latter can also be verified by using the Cauchy-Schwarz inequality:
\begin{align}
&\sum_{x}\frac{\text{Re}\{\langle\xi_{\perp}|M_x|\xi\rangle\}^{2}}{\langle\xi|M_x|\xi\rangle}\leq\sum_{x}\frac{\langle\xi_{\perp}|M_x|\xi_{\perp}\rangle\langle\xi|M_x|\xi\rangle}{\langle\xi|M_x|\xi\rangle} \nonumber\\
&=\sum_{x}\langle\xi_{\perp}|M_x|\xi_{\perp}\rangle=1.
\label{eq:CS_ineq}
\end{align}

Hence, $0 \leq \gamma_\cM \leq 1$, and it depends solely on the imperfect measurement $\cM$; or, in the common cases discussed in the main text, the noisy detection channel $\cP$ that determines the effective $\cM$, i.e., with all measurement elements then given by $M_x=\sum_i p\!\left(x|i\right)\Pi_{i}$, where $\Pi$ is some fixed (perfect) projective measurement, so that the imperfect measurement is completely specified by the stochastic map, $\cP\sim\{p(x|i)\}$, representing the readout noise. In such a situation, see also the classical interpretation below, we refer to $\gamma_\cM$ in \eqnref{eq:gamma_M} as $\gamma_\cP$. \hfill $\blacksquare$

\subsection{Properties of $\gamma_\cM$}

\noindent In this section we discuss several properties of $\gamma_{\cM}$.

\paragraph{Classical interpretation of $\gamma_\cP$.}
For commuting $\{ M_x \}_{x}$, namely when the imperfect measurement is specified by the noisy detection channel $\cP\sim\{p(x|i)\}$ such that $\forall_x:\;M_x=\sum_i p\!\left(x|i\right)\Pi_{i}$,
$\gamma_\cP\equiv\gamma_\cM$ has a classical interpretation. Note that in this case \eqnref{eq:gamma_M} reads:
\begin{align}
\gamma_\cP
&=\underset{{\bf a},{\bf b}}{\max}\sum_{x}\frac{\left(\sum_{i}p\!\left(x|i\right)a_{i}b_{i}\right)^{2}}{\sum_{i}p\!\left(x|i\right)a_{i}^{2}},
\label{eq:gamma_P}
\end{align}
where ${\bf a},{\bf b}$ are real normalised, orthogonal vectors: ${\bf a} \cdot {\bf b}=0$, $|{\bf a}|^2=|{\bf b}|^2=1$. It is simple to see that ${\bf a},{\bf b}$ can be assumed to be real vectors. In order to gain further intuition, we can define $p_{i}:=a_{i}^{2}$ and a `derivative' $\upd p_{i}:=a_{i}b_{i}$ such that then $a_{i}=\sqrt{p_{i}}$ and $b_{i}=2\upd(\sqrt{p_{i}})$. As a result, it can be seen that the constraints of $|{\bf a}|^2=1$, ${\bf a} \cdot {\bf b}=0$, $|{\bf b}|^2=1$ are equivalent to the constraints:~$\sum_{i}p_{i}=1$,  $\sum_{i} \upd p_{i}=0$, $\sum_{i}\frac{\upd p_{i}^{2}}{p_{i}}=1$. In nuce, $\left\{ p_{i}\right\} _{i}$ is a probability distribution, $\left\{ \upd p_{i}\right\} _{i}$ is the derivative vector of the probability distribution and the constraint of $\sum_{i}\frac{\upd p_{i}^{2}}{p_{i}}=1$ is a normalization constraint:~the original FI equals to $1$. In this new notation, \eqnref{eq:gamma_P} reads: 
\begin{equation}
\gamma_{\cP}=\underset{\boldsymbol{p},\boldsymbol{\upd p}}{\max}
\sum_x
\frac{\left(\sum_{i}p\!\left(x|i\right)\upd p_{i}\right)^{2}}{\sum_{i}p\!\left(x|i\right)p_{i}},
\label{eq:classical_interp}
\end{equation}      
with the constraint of $\sum_{i}\frac{\upd p_{i}^{2}}{p_{i}}=1$. 
Hence, $\gamma_\cP$ can interpreted as the \emph{optimal noisy classical FI}, optimised over all $\left\{ p_{i}\right\} _{i},\left\{ \upd p_{i}\right\} _{i}$
with the original FI of 1.

\paragraph{Data processing inequality $\gamma_{\cP_2 \circ \cP_1} \leq \gamma_{\cP_1}$.}
Let us consider again the setting of $\gamma_{\cP}$ in \eqnref{eq:gamma_P} specified by the noisy detection channel, $\cP$, that constitutes a stochastic map with entries $[\cP]_{xi}=p(x|i)$ and yields the imperfect measurement $\{M_x=\sum_i [\cP]_{xi}\Pi_i\}_x$. Now, considering a composition of any two noisy detection channels $\cP_2 \circ \cP_1$ as an effective stochastic map, the corresponding imperfect measurement simply reads $\{\tilde{M}_x=\sum_i [\cP_2\cP_1]_{xi}\Pi_i\}_x$. We demonstrate that the resulting $\gamma$-coefficient \eref{eq:gamma_P} obtained via such a composition must be contractive, i.e.~$\gamma_{\cP_2 \circ \cP_1} \leq \gamma_{\cP_1}$. 

We prove this by first observing that \eqnref{eq:classical_interp} satisfies $\gamma_{\cP}=\underset{\boldsymbol{p},\boldsymbol{\upd p}}{\max}\sum_{x}\frac{\left(\sum_{i}p\!\left(x|i\right)\upd p_{i}\right)^{2}}{\sum_{i}p\!\left(x|i\right)p_{i}}\leq\underset{\boldsymbol{p},\boldsymbol{\upd p}}{\max}
\sum_{i}\frac{\upd p_{i}^{2}}{p_{i}}=1$, which follows from the Cauchy-Schwarz inequality $\sum_{x}\frac{\left(\sum_{i}p\!\left(x|i\right)\upd p_{i}\right)^{2}}{\sum_{i}p\!\left(x|i\right)p_{i}}\leq \sum_{x}p\!\left(x|i\right)\left(2\upd\sqrt{p_{i}}\right)^{2}=\sum_{i}\frac{\upd p_{i}^{2}}{p_{i}}$. Now, if we consider \eqnref{eq:classical_interp} but for $\cP=\cP_{2}\circ\cP_{1}$, we have $\gamma_{\cP}=\underset{\boldsymbol{p},\boldsymbol{\upd p}}{\max}\sum_{x}\frac{\left(\sum_{j,k}p_{2}\left(x|k\right)p_{1}\left(k|j\right)\upd p_{j}\right)^{2}}{\sum_{j,k}p_{2}\left(x|k\right)p_{1}\left(k|j\right)p_{j}}$. Hence, defining $p'_{k}=\sum_j p_1\!\left(k|j\right)p_{j}$, we similarly obtain by Cauchy-Schwarz inequality:~$\gamma_{\cP}=\underset{\boldsymbol{p},\boldsymbol{\upd p}}{\max}\sum_{x}\frac{\left(\sum_{k}p_{2}\left(x|k\right)\upd p'_{k}\right)^{2}}{\sum_{k}p_{2}\left(x|k\right)p'_{k}}\leq\underset{\boldsymbol{p},\boldsymbol{\upd p}}{\max}\sum_{k}\frac{\left(\upd p'_{k}\right)^{2}}{p'_{k}}=\gamma_{\cP_{1}}$, which completes the proof.

\paragraph{Monotonically increasing with the number of probes, i.e.~$\forall_\cM:\;\gamma_{\cM}\leq\gamma_{\cM\otimes I}\leq\gamma_{\cM\otimes\cM}$}.
Let us first prove that $\gamma_{\cM\otimes I}\leq\gamma_{\cM\otimes\cM}$ for any imperfect measurement $\cM$, which follows from convexity. In general, for any two positive semidefinite operators $M_{1},M_{2}\ge0$:
\begin{align}
&\frac{\left(\text{Re}\langle\xi_{\perp}|M_{1}|\xi\rangle\right)^{2}}{\langle\xi|M_{1}|\xi\rangle}+\frac{\left(\text{Re}\langle\xi_{\perp}|M_{2}|\xi\rangle\right)^{2}}{\langle\xi|M_{2}|\xi\rangle}\geq \nonumber \\
&\frac{\left(\text{Re}\langle\xi_{\perp}|M_{1}+M_{2}|\xi\rangle\right)^{2}}{\langle\xi|M_{1}+M_{2}|\xi\rangle},    
\end{align}
which follows from the convexity of the function $\frac{x^{2}}{y}$, where we take $x_{i}=\text{Re}\langle\xi_{\perp}|M_{i}|\xi\rangle$, $y_{i}=\langle\xi|M_{i}|\xi\rangle$. This convexity implies:
\begin{align}
\sum_{j,k}\frac{\left(\text{Re}\langle\xi_{\perp}|M_{j}\otimes M_{k}|\xi\rangle\right)^{2}}{\langle\xi|M_{j}\otimes M_{k}|\xi\rangle}
&\geq
\sum_{j}\frac{\left(\text{Re}\langle\xi_{\perp}|\sum_{k}M_{j}\otimes M_{k}|\xi\rangle\right)^{2}}{\langle\xi|\sum_{k}M_{j}\otimes M_{k}|\xi\rangle} \nonumber\\
&=\sum_{j}\frac{\left(\text{Re}\langle\xi_{\perp}|M_{j}\otimes I|\xi\rangle\right)^{2}}{\langle\xi|M_{j}\otimes I|\xi\rangle}    
\end{align}
and, hence, $\gamma_{\cM\otimes I}\leq\gamma_{\cM\otimes\cM}$.

Now, $\gamma_{\cM}\leq\gamma_{\cM\otimes I}$ is assured by the fact that the maximisation over all orthogonal $\ket{\xi}$ and $\ket{\xi_\perp}$ in \eqnref{eq:gamma_M} defining 2-dimensional subspace in the support of $\cM$, is trivially contained within the maximisation over all orthogonal $\ket{\tilde{\xi}},\ket{\tilde{\xi}_\perp}$ lying in the support of $\cM\otimes I$. In particular, for any two $\ket{\xi}$ and $\ket{\xi_\perp}$ one may choose $\ket{\tilde{\xi}}=\ket{\xi}\ket{\chi}$ and $\ket{\tilde{\xi}_\perp}=\ket{\xi_\perp}\ket{\chi}$ with any $\ket{\chi}$, so that
\begin{align}
\sum_{j}\frac{\left(\text{Re}\langle\xi_{\perp}|M_{j}|\xi\rangle\right)^{2}}{\langle\xi|M_{j}|\xi\rangle}=\sum_{j}\frac{\left(\text{Re}\langle\chi|\langle\xi_{\perp}|M_{j}\otimes I|\xi\rangle|\chi\rangle\right)^{2}}{\langle\chi|\langle\xi|M_{j}\otimes I|\xi\rangle|\chi\rangle}.    
\end{align}

\paragraph{Sufficient condition for $\gamma_{\cM}=1$.} 
We mention in the main text that a sufficient condition for a perfect QFI, i.e.~$\gamma_\cM=1$, is perfect distinguishability between two states:~there exist two orthogonal states, $\ket{\xi}$ and $\ket{\xi_{\perp}}$, such that for every $M_x$ either $M_x \ket{\xi}=0 $ or $M_x\ket{\xi_\perp}=0$. In order to see this, observe  from the Cauchy-Schwarz inequality \eref{eq:CS_ineq} that $\gamma_\cM=1$ if and only if 
there exist $\ket{\xi}, \ket{\xi_\perp}$ such that $\sqrt{M_x}|\xi\rangle\propto\sqrt{M_x}|\xi_{\perp}\rangle$ for every $x$. It is straightforward to see that given perfect distinguishability this condition is satisfied, with the proportionality constant being exactly zero for all $x$.

\subsection{Unitary encoding with two-outcome measurement for a qubit:~Optimal state and measurement}
\label{sec:optimal_meas_qubit}
Using Bloch representation, our initial probe state is $\rho=\frac{1}{2}\big(\id+\rvec_0\cdot\boldsymbol{\sigma}\big)$, where the real Bloch vector has the usual constraint $|\rvec_0|^2=\mathrm{r}_{0\mathrm{x}}^2+\mathrm{r}_{0\mathrm{y}}^2+\mathrm{r}_{0\mathrm{z}}^2\leq1$ with equality for pure state. After the encoding with $U_\theta=\upe^{\upi h\theta}$, $h=\pauliz/2$,  $\rho$ evolves to $\rho(\theta)=\frac{1}{2}\big(\id+\rvec(\theta)\cdot\boldsymbol{\sigma}\big)$, where $\mathrm{r}_{\mathrm{x}}(\theta)=\mathrm{r}_{0\mathrm{x}}\cos(\theta)+\mathrm{r}_{0\mathrm{y}}\sin(\theta), \mathrm{r}_{\mathrm{y}}(\theta)=-\mathrm{r}_{0\mathrm{x}}\sin(\theta)+\mathrm{r}_{0\mathrm{y}}\cos(\theta)$, and $\mathrm{r}_{\mathrm{z}}(\theta)=\mathrm{r}_{0\mathrm{z}}$. Moreover, $\dot{\rvec}(\theta)=\partial_\theta \rvec(\theta)$ is perpendicular to $\rvec(\theta)$, with $|\dot{\rvec}|^2=\mathrm{r}_{0\mathrm{x}}^2+\mathrm{r}_{0\mathrm{y}}^2\leq1$. The two-outcome measurement prior to the stochastic mapping are described by the operators $\Pi_{1,\phivec}=\frac{1}{2}\big(s\id+\mvec(\phivec)\cdot\boldsymbol{\sigma}\big)$ and $\Pi_{2,\phivec}=\frac{1}{2}\big((2-s)\id-\mvec(\phivec)\cdot\boldsymbol{\sigma}\big)$, with the positive constraints $0<m\equiv|\mvec|\leq s^*\equiv\min\{s,2-s\}\leq1$. In the main text we consider from onset projective measurements with $m=1$, but for the sake of mathematical completeness, let us for now allow any two-outcome measurement, and show later $m=1$ is optimal indeed.

With $\phivec=\{m,\varphi,\vartheta\}$, we parametrize $\mvec(\phivec)=m\big[\cos\varphi(\cos\vartheta\,\bvec_1+\sin\vartheta\,\bvec_2)+\sin\varphi\,\bvec_3\big]$ in the local Cartesian basis $\{\bvec_1=\rvec(\theta), \bvec_2=\dot{\rvec}/|\dot{\rvec}|, \bvec_3=\bvec_1\times\bvec_2\}$. The respective outcome probabilities are $\pphitheta(1)=\frac{1}{2}\big(s+\mvec\cdot\rvec\big)=\frac{1}{2}(s+m\cos\varphi\cos\vartheta)$ and $\pphitheta(2)=1-\pphitheta(1)$, while $\dotpphitheta(1)=-\dotpphitheta(2)=\frac{1}{2}m\cos\varphi\sin\vartheta|\dot{\rvec}|$. Note that, while the $\theta$ dependence are not seen explicit here, they are present still, as $\varphi$ and $\vartheta$ are defined with respect to $\theta$. For noisy detection channel that is specified by the stochastic mapping $\cP\sim\{p(x|i)\}$, i.e. $\cM\sim\{M_x=\sum_i p(x|i)\Pi_i\}$, we have $\qxphitheta=\sum_i p(x|i)\piphitheta$, and the FI, $F=\sum_x f_x$, with 
\begin{align}
f_x&=\dotqxphitheta^2/\qxphitheta=\frac{\frac{1}{2}a_x^2}{b_xy^2+c_xy}\sin^2\vartheta|\dot{\rvec}|^2 \label{eq:Asmallfx}
\end{align}
where  $y\equiv(m\cos\varphi)^{-1}\geq y^*\equiv1/s^*\geq1$, 
$a_x=p(x|1)-p(x|2), b_x=p(x|1)+(2-s)p(x|2)$, and  $c_x= a_x\cos\vartheta$. 

Consider now maximization of $f_x$ over the input state and the measurement. Evidently, we should choose the input state such that $|\dot{\rvec}|=1$, i.e., pure state that lies in the equatorial plane of the Bloch sphere. This choice also means that $\bvec_3$ is now $\evec_\mathrm{z}$. Moreover, we should minimize the function $g_x(y)=b_xy^2+c_xy$ in the denominator, subject to $y\geq y^*$. First then, we should choose $2-s=s^*$ in $b_x$. Next, since $g_x$ is a convex function, and the roots of $g_x$ are 0 and $-c_x/b_x$ for which $|c_x/b_x|\leq y^*$, we have $\min_{y\geq y^*} g_x(y)=g_x(y^*)$. That is, we have $\varphi=0$ and $m=s^*$, such that $\mvec$ has no $\bvec_3=\evec_\mathrm{z}$ component.

To confirm that we should always choose projective measurement before the noisy detection channel whenever possible, i.e., $m=s^*=1$, we put in $s^*=1/y^*$ into $g_x(y^*)$ for an explicit convex function of $y^*$. One can then verify readily that the roots are now not greater than 1, and therefore, the optimal choice of $y^*$ is 1. Finally, as all the above optimizations hold for all $f_x$, it follows that they apply to the total FI, and therefore
\begin{align}\label{eq:AFmax}
\bar{\mathcal{F}}^{\mathrm{(im)}}&=\max_\rho \max_{\phivec} F \nonumber\\
&=\max_{\vartheta} \sum_x \frac{\frac{1}{2}\big(p(x|1)-p(x|2)\big)^2 \sin^2\vartheta}{p(x|1)+p(x|2)+\big(p(x|1)-p(x|2)\big)\cos\vartheta}.
\end{align} 

Note that as \eqnref{eq:AFmax} depends on $\vartheta$, which is an angle defined relative to $\rvec(\theta)$, it follows that in this case here we have a freedom to fix either the measurement or the input state and optimize over the other, as long as both are restricted to the equatorial plane. In particular, we may fix $\rho=\ket{+_\mathrm{y}}\bra{+_\mathrm{y}}$, i.e., $\rvec_0=\evec_\mathrm{y}$, and optimize over $\{\Pi_{i,\phivec}=\ket{\Pi_i}\bra{\Pi_i}\}$ with $\ket{\Pi_{1,2}}=(\ket{0}\pm\upe^{\upi\phi}\ket{1})/\sqrt{2}$, i.e., $\mvec=\cos\phi\,\evec_\mathrm{x}+\sin\phi\,\evec_\mathrm{y}$. Equivalently, we may optimize over $\rho=\ket{\psi}\bra{\psi}$ with $\ket{\psi}=\upe^{\upi\phi\pauliz/2}\ket{+_\mathrm{y}}=(\ket{0}+\upi\upe^{\upi\phi}\ket{1})/\sqrt{2}$, i.e., $\rvec_0=-\sin\phi\,\evec_\mathrm{x}+\cos\phi\,\evec_\mathrm{y}$, with fixed $\Pi_{1,\phivec}=\ket{+}\bra{+}, \Pi_{2,\phivec}=\ket{-}\bra{-}$, i.e., $\mvec=\evec_\mathrm{x}$. Both give the same expression with $\vartheta=\theta+\phi-\pi/2$ in \eqnref{eq:AFmax}, which turns into \refmain{Eq.~(9) in the main text}.


\subsection{Relations of quantum metrology with imperfect measurements to the ``standard'' setting of noisy parameter encoding}
In this section, we discuss in detail the differences and relation between our quantum metrology with imperfect measurement model and the ``standard" noisy metrology model. In the latter, the measurement is taken to be perfect, and the noise is described by a CPTP channel that acts before the measurement stage. 

\begin{figure}[t!]
	\centering
	\includegraphics[width=0.35\textwidth]{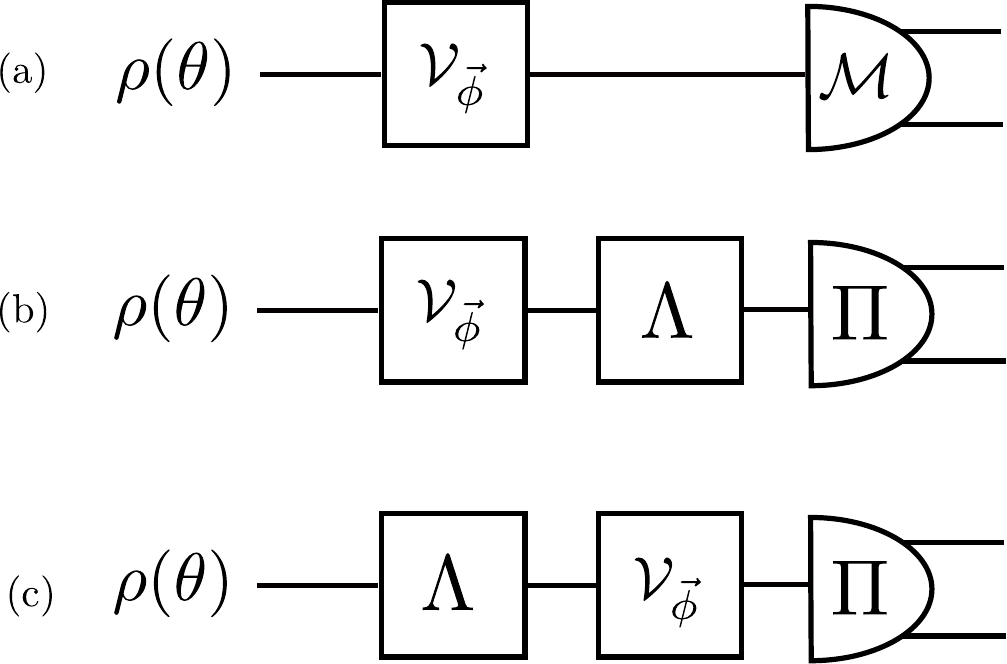}
	\caption{\textbf{Different models for noisy quantum metrology.} (a) Quantum metrology with imperfect measurement $\cM$. $\cV_{\vec{\phi}}$ is some control unitary operation for optimizing the measurement basis. (b) Equivalent picture of (a), with $\Pi$ being a projective measurement which can be chosen and fixed arbitrarily, and $\Lambda$ is a CPTP channel. (c) Quantum metrology with ``standard" noisy parameter encoding, where the noise, described by a CPTP channel $\Lambda$, is \emph{independent} of the choice of the measurement settings $\phivec$.}
	\label{fig:QMwIM-note-fig-1}
\end{figure}

Firstly, let us denote $\cB(\cH_d)$ as the set of bounded linear
operators on the Hilbert space $\cH_d$ with dimension  $d$. Then, recall that our quantum metrology with imperfect measurement on the physical level is as depicted in \figref{fig:QMwIM-note-fig-1}(a): an encoded qudit described by the state $\rho(\theta)\in \cB(\cH_d)$, undergoes a control unitary operation $\cV_{\phivec}\sim\{\Vphi\}$ such that $\rho(\theta)\rightarrow \cV_{\phivec}[\rho(\theta)]=\Vphi\rho(\theta)\Vphi^\dagger$, which is then subject to an imperfect measurement $\cM\sim\{M_x\}_{x=1}^{|X|}$. Here, all the $M_x$ and the $\Vphi$ are elements of $\cB(\cH_{d})$ as well. The probability of getting the outcome $x$ is given by the Born's rule, $\qxphitheta=\Tr\{M_x \Vphi\rho(\theta)\Vphi^\dagger\}$. 

We can also think of it in a mathematically equivalent picture, as in \figref{fig:QMwIM-note-fig-1}(b), 
where we describe the imperfect measurement $\cM\sim\{M_x\}_{x=1}^{|X|}$ rather by a noisy CPTP channel $\Lambda$ that is followed by a perfect projective measurement $\Pi\sim\{\Pi_x=\proj{x}\}_{x=1}^{|X|}$. That is, we can always find some quantum (CPTP) map $\Lambda\sim\{k_\ell\}_\ell$ that allows for the \emph{conjugate-map decomposition} of $\cM$, i.e.:
\begin{equation}\label{eq:ALambdachanneldef}
\forall_{x=1,\cdots,|X|}:\quad
M_x = \Lambda^\dagger [\Pi_x] = \sum_{\ell} k_\ell^\dagger \Pi_x k_\ell,
\end{equation}
or, rewriting the above in a compact way:
\begin{equation}\label{eq:ALambdachanneldef2}
\cM = \Lambda^\dagger [\Pi].
\end{equation}

Note that:~(i) the channel $\Lambda:\cB(\cH_{d})\rightarrow\cB(\cH_{|X|})$ acts on operators of $d$ dimension and outputs operators of $|X|$ dimension;~(ii) given $\cM$, there are multiple $\Lambda$ (and $\Pi$) satisfying \eqnref{eq:ALambdachanneldef2};~and (iii) $\Lambda$ is defined \emph{independently} of the control unitary $\cV_{\phivec}$---one can consider the l.h.s.~of \eqnref{eq:ALambdachanneldef} as $\Vphi^{\dagger} M_{x}\Vphi$ instead, and define $\phivec$-dependent $\Lambda$, but doing so is neither necessary nor helpful.

One can prove that the condition \eqref{eq:ALambdachanneldef}, or \eqref{eq:ALambdachanneldef2}, can always be satisfied, by considering as an example the \emph{quantum-classical channel} that we used as well in the main text (Lemma 3 and Corollary 2)~%
\footnote{Note that, in the main text, the quantum-classical channel $\Lambda_{\theta,\phivec}$ is defined together with the encoding $\cE_{\theta}$ as well as the control operation, i.e. $\Lambda_{\theta,\phivec}=\Lambda\circ\cV_{\phivec}\circ\cE_{\theta}$}%
, namely,
\begin{align}\label{eq:AQCchannel}
\Lambda\sim \Big\{\ket{x}\bra{i} \sqrt{M_x}\Big\}_{x,i} \quad x=1,\dots,|X|; \quad i=1,\cdots,d.
\end{align}
Another example of a CPTP map that also always satisfies \eqnref{eq:ALambdachanneldef2} is
\begin{align}\label{eq:ATuviachannel}
\Lambda\sim \Big\{\sum_{x=1}^{|X|} \ket{x}\bra{i} \sqrt{M_x}\Big\}_{i=1}^{d},
\end{align}
whose rank is now $d$, rather than $d\cdot\dimM$ in \eqnref{eq:AQCchannel}.

Note that, both the orthogonal sets of basis ket $\{\ket{x}\}_{x}$ and bra $\{\bra{i}\}_{i}$ in \eqnsref{eq:AQCchannel}{eq:ATuviachannel} can essentially be understood as flags, and can be chosen arbitrarily. Importantly, despite this equivalent picture in Fig.\ref{fig:QMwIM-note-fig-1}(b), it is however still not the same as the ``standard" noisy metrology scheme as depicted in Fig.\ref{fig:QMwIM-note-fig-1}(c), where the noise, described by a CPTP map $\Lambda$, acts \emph{independently} of the choice of the control unitary $\cV_{\phivec}$. The crucial difference is that, in the former the control operation $\cV_{\phivec}$ is applied \emph{before} the channel $\Lambda$, whereas in the latter it is applied \emph{after} the channel $\Lambda$. Then, unless $\cV_{\phivec}$ commutes with $\Lambda$, which is hardly ever true, the two noise model will not be equivalent.

Let us still study more closely the two models from the perspective of parameter estimation, whereby the focus is on the QFI and channel QFI (perfect versus imperfect). Denote $F[\rho(\theta),\cM]$ as the classical Fisher information for $\theta$ with the state $\rho(\theta)$ and measurement $\cM\sim\{M_x\}_x$, i.e., $F[\rho(\theta),\cM]=\sum_x \frac{(\partial_\theta \Tr\{\rho(\theta)M_x\})^2}{\Tr\{\rho(\theta)M_x\}}$. Then, for the imperfect measurement model, the imperfect QFI is given by \refmain{(see Eq.~(5) in the main text)}:
\begin{align}\label{eq:AimQFIbound1}
\FQim&=\max_{\phivec} F[\cV_{\phivec} [\rho(\theta)], \Lambda^\dagger [\Pi]] \nonumber\\
&:=F[\opt{\cV}[\rho(\theta)], \Lambda^\dagger [\Pi]] =F[(\Lambda\circ\opt{\cV})[\rho(\theta)], \Pi].
\end{align}
Note that the optimal control unitary $\opt{\cV}\sim\{\Vopt\}$ generally depends on the given $\rho(\theta)$, though for simplicity of the notation we will not write out this dependence explicitly for now. It then follows that 
\begin{align}\label{eq:AimQFIbound2}
	\FQim&\leq\max_{W} F[(\Lambda\circ\opt{\cV})[\rho(\theta)], W\Pi W^\dagger] \nonumber\\
&:=\FQ[(\Lambda\circ\opt{\cV})[\rho(\theta)]],
\end{align}
where $W$ is a unitary operator in $\cB(\cH_{|X|})$, and so we may formally upper bound the imperfect QFI by the QFI of $(\Lambda\circ\opt{\cV})[\rho(\theta)]$, which can be interpreted as a state that undergoes a ``standard" noisy channel $\Lambda\circ\opt{\cV}$. Moreover, given the imperfect measurement $\cM$, one can further optimize over different possible $\Lambda$ that satisfies \eqnref{eq:ALambdachanneldef2}, and get
\begin{align}\label{eq:AimQFIbound3}
	\FQim&\leq\min_{\substack{\Lambda\\\Lambda^\dagger[\Pi]=\cM}}\FQ[(\Lambda\circ\opt{\cV})[\rho(\theta)]].
\end{align}
Despite the established formal relations \eqnsref{eq:AimQFIbound2}{eq:AimQFIbound3}, note that $\Lambda\circ\opt{\cV}$ is defined using the knowledge about optimal control unitary. However, if we had known
$\opt{\cV}$, we would have already in fact obtained $\FQim$, and there is no need for the upper bounds. In other words, the formal bounds \eqnsref{eq:AimQFIbound2}{eq:AimQFIbound3} are not that meaningful in practice.

\subsubsection{Proof of the Observation 1 in Methods}
\label{sec:proof_Obs1}
Still, there is a special case worth mentioning, where we can indeed meaningfully evaluate a QFI-based upper bound without solving exactly for the $\opt{\cV}$. Suppose that from the symmetry of the estimation problem we know that the optimal control operation must also carry it, so that its optimisation may be restricted to elements in some compact group $G$, i.e., $\opt{\cV}\in G$. Then, if $\Lambda$ satisfying \eqnref{eq:ALambdachanneldef2} is known to be \emph{$G$-covariant}, i.e., 
\begin{equation}
    \forall_{g\in G}:\;\Lambda\circ\cV_{g}=\cW_g\circ\Lambda,
    \label{eq:comm_Lambda_V}
    \end{equation}
where $\cW_g$ is some unitary representation of $G$ in $\cH_\dimM$--- in particular, for $\cW_{\mathrm{opt}}$ correspondingly to $\opt{\cV}$ in \eqnref{eq:comm_Lambda_V},
we have from \eqnref{eq:AimQFIbound2} that
\begin{align}\label{eq:AimQFIbound4}
\FQim&\leq \FQ[(\opt{\cW}\circ\Lambda)[\rho(\theta)] ] =\FQ[\Lambda[\rho(\theta)]],
\end{align}
where the equality in \eqref{eq:AimQFIbound4} follows from the fact that QFI is invariant under parameter-independent unitary transformation.  \hfill $\blacksquare$  \bigskip 

Extension to channel QFI is straightforward. Let us denote the (perfect) parameter-encoding channel as $\cE_\theta$, such that $\rho(\theta)=\cE_\theta [\rho]$ for the input probe state $\rho$, which we will eventually optimize over. For the imperfect measurement case, we have by definition \refmain{(see Eq.~(5) in the main text)}:\begin{align}\label{eq:AimcQFIbound1}
\FQimbar&=\max_{\rho}\max_{\phivec} F[(\cV_{\phivec}\circ \cE_\theta) [\rho], \Lambda^{\dagger}[\Pi]] \nonumber \\
&:= F[(\Lambda\circ\opt{\cV}\circ \cE_\theta) [\rhoopt] , \Pi] \nonumber\\
&=\sum_x \frac{(\Tr\{(\Lambda\circ\opt{\cV}\circ \dot{\cE}_\theta) [\rhoopt] \Pi_x\})^2}{\Tr\{(\Lambda\circ\opt{\cV}\circ \cE_\theta) [\rhoopt] \Pi_x\}},
\end{align}
where $\opt{\cV}$ and $\rhoopt$ are respectively the optimal control unitary and input state, and $\dot{\cE}_{\theta}=\partial_{\theta}\cE_{\theta}$ is the derivative of the encoding channel w.r.t. the parameter, such that for any $\rho(\theta)=\cE_{\theta}[\rho]$, $\partial_{\theta}\rho(\theta)=\dot{\cE}_{\theta}[\rho]$. Similarly to \eqnref{eq:AimQFIbound2}, one then obtains
\begin{align}\label{eq:AimcQFIbound2}
	\FQimbar&\leq \max_{\sigma} \max_{W} F[(\Lambda\circ\opt{\cV}\circ \cE_\theta) [\sigma] , W\Pi W^{\dagger}] \nonumber\\ 
	&:=\FQbar[(\Lambda\circ\opt{\cV}\circ\cE_\theta)],
\end{align}
where $\sigma$ is some state in $\cH_{d}$, $W$ is a unitary operator in $\cB(\cH_{\dimM})$, and $\FQbar$ is the channel QFI. Again, although \eqnref{eq:AimcQFIbound2} now provides a formal relation between the imperfect channel QFI and the channel QFI of a ``standard" noisy encoding $(\Lambda\circ\opt{\cV}\circ\cE_\theta)$, it requires knowledge of the optimal control unitary $\opt{\cV}$ (which implicitly depends on the optimal state $\rhoopt$), and is hence in general not immediately applicable. 

Still, when we know that, thanks to some symmetry of the problem, the optimisation over control $\cV_{\phivec}$ can be restricted to elements of some compact group $G$, meaningful upper bound on the imperfect channel QFI can be formulated. Firstly, following directly from the Observation 1, in case the conjugate map $\Lambda$ satisfies not only \eqnref{eq:ALambdachanneldef2} but also the $G$-covariant condition \eqnref{eq:comm_Lambda_V}, we can immediately conclude by maximising \eqnref{eq:AimQFIbound4} over the input probe-states that $\FQimbar\leq\FQbar[(\Lambda\circ\cE_\theta)]$. 

\subsubsection{Proof of the Observation 2 in Methods}
\label{sec:proof_Obs2}
Secondly, suppose that the encoding channel $\cE_{\theta}$, as well as its derivate, $\dot{\cE}_{\theta}=\partial_{\theta}\cE_{\theta}$, are both $G$-covariant locally around the parameter value $\theta$, i.e., 
\begin{align}
    \forall_{g\in G}:\;\cV_g\circ\cE_{\theta}&=\cE_{\theta}\circ\cW_{g}, \nonumber\\
    \cV_g\circ\dot{\cE}_{\theta}&=\dot{\cE}_{\theta}\circ\cW_{g},
    \label{eq:comm_Lambda_VI}
    \end{align}
where $\cW_g$ is some unitary representation of $G$ in $\cH_d$--- in particular, $\cW_{\mathrm{opt}}$ correspondingly for $\opt{\cV}$.
In this case, from \eqref{eq:AimcQFIbound2}, 
\begin{align}\label{eq:AimcQFIbound4}
\FQimbar&\leq\FQbar[(\Lambda\circ\cE_\theta\circ\opt{\cW})] = \FQbar [(\Lambda\circ\cE_\theta)],
\end{align}
where the equality in \eqref{eq:AimcQFIbound4} follows from the fact that channel QFI is invariant under parameter-independent unitary transformation on the input probe state. Note that it is necessary to include the local $G$-covariant condition for $\dot{\cE}_{\theta}$ here, as FI is not a function of the state alone but also its derivative, c.f.~explicitly \eqnref{eq:AimcQFIbound1} and \eqnref{eq:QFI_def} below. \hfill $\blacksquare$

\subsubsection{Computing $\FQbar [(\Lambda\circ\cE_\theta)]$ in \eqnref{eq:AimcQFIbound4} as an SDP via a `seesaw' method}
\label{sec:seesaw}
The QFI is defined as a function of a quantum state $\rho\equiv\rho(\theta)$ and its derivative $\dot{\rho}\equiv\partial_\theta\rho(\theta)$, as follows~\cite{Braunstein1994}:
\begin{align}
\FQ[\rho,\dot{\rho}]:=
\begin{array}{c}
\Tr\!\left\{ \rho L^{2}\right\} =\Tr\!\left\{ \dot{\rho}L\right\} \\
\textnormal{s.t.}\;\dot{\rho}=\frac{1}{2}\left\{ \rho,L\right\} 
\end{array} &.
\label{eq:QFI_def}
\end{align}
Equivalently, it may be expressed as the maximisation of the \emph{error propagation formula} over all the quantum observables, i.e., Hermitian operators $O=O^{\dagger}$, as~\citep{Escher2012}:
\begin{equation}
\FQ[\rho,\dot{\rho}]=\max_{O=O^{\dagger}}\frac{\left|\left\langle \dot{O}\right\rangle \right|^{2}}{\Delta^{2}O}=\max_{O=O^{\dagger}}\frac{\left|\Tr\!\left\{ \dot{\rho}O\right\} \right|^{2}}{\Tr\!\left\{ \rho O^{2}\right\} -\Tr\!\left\{ \rho O\right\} ^{2}},
\label{eq:QFI_max_err_prop}
\end{equation}
which is always maximised by $\opt{O}=L-\Tr\!\left\{ \rho L\right\}$
with $L$ being the SLD operator defined implicitly in \eqnref{eq:QFI_def}.

However, the fraction in \eqnref{eq:QFI_max_err_prop} can always be rewritten by introducing another maximisation, i.e.:
\begin{equation}
\frac{\left|\left\langle \dot{O}\right\rangle \right|^{2}}{\Delta^{2}O}=\max_{\alpha\in\mathbb{R}}\left\{ -\alpha^{2}\Delta^{2}O+2\alpha\left|\left\langle \dot{O}\right\rangle \right|\right\} ,
\end{equation}
with the maximum occurring at $\opt{\alpha}=\frac{\left|\left\langle \dot{O}\right\rangle \right|}{\Delta^{2}O}$.
Hence, we may write again \eqnref{eq:QFI_max_err_prop} as
\begin{align}
\FQ[\rho,\dot{\rho}] & =\max_{O=O^{\dagger}}\max_{\alpha\in\mathbb{R}}\left\{ -\alpha^{2}\Delta^{2}O+2\alpha\left|\left\langle \dot{O}\right\rangle \right|\right\} \\
& =\max_{O'=O'^{\dagger}}\left\{ -\Delta^{2}O'+2\left|\left\langle \dot{O}'\right\rangle \right|\right\} ,
\label{eq:QFI_maxO'}
\end{align}
while noticing that the two maximisations can be recast into one by defining $O':=\alpha O$. Moreover, for any $O'$ above we may define a shifted operator $X:=O'-\left\langle O'\right\rangle$, so that substituting $O'=X+\Tr\!\left\{ \rho O'\right\} $
into \eqnref{eq:QFI_maxO'}, we obtain~\citep{macieszczak2013}:
\begin{align}
\FQ[\rho,\dot{\rho}] 
 & =\max_{X=X^{\dagger}}\left\{ -\left\langle X^{2}\right\rangle +2\left|\left\langle \dot{X}\right\rangle \right|\right\} \\
 & =\max_{X=X^{\dagger}}\left\{ -\left\langle X^{2}\right\rangle +2\left\langle \dot{X}\right\rangle \right\} ,\label{eq:QFI_maxX}
\end{align}
where the maximum is now performed over all Hermitian operators $X=X^{\dagger}$. We have also dropped the absolute value, as $\left\langle \dot{X}\right\rangle =\Tr\!\left\{ \dot{\rho}X\right\} =\partial_{\theta}\Tr\!\left\{ \rho X\right\}$ is real, while the first (quadratic) term above is unaffected by the $X\to-X$ transformation---and so must be the maximal value attained in \eqnref{eq:QFI_maxX}.

Let us note that \eqnref{eq:QFI_maxX} constitutes a valid lower bound on the QFI for any fixed $X$, i.e.
\begin{equation}
\forall_{X=X^{\dagger},\rho,\dot{\rho}}:\;-\Tr\!\left\{ \rho X^{2}\right\} +2\Tr\!\left\{ \dot{\rho}X\right\} \;\le\;\FQ[\rho,\dot{\rho}],
\label{eq:QFI_LB_fixedX}
\end{equation}
while the optimal $\opt{X}$, yielding the maximum in \eqnref{eq:QFI_maxX}, is related to the actual optimal observable $\opt{O}$ in \eqnref{eq:QFI_max_err_prop} via
\begin{align}
\opt{X} &= O'_{\mathrm{opt}}-\left\langle O'_{\mathrm{opt}}\right\rangle = \opt{\alpha}\left(\opt{O}-\left\langle \opt{O}\right\rangle \right) \\
& = \frac{\left|\left\langle \dot{O}_{\mathrm{opt}}\right\rangle \right|}{\Delta^{2}\opt{O}}\left(\opt{O}-\left\langle \opt{O}\right\rangle \right).
\label{eq:X_opt}
\end{align}
Hence, by substituting further $\opt{O}=L-\Tr\!\left\{ \rho L\right\}$, we can explicitly relate $\opt{X}$
to the SLD and the QFI, i.e.:
\begin{equation}
\opt{X}
=\frac{\FQ[\rho,\dot{\rho}]}{\left|\Tr\!\left\{ \dot{\rho}L\right\} \right|}\left(L-\Tr\!\left\{ \rho L\right\} \right).
\label{eq:X_opt(F,L)}
\end{equation}

Let us consider for our purposes the case of unitary parameter encoding, i.e.~$\cE_{\theta}=\cU_{\theta}\sim\{\upe^{-\upi\theta H}\}$ so that $\dot{\rho}=\upi[\rho,H]$, but the following analysis can be directly generalised to allow for arbitrary $\cE_\theta$ and $\dot{\cE}_\theta$. Then, using the expression \eref{eq:QFI_def} for the QFI, we may rewrite the potentially valid---e.g.~given the $G$-covariance \eref{eq:comm_Lambda_V} or \eref{eq:comm_Lambda_VI}---upper bound on the imperfect QFI $\FQimbar$ in \eqnref{eq:AimcQFIbound4} for the encoding $U_\theta=\upe^{-\upi \theta H}$ as:
\begin{eqnarray}
\FQbar[H,\Lambda]  
& := & \FQbar [(\Lambda\circ\cU_\theta)] \\
&=& \max_{\sigma\ge0}\;\FQ\!\left[\Lambda[\sigma],\Lambda\!\left[\upi[\sigma,H]\right]\right] \\
&=& \max_{\sigma\ge0}
\begin{array}{c}
\upi\,\Tr\!\left\{ \left[\sigma,H\right]\Lambda^{\dagger}\!\left[L\right]\right\} \\
\textnormal{s.t.}\;\Lambda\!\left[\upi[\sigma,H]\right]=\frac{1}{2}\left\{ \Lambda[\sigma],L\right\} 
\end{array}.
\label{eq:channelQFI_after}
\end{eqnarray}
Moreover, we may now use \eqnref{eq:QFI_maxX} to further re-express the above channel QFI (for $\Lambda\circ\cU_\theta$) as
\begin{equation}
\FQbar[H,\Lambda]=\max_{\sigma\ge0}\max_{X=X^{\dagger}}\Tr\!\left\{ \sigma\left(-\Lambda^{\dagger}\!\left[X^{2}\right]+2\mathrm{i}\left[H,\Lambda^{\dagger}[X]\right]\right)\right\} .
\label{eq:channelQFI_after_maxX}
\end{equation}

As a result, we can formulate a numerical \emph{`seesaw' algorithm} that allows us to compute the channel QFI \eref{eq:channelQFI_after} by exploiting \eqnref{eq:channelQFI_after_maxX}, as follows~\citep{macieszczak2013}:
\begin{enumerate}
\item Select (randomly) a starting state $\sigma_{0}$ and calculate the corresponding QFI $\FQ$ as well as the SLD $L$ it would lead to in \eqnref{eq:channelQFI_after}, i.e.~without performing the maximisation over the input states $\sigma$.
\item Use the obtained $\FQ$ and $L$ to compute the optimal operator $X_{0}=\opt{X}$ according to \eqnref{eq:X_opt(F,L)}, which then maximises \eqnref{eq:channelQFI_after_maxX} for the fixed state $\sigma_{0}$.
\item Maximise the expression \eref{eq:channelQFI_after_maxX} over the input states with the operator $X_{0}$ being now fixed, in order to determine the best state $\sigma_{1}$ that then yields
the tightest lower bound \eref{eq:QFI_LB_fixedX} on QFI for $X_{0}$.
\item Return to step 1 and use $\sigma_{1}$ as the new starting state. 
\end{enumerate}
The above procedure is computationally efficient, as $\FQ$ and $L$ in step 1 are obtained solving a linear
programme, as in \eqnref{eq:QFI_def}, while finding the optimal input state in step 3 corresponds to solving the maximal eigenvalue of a Hermitian operator defined within $(\dots)$ of \eqnref{eq:channelQFI_after_maxX}. Although the convergence of the algorithm is generally assured~\citep{macieszczak2013}, even if its rate is slow, at any stage it yields a valid lower bound \eref{eq:QFI_LB_fixedX} on the channel QFI \eref{eq:channelQFI_after_maxX}.

\subsubsection{Maximising $\FQbar[(\Lambda\circ\cE_\theta)]$ in \eqnref{eq:AimcQFIbound4} further, over all conjugate-map decompositions of an imperfect measurement}
\label{sec:seesaw_max_CPTPmaps}
Recall that $\Lambda$ in \eqnref{eq:AimcQFIbound4}, and so in \eqnref{eq:channelQFI_after_maxX}, corresponds to some valid conjugate-map decomposition of a given imperfect measurement $\cM$, for which it must fulfil the condition \eref{eq:ALambdachanneldef2}. As we now show, the above `seesaw' formulation allows naturally to incorporate in \eqnref{eq:channelQFI_after_maxX} also the maximisation over all such conjugate maps, i.e.~all quantum (CPTP) channels $\Lambda$ satisfying $\mathcal{M}=\Lambda^{\dagger}[\Pi]$ for some projective measurement $\Pi$.

Let us define the corresponding maximum as 
\begin{eqnarray}
\FQbar[H,\cM] 
&:=&
\max_{\Lambda\in\mathrm{CPTP}}\begin{array}{c}
\FQbar[H,\Lambda]\\
\mathrm{s.t.}\;\Lambda^{\dagger}[\Pi]=\mathcal{M}
\end{array}, \\
&=&
\max_{\substack{\Lambda\in\mathrm{CPTP}\\\sigma\ge0}}
\begin{array}{c} 
\upi\,\Tr\!\left\{ \left[\sigma,H\right]\Lambda^{\dagger}\!\left[L\right]\right\} \\
\mathrm{s.t.}\;\begin{array}{c}
\Lambda^{\dagger}[\Pi]=\mathcal{M}\\
\Lambda\left[\mathrm{i}[\sigma,H]\right]=\frac{1}{2}\left\{ \Lambda\left[\sigma\right],L\right\} 
\end{array}
\end{array}, 
\label{eq:imp_ch_QFI_UB_max}
\end{eqnarray}
where have substituted for $\FQbar[H,\Lambda]$ according to \eqnref{eq:channelQFI_after}. Note that in the above expression any projective measurement, $\Pi$, can be used as a reference, because $\FQbar[H,\cM]$ is invariant under the transformation $\Pi\to W\Pi W^{\dagger}$, $L\to WLW^{\dagger}$ and $\Lambda[\bullet]\to W\Lambda[\bullet]W^{\dagger}$, which implies $\Lambda^{\dagger}[\bullet]\to\Lambda^{\dagger}[W^{\dagger}\bullet W]$, for any unitary $W$. 

Now, similarly to \eqnref{eq:channelQFI_after_maxX}, we rewrite \eqnref{eq:imp_ch_QFI_UB_max} as
\begin{equation}
\FQbar[H,\cM]
=\max_{
\substack{\Lambda\in\mathrm{CPTP}\\\sigma\ge0,X=X^{\dagger}}
}
\begin{array}{c}
\Tr\!\left\{ \sigma\left(-\Lambda^{\dagger}\!\left[X^{2}\right]+2\upi\!\left[H,\Lambda^{\dagger}[X]\right]\right)\right\} \\
\mathrm{s.t.}\;\Lambda^{\dagger}[\Pi]=\mathcal{M}
\end{array},
\label{eq:imp_ch_QFI_UB_maxX}
\end{equation}
which we simplify further by denoting the action of any CPTP map $\Lambda$ via its Choi-Jamio\l{}kowski (CJ) state~\citep{Bengtsson2006}. 

In particular, for any input state $\sigma$ it is true that $\Lambda[\sigma]=\Tr_{\mathrm{B}}\!\left\{ \varrho_{\Lambda}(\mathbb{I}\otimes\sigma^{T})\right\}$, where the CJ state of the map $\Lambda$ is defined as $\varrho_{\Lambda}:=\Lambda\otimes\mathcal{I}\left[\left|\!\left.\mathbb{I}\right\rangle \right\rangle\!\!\left\langle\left\langle\mathbb{I}\right.\!\right|\right]$. Here, any square matrix $A$ defines a bipartite state $\left|\!\left.A\right\rangle \right\rangle =\sum_{ij}A_{ij}\left|i,j\right\rangle _{\mathrm{AB}}=\sum_{ij}A_{ij}\left|i\right\rangle _{\mathrm{A}}\left|j\right\rangle _{\mathrm{B}}$, so that $\left|\!\left.\mathbb{I}\right\rangle \right\rangle =\sum_{i}\left|i,i\right\rangle _{\mathrm{AB}}$ is the (unnormalised) maximally entangled state. Then, it is straightforward to prove that the CJ state for the conjugate map of $\Lambda$, i.e.~$\Lambda^{\dagger}$, is nothing but
$\varrho_{\Lambda^{\dagger}} 
=\mathbb{S}_{\mathrm{AB}}\varrho_{\Lambda}^{*}\mathbb{S}_{\mathrm{AB}}$,
where $\mathbb{S}_{\mathrm{AB}}$ is the swap operator such that $\mathbb{S}_{\mathrm{AB}}\left|\psi\right\rangle _{\mathrm{A}}\left|\phi\right\rangle _{\mathrm{B}}=\left|\phi\right\rangle _{\mathrm{A}}\left|\psi\right\rangle _{\mathrm{B}}$
for any states $\left|\phi\right\rangle $ and $\left|\psi\right\rangle $.
Consequently, the action of $\Lambda^{\dagger}$ corresponds to
\begin{eqnarray}
\forall_{X=X^{\dagger}}:\;\Lambda^{\dagger}[X]
&=& \Tr_{\mathrm{B}}\!\left\{ \varrho_{\Lambda^{\dagger}}(\mathbb{I}\otimes X^{T})\right\} \\
&=& \Tr_{\mathrm{B}}\!\left\{ \mathbb{S}_{\mathrm{AB}}\varrho_{\Lambda}^{*}\mathbb{S}_{\mathrm{AB}}(\mathbb{I}\otimes X^{T})\mathbb{S}_{\mathrm{AB}}\mathbb{S}_{\mathrm{AB}}\right\} \nonumber\\
&=& \Tr_{\mathrm{A}}\!\left\{ \varrho_{\Lambda}^{*}(X^{T}\otimes\mathbb{I})\right\},
\label{eq:conj_map_CJrep}
\end{eqnarray}
and the TP-property of $\Lambda$, imposing $\Tr_{\mathrm{A}}\varrho_{\Lambda}=\mathbb{I}$, ensures consistently that $\Tr_{\mathrm{B}}\varrho_{\Lambda^{\dagger}}=\Tr_{\mathrm{B}}\!\left\{ \mathbb{S}_{\mathrm{AB}}\varrho_{\Lambda}^{*}\mathbb{S}_{\mathrm{AB}}\right\} =\left(\Tr_{\mathrm{A}}\!\left\{ \varrho_{\Lambda}\right\} \right)^{\star}=\mathbb{I}$,
so that the conjugate map is indeed always unital, i.e.~$\Lambda^{\dagger}[\mathbb{I}]=\mathbb{I}$. 

Finally, using \eqnref{eq:conj_map_CJrep} to replace the maximisation over quantum maps $\Lambda$ in \eqnref{eq:imp_ch_QFI_UB_maxX} by the corresponding CJ states $\varrho_\Lambda$, we obtain
\begin{widetext}
\begin{align}
\FQbar[H,\cM]
& =
\max_{\varrho_{\Lambda}\ge0}\max_{\sigma\ge0}\max_{X=X^{\dagger}}
\quad
\begin{array}{l}
-\Tr\!\left\{ \sigma\,\Tr_{\mathrm{A}}\!\left\{ \varrho_{\Lambda}^{*}\left((X^{2})^{T}\otimes\mathbb{I}\right)\right\} \right\} +2\mathrm{i}\Tr\!\left\{ \sigma\left[H,\Tr_{\mathrm{A}}\!\left\{ \varrho_{\Lambda}^{*}(X^{T}\otimes\mathbb{I})\right\} \right]\right\} \\
\mathrm{s.t.}\;\forall_{x}:\,\Tr_{\mathrm{A}}\!\left\{ \varrho_{\Lambda}^{*}\left((\Pi_{x})^{T}\otimes\mathbb{I}\right)\right\} =M_{x},\;\Tr_{\mathrm{A}}\varrho_{\Lambda}=\mathbb{I}
\end{array}\nonumber \\
& =
\max_{\varrho_{\Lambda}\ge0}\max_{\sigma\ge0}\max_{X=X^{\dagger}}
\quad
\begin{array}{l}
-\Tr\!\left\{ \varrho_{\Lambda}^{*}\left((X^{2})^{T}\otimes\sigma\right)\right\} +2\mathrm{i}\Tr\!\left\{ \varrho_{\Lambda}^{*}(X^{T}\otimes\left[\sigma,H\right])\right\} \\
\mathrm{s.t.}\;\forall_{x}:\,\Tr_{\mathrm{A}}\!\left\{ \varrho_{\Lambda}^{*}\left((\Pi_{x})^{T}\otimes\mathbb{I}\right)\right\} =M_{x},\;\Tr_{\mathrm{A}}\varrho_{\Lambda}=\mathbb{I}
\end{array},
\label{eq:imp_ch_QFI_UB_max_CJ}
\end{align}
\end{widetext}
which we evaluate by adding to the aforementioned `seesaw' algorithm one more step in which we maximise over $\varrho_\Lambda\ge0$ (given the linear constraints to reproduce the elements of the imperfect measurement), while fixing the input state, $\sigma$, and the Hermitian operator, $X$.

\subsubsection{Example 1: Single-qubit phase sensing with bit-flip noise}
Let us first illustrate the subtleties of $G$-covariance, and limited applicability of the Observation 1 discussed in \secref{sec:proof_Obs1} for the \emph{imperfect QFI}, $\FQim$, by considering the unitary encoding $\cE_{\theta}\sim\{\upe^{\upi\theta\sigma_z/2}\}$ on a qubit, with imperfect measurement $\cM\sim\{M_1=\pp\proj{+}+(1-\qq)\proj{-}, M_2=\qq\proj{+}+(1-\pp)\proj{-}\}$ with $0\leq \pp,\qq\leq1$. Consider also the input probe state in the equatorial Bloch plane, such that for example the encoded state is $\rho(\theta)=\proj{\psi(\theta)}, \ket{\psi(\theta)}=\frac{1}{\sqrt{2}}(\ket{0}+\upe^{\upi\theta}\ket{1})$. 

\begin{figure}[t!]
  \centering
  \includegraphics[width=0.41\textwidth]{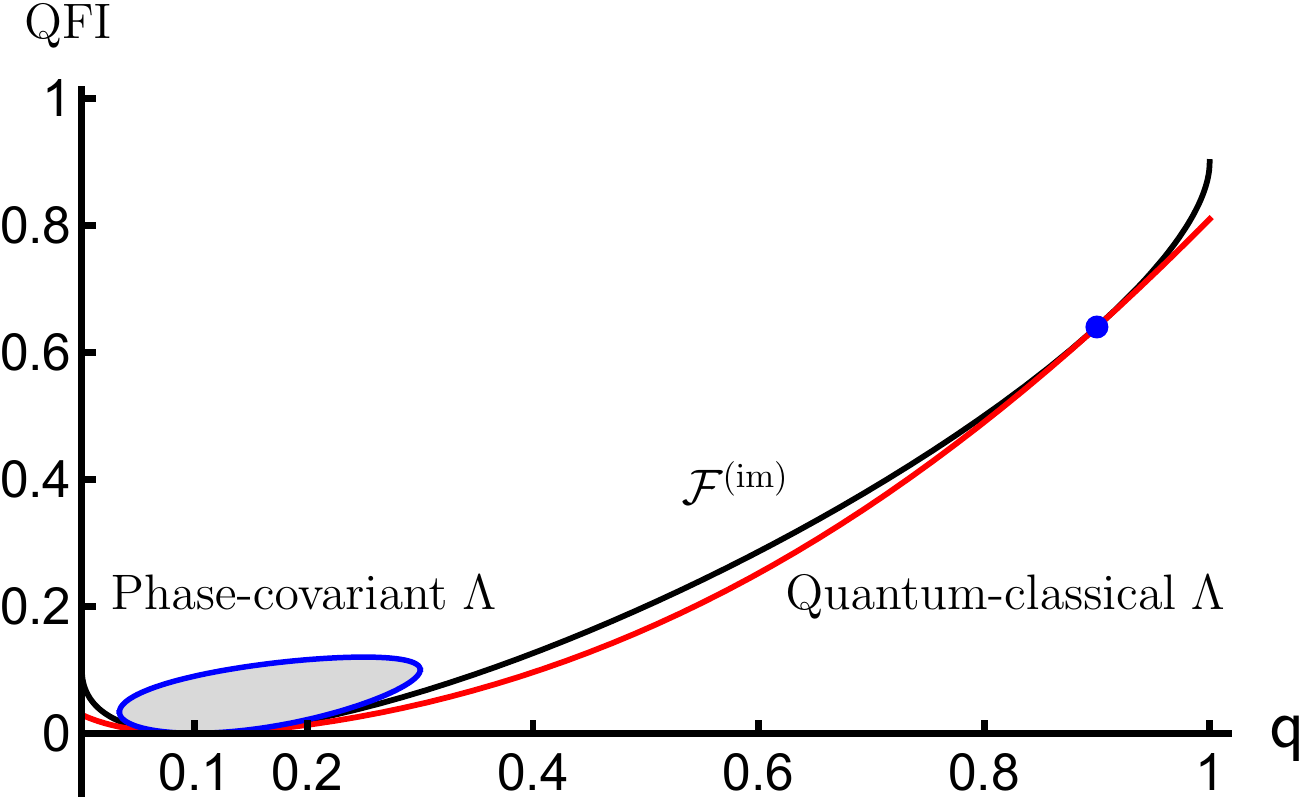}
  \caption{
  \textbf{Imperfect QFI $\FQim$ vs $G$-covariance-based bounds $\FQ[\Lambda[\rho(\theta)]]$ for a single qubit} ($N=1$), when estimating the phase encoded onto the input state $\ket{+}$, with the projective measurement affected by asymmetric bit-flip errors parametrised by $\pp=0.9$ and $\qq$ (horizontal axis). For any phase-covariant channel $\Lambda$ that commutes with the U(1)-representation $\cV_{g}\sim\{\upe^{\upi g \sigma_z}\}$---and, hence, the optimal unitary $\opt{\cV}$, see \secref{sec:optimal_meas_qubit}---the $G$-covariance condition \eref{eq:comm_Lambda_V} applies and the Observation 1 implies that $\FQim\leq\FQ[\Lambda[\rho(\theta)]]$. Hence, the QFI \eref{eq:APhasecovariantQFI} applicable to all phase-covariant maps $\Lambda$ satisfying \eqnref{eq:ALambdachanneldef2} yields upper bounds marked with the \emph{blue solid line} encircling the shaded region, as well as the \emph{blue dot} at $\pp=\qq$, that consistently lie above the \emph{black solid line} denoting the true $\FQim$, as given by \refmain{Eq.~(11) in the main text}. Still, the phase-covariance can be ensured only for a narrow range of $\qq$---beyond which any $\Lambda$ that fulfils \eqnref{eq:ALambdachanneldef2} may yield $\FQ[\Lambda[\rho(\theta)]]<\FQim$, e.g.~the quantum-classical channel defined in \eqnref{eq:AQCchannel} that leads to the \emph{red curve} above. However, we also observe that the channel defined in \eqnref{eq:ATuviachannel} yields $\FQ[\Lambda[\rho(\theta)]]=\FQim$ despite not being phase-covariant (the resulting QFI coincides with the \emph{black line}).
  }
  \label{fig:QMwIM-note-fig-2}
\end{figure}

As we have established above in \secref{sec:optimal_meas_qubit}, the optimal control unitary in this case must have the structure $\Vopt=\upe^{\upi\varphi_{\mathrm{opt}}\sigma_z}$ for some $\varphi_{\mathrm{opt}}$ and, hence, belongs to the U(1) group with unitary representation:~$\cV_{g}\sim\{\upe^{\upi g \sigma_z}\}$ and $g\in[0,2\pi)$. Consequently, when $\Lambda$ can be chosen to not only fulfil \eqref{eq:ALambdachanneldef2} but also be \emph{phase-covariant}~\cite{Holevo1993, holevo_covariant_1996, Smirne2016}, i.e. $\forall_{g}\,:\;[\Lambda,\cV_{g}]=0$, the $G$-covariance condition \eref{eq:comm_Lambda_V} is satisfied. For the special case of symmetric mixing with $\pp=\qq$ and any $0\leq\pp\leq1$, one can show that the dephasing channel fulfilling \eqref{eq:ALambdachanneldef2}, $\Lambda_{\mathrm{dep}}\sim \{\sqrt{\pp}\id,\sqrt{1-\pp}\sigma_z\}$, is phase-covariant and, hence, the upper bound \eqref{eq:AimQFIbound4} is applicable. Moreover, in this case the bound \eref{eq:AimQFIbound4} is tight:~by taking the limit $\delta\rightarrow0$ in \refmain{Eq.~(11) in the main text}, $\FQim$ coincides with $\FQ[\Lambda_{\mathrm{dep}}[\rho(\theta)]]=(2\pp-1)^2$. This is demonstrated by the blue dot in \figref{fig:QMwIM-note-fig-2} for $\pp=\qq=0.9$.

For asymmetric mixing $\pp\neq\qq$, however, one can show that for a wide range of $\pp$ and $\qq$ any channel $\Lambda$ satisfying \eqnref{eq:ALambdachanneldef2} cannot exhibit phase-covariance. In particular, $\Lambda$ can only be phase-covariant when there exists $\phi\in[0,2\pi)$ such that 
\begin{equation}
|\cos\phi|\geq |\delta| 
\quad\text{and}\quad 
1\geq\frac{4\eta^{2}}{\sin^{2}\phi}+\frac{\delta^{2}}{\cos^{2}\phi}
\label{eq:ACPconditions},
\end{equation}
where $\eta:=\pp+\qq-1$ and $\delta:=\pp-\qq$ as in the main text, and the above conditions originate from the CP-constraints on any phase-covariant channel (see e.g.~Ref.~\cite{Smirne2016}). Nonetheless, note that if conditions \eref{eq:ACPconditions} can be fulfilled for a given pair of $\pp\ne\qq$, there may exist more than one valid phase-covariant conjugate map $\Lambda^\dagger$ in \eqnref{eq:ALambdachanneldef2}, i.e.~there can be multiple solutions for $\phi$ satisfying the inequalities \eref{eq:ACPconditions}. 

In general, for any phase-covariant $\Lambda$ satisfying \eqnref{eq:ALambdachanneldef2}, $\FQim\leq\FQ[\Lambda[\rho(\theta)]]$ holds with $\FQim$ given by \refmain{Eq.~(11) in the main text}, while
\begin{align} \label{eq:APhasecovariantQFI}
\FQ[\Lambda[\rho(\theta)]]=\frac{\eta^{2}}{\sin^{2}\phi}
\end{align}
depending on the choice of $\phi$ satisfying constraints \eref{eq:ACPconditions}. In \figref{fig:QMwIM-note-fig-2}, we mark such a region by grey shading (blue oval boundary) with all values of $\FQ[\Lambda[\rho(\theta)]]$ in \eqnref{eq:APhasecovariantQFI} lying consistently above the true $\FQim$ (black line). Note that this is possible \emph{only} for relatively small $\qq$ (apart from special $\qq=\pp$), given the value of $\pp=0.9$ chosen.

On the contrary, within the range of $\pp\neq\qq$ in \figref{fig:QMwIM-note-fig-2} that yield imperfect measurements $\cM$ whose valid conjugate-map decompositions in \eqnref{eq:ALambdachanneldef2} may \emph{not} exhibit phase-covariance (or more generally, $G$-covariance)---range of $\qq$ without any solution marked in blue---the Observation 1 is not applicable and the upper bound \eqref{eq:AimQFIbound4} can no longer be taken for granted. Indeed, as demonstrated by the red solid line in \figref{fig:QMwIM-note-fig-2} for the quantum-classical channel \eqref{eq:ALambdachanneldef}, which yields a valid conjugate-map decomposition for any $\qq$, its QFI no longer upper-bounds the imperfect QFI with, in fact, $\FQim\geq\FQ[\Lambda[\rho(\theta)]]$. It is so, as $\FQ[\Lambda[\rho(\theta)]]$ equals then the classical FI for the imperfect measurement $\cM$ with no control ($\cV_{\phivec}=\id$) and, hence, by definition is always smaller or equal to $\FQim$. Interestingly, in this particular qubit example with asymmetric bit-flip noise, $\FQ[\Lambda[\rho(\theta)]]$ coincides with the imperfect QFI, $\FQim$, when $\Lambda$ is chosen to be the channel defined in \eqnref{eq:ATuviachannel}.

\begin{figure}[t!]
	\centering
	\includegraphics[width=0.41\textwidth]{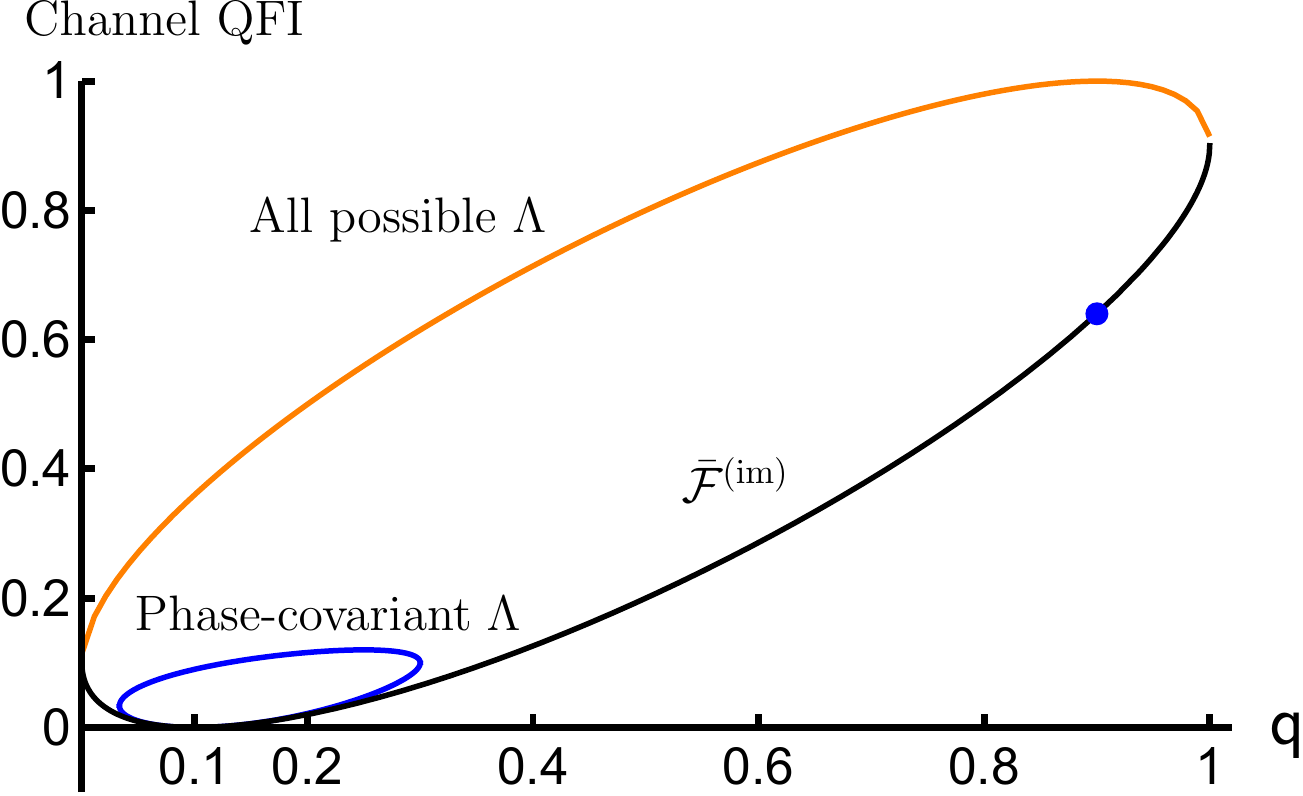}
	\caption{
\textbf{Imperfect channel QFI $\FQimbar$ vs $G$-covariance-based bounds $\FQbar[(\Lambda\circ\cE_\theta)]$ in phase sensing with a single qubit} ($N=1$), with the projective measurement affected by asymmetric bit-flip errors parametrised by $\pp=0.9$ and $\qq$ (horizontal axis). As the phase encoding $\cE_{\theta}\sim\{\upe^{\upi\theta\sigma_z/2}\}$ commutes with the U(1)-representation $\cV_{g}\sim\{\upe^{\upi g\sigma_z}\}$---and, hence, the optimal unitary $\opt{\cV}$, see \secref{sec:optimal_meas_qubit}---the $G$-covariance condition \eqref{eq:comm_Lambda_VI} is satisfied and the Observation 2 implies that $\FQimbar\leq \FQbar[(\Lambda\circ\cE_\theta)]$ for \emph{any} choice of $\Lambda$ satisfying \eqnref{eq:ALambdachanneldef2}, For any such $\Lambda$, the upper bound can then always be efficiently computed by resorting to the `seesaw' method via \eqnref{eq:channelQFI_after_maxX}. This is illustrated by choosing $\Lambda$ to represent phase-covariant channels (\emph{blue line}, see \eqnref{eq:APhasecovariantQFI}, and \emph{blue dot} at $\qq=\pp$), the quantum-classical channel \eref{eq:AQCchannel} and the channel \eref{eq:ATuviachannel}, for both of which we observe $\FQbar [(\Lambda\circ\cE_\theta)]=\FQimbar$ (coinciding with the \emph{black line} depicting the true imperfect channel QFI). However, by resorting to \eqnref{eq:imp_ch_QFI_UB_max_CJ}, we also perform optimisation over \emph{all possible} conjugate-map decompositions, $\Lambda$ satisfying \eqnref{eq:ALambdachanneldef2}, in order to observe that the so-determined $\FQbar[(\Lambda\circ\cE_\theta)]$ (\emph{orange line}) is not even convex in $\qq$, being tight only at $\qq=0,1$.
  }
	\label{fig:QMwIM-note-fig-3}
\end{figure}

The same exemplary model can also be used to illustrate the applicability of the Observation 2 discussed in \secref{sec:proof_Obs1}, which applies rather to the \emph{imperfect channel QFI}, $\FQimbar$, incorporating maximisation over the input states in \eqnsref{eq:channelQFI_after}{eq:channelQFI_after_maxX}. As explained in the main text, see particularly \refmain{Fig.~2 therein}, when allowing for arbitrary control any input state lying on the equator of the Bloch sphere is optimal, so that $\FQimbar=\FQim$ for the input $\ket{+}$ and the corresponding curve (\emph{black solid line}) in \figref{fig:QMwIM-note-fig-3} is just the same as the one for $\FQim$ in \figref{fig:QMwIM-note-fig-2} (given by \refmain{Eq.~(11) in the main text}). Now, as the phase encoding $\upe^{\upi\theta\sigma_z/2}$ commutes with the group of control unitaries $\upe^{\upi\varphi\sigma_z}$ for any $\varphi$, the $G$-covariance condition \eqref{eq:comm_Lambda_VI} is satisfied instead, and at the level of the imperfect channel QFI the inequality $\FQimbar\leq\FQbar[(\Lambda\circ\cE_{\theta})]$ holds for \emph{any} channel $\Lambda$ satisfying \eqnref{eq:ALambdachanneldef2}.

In Fig.~\ref{fig:QMwIM-note-fig-3}, with help of the `seesaw' algorithm introduced in \secref{sec:seesaw}, we compute $\FQbar[(\Lambda\circ\cE_{\theta})]$ in \eqnref{eq:imp_ch_QFI_UB_maxX} with $\Lambda$ representing:~phase-covariant channels, quantum-classical channel \eref{eq:AQCchannel} and the channel \eref{eq:ATuviachannel}. In the first case, as any state from the equator is still optimal thanks to the phase-covariance property, we can equivalently utilise \eqnref{eq:APhasecovariantQFI} and recover the same upper bounds as in \figref{fig:QMwIM-note-fig-2} (\emph{blue oval and dot}). However, in the latter two cases, we find the resulting upper bounds to coincide with the imperfect channel QFI, i.e.~$\FQimbar=\FQbar[(\Lambda\circ\cE_{\theta})]$ with $\Lambda$ of \eqnref{eq:AQCchannel} or \eqnref{eq:ATuviachannel}. Note that this contrasts the case of quantum-classical channel in \figref{fig:QMwIM-note-fig-2}, which thanks to the optimisation of the input state (for each value of $\qq$) yields now a valid ``upper-bound''---actually reproducing the exact $\FQimbar$.

Furthermore, we resort to the generalisation of the `seesaw' algorithm that includes maximisation over \emph{all possible} conjugate-map decompositions of the imperfect measurement, as discussed in \secref{sec:seesaw_max_CPTPmaps}. In this way, we obtain $\FQbar[(\Lambda\circ\cE_{\theta})]$ maximised over all $\Lambda$ satisfying \eqnref{eq:ALambdachanneldef2} for each bit-flip error values $\pp$ and $\qq$---\emph{orange line} in \figref{fig:QMwIM-note-fig-3}. We observe that such an upper is not even convex as the true imperfect channel QFI, $\FQimbar$ (\emph{black line}) and, hence, tight only at the extremal values of $\qq=0$ or $1$.

\subsubsection{Example 2: $N=2$ qubits phase sensing with bit-flip noise}

\begin{figure}[t!]
	\centering
	\includegraphics[width=0.48\textwidth]{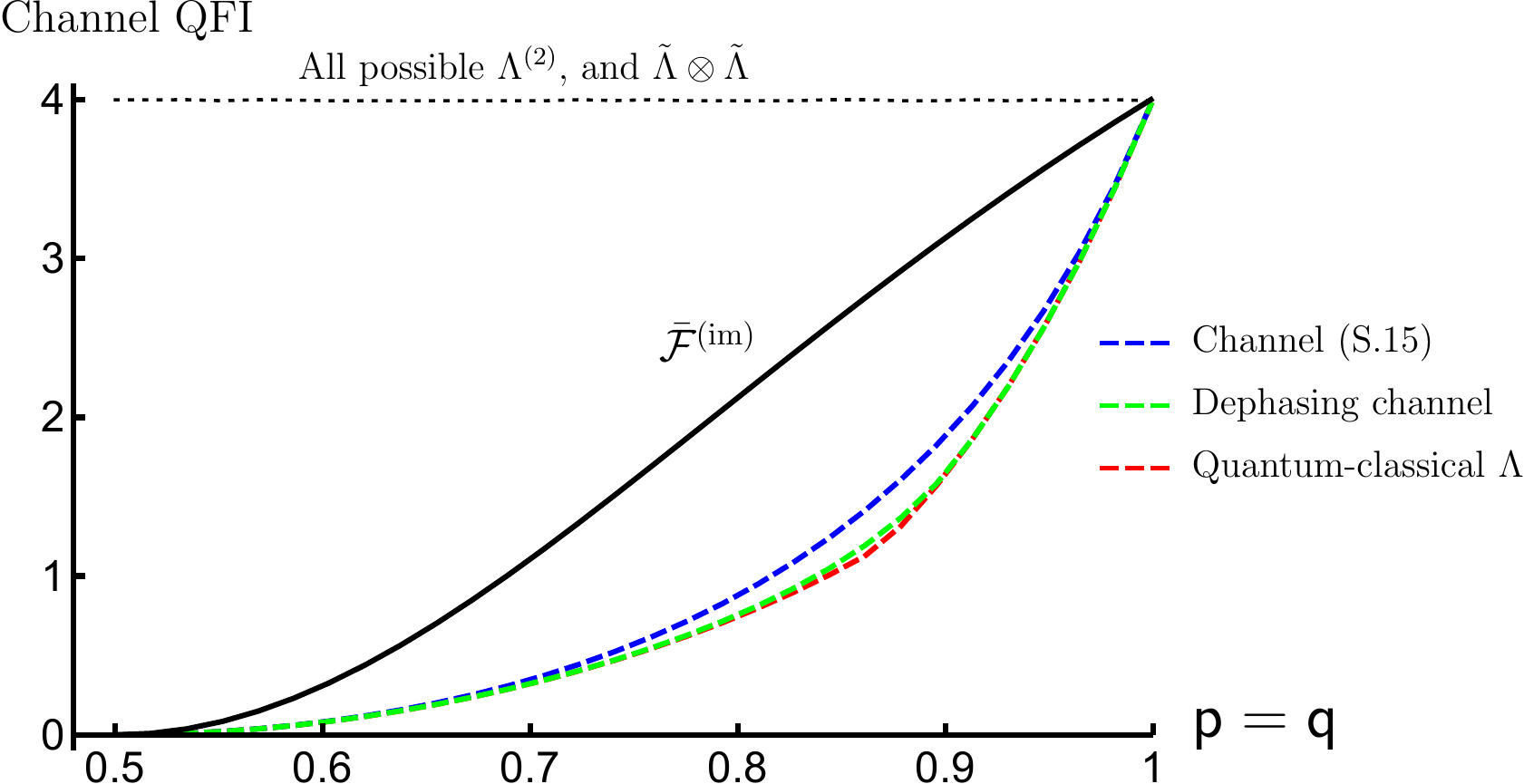}
	\caption{
\textbf{Imperfect channel QFI $\FQimbar$ vs channel QFIs $\FQbar[(\Lambda\circ\cE_\theta)^{\otimes 2}]$ in phase sensing with two qubits} ($N=2$), as a function of the probability for symmetric bit-flip errors, $\pp=\qq\geq0.5$, affecting qubit projective-measurement outcomes, where $\Lambda$ are some valid conjugate-map decompositions \eref{eq:ALambdachanneldef2} of the resulting (local) imperfect measurements.   
The \emph{black solid line} depicts the true imperfect channel QFI, $\FQimbar$, obtained by determining numerically for each $\pp=\qq$ the optimal control unitary, $\opt{\cV}$, which now crucially acts globally on both qubits and may not commute with neither the conjugate-map decompositions considered, $\Lambda^{\otimes{2}}$, nor the encoding, $\cE_{\theta}^{\otimes{2}}$. As result, both the $G$-covariance conditions \eref{eq:comm_Lambda_V} and \eref{eq:comm_Lambda_VI} need not be satisfied. This is confirmed by observing that \eqnref{eq:AimcQFIbound4} is explicitly invalidated, i.e.~$\FQimbar\not\leq\FQbar[(\Lambda\circ\cE_{\theta})^{\otimes2}]$, for conjugate-map decompositions with $\Lambda$ corresponding to either:~the dephasing channel (\emph{green}), quantum-classical channel \eqref{eq:AQCchannel} (\emph{red}) or the channel defined in \eqnref{eq:ATuviachannel} (\emph{blue})---observe all the corresponding curves lying below the one of $\FQimbar$.
On the other hand, the `seesaw' algorithm allows us to find also the local and global channels $\tilde{\Lambda}^{\otimes 2}$ and $\Lambda^{(2)}$, respectively, that constitute valid conjugate-map decompositions \eref{eq:ALambdachanneldef2} at each $\pp=\qq$ but are \emph{uninformative}---yield $\FQbar[(\tilde{\Lambda}\circ\cE_{\theta})^{\otimes2}]=\FQbar[\tilde{\Lambda}^{(2)}\circ\cE_{\theta}^{\otimes2}]=\FQbar[\cE_{\theta}^{\otimes2}]=4$ (\emph{black dots}), i.e.~the perfect quantum channel QFI. This emphasises further that, already for $N=2$, results of quantum metrology with noisy encoding may not be directly used to estimate performance with imperfect measurements.}
	\label{fig:QMwIM-note-fig-4}
\end{figure}

In this example, we demonstrate a scenario where now the $G$-covariance condition \eqref{eq:comm_Lambda_VI} is not satisfied and, hence, also Observation 2 is not applicable. We consider generalisation of the above phase-sensing example to the case of two ($N=2$) qubits, in which each qubit probe state undergoes the same encoding and imperfect measurement. For our purposes, we consider only the case of symmetric ($\pp=\qq$) bit-flip as the detection noise, as it is sufficient to explore the possibility of the optimal input probe state now being entangled and, crucially, the control operation $\cV_{\phivec}$ now acting globally on both qubits, i.e.~constituting an entangling gate.

First, from Lemma $1$, we know that the imperfect measurement does not change the optimal input state,
which is thus: 
$\ket{\Phi_{+}}=(\ket{00}+\ket{11})/\sqrt{2}$. Then, one may prove that the optimal two-qubit control unitary must take the form (see also the discussion below in \secref{sec:proof_of_thm1}):
$\opt{\cV}\sim \{\Vopt=W\upe^{\upi \varphi_{\mathrm{opt}}(\sigma_{z}\otimes\id+\id\otimes\sigma_{z})}\}$ for some optimal real coefficient $\varphi_{\mathrm{opt}}$, where $W$ is the unitary that transforms the Bell basis onto the product basis $\{\ket{++},\ket{+-},\ket{-+},\ket{--}\}$. That is, $W=\ketbra{++}{\Phi_{+}}+\ketbra{+-}{\Psi_{+}}+\ketbra{-+}{\Psi_{-}}+\ketbra{--}{\Phi_{-}}$, where $\ket{\Phi_{\pm}}=(\ket{00}\pm\ket{11})/\sqrt{2}$ and $\ket{\Psi_{\pm}}=(\ket{01}\pm\ket{10})/\sqrt{2}$. However, one can also verify that unitaries of such a form generally do not commute with the encoding $\cE_{\theta}^{\otimes 2}$, so there is no straightforward identification that the $G$-covariance condition \eqref{eq:comm_Lambda_VI} can be satisfied. In fact, since satisfying the $G$-covariance condition \eqref{eq:comm_Lambda_VI} implies that \eqnref{eq:AimcQFIbound4} must hold for any $\Lambda$, finding an example of $\Lambda$ that violates \eqnref{eq:AimcQFIbound4} and, hence, the Observation 2, would imply that \eqref{eq:comm_Lambda_VI} \emph{cannot} be fulfilled.

We illustrate this explicitly in \figref{fig:QMwIM-note-fig-4}, where now---contrastingly to the single-qubit case---all the channel QFIs $\FQbar[(\Lambda\circ\cE_{\theta})^{\otimes2}]$ evaluated for the conjugate-map decompositions \eref{eq:ALambdachanneldef2} with $\Lambda$ representing:~the dephasing channel (\emph{green dashed line}), the quantum-classical channel \eqref{eq:AQCchannel} (\emph{red dashed line}) and the channel \eref{eq:ATuviachannel} (\emph{blue dashed line});~lie below the true imperfect channel QFI, $\FQimbar$, for any $\tfrac{1}{2}<\pp=\qq<1$. Hence, all these three examples invalidate the upper bounds \eref{eq:AimQFIbound4} and \eref{eq:AimcQFIbound4}, and prove that at any such $\pp=\qq$, $\opt{\cV}$ must not commute with neither $\Lambda^{\otimes 2}$ nor $\cE_{\theta}^{\otimes 2}$, i.e.~none of the $G$-covariance conditions \eref{eq:comm_Lambda_V} and \eref{eq:comm_Lambda_VI}, respectively, may hold.

We evaluate the aforementioned channel QFIs in \figref{fig:QMwIM-note-fig-4} for the three types of conjugate-map decompositions by resorting to the `seesaw' algorithm described in \secref{sec:seesaw}. However, one should also recall from \figref{fig:QMwIM-note-fig-3} (see the maximum of the \emph{orange line}) that in the single-qubit case there exists $\tilde{\Lambda}$ satisfying \eref{eq:ALambdachanneldef2} at $\pp=\qq$, such that $\FQbar[(\tilde{\Lambda}\circ\cE_{\theta})]=\FQbar[\cE_{\theta}]=1$, i.e.~its corresponding channel QFI actually equals the perfect channel QFI and, hence, is useless in estimating the impact of the detection (bit-flip) noise. We verify that such a conjugate-map decomposition remains \emph{uninformative} also in the two-qubit scenario, in which it leads to $\FQbar[(\tilde{\Lambda}\circ\cE_{\theta})^{\otimes 2}]=\FQbar[\cE_{\theta}^{\otimes 2}]=4$ for any $\pp=\qq$---see the \emph{black dotted line} in \figref{fig:QMwIM-note-fig-4}. For consistency, by resorting to the `seesaw' algorithm of \secref{sec:seesaw_max_CPTPmaps} that includes maximisation over conjugate-map decompositions, we also verify that there exists a global $\Lambda^{(2)}$ that satisfies the condition \eref{eq:ALambdachanneldef2} for $N=2$, i.e.~$\cM^{\otimes 2} = \Lambda^{(2)\dagger} [\Pi^{\otimes 2}]$, and similarly leads to $\FQbar[\tilde{\Lambda}^{(2)}\circ\cE_{\theta}^{\otimes 2}]=\FQbar[\cE_{\theta}^{\otimes 2}]=4$. Interestingly, we observe that neither $\tilde{\Lambda}^{\otimes 2}$ nor $\Lambda^{(2)}$ commute in general with $\opt{\cV}$, so that $G$-covariance condition \eref{eq:comm_Lambda_V} still does not apply---making the connection between metrological protocols with imperfect (local) measurements and the ``standard'' setting of quantum metrology with noisy encoding on multiple probes even less apparent.

\subsection{Hierarchy of moment-based lower bounds on the FI}
Using only partial information from a full probability distribution, such as considering only up to certain finite moments, one obtains a lower bound for the FI. For the case of univariate and single-parameter estimation, we provide here a simple ``physicist's" reformulation for constructing such a lower bound, which consistently agrees with more abstract considerations~\cite{Sankaran1964, Jarrett1984, Stein2021}.

We first rewrite the FI $F= \sum_x \dotqxphitheta^2/\qxphitheta$ by making use of essentially a simple identity:~any real quadratic function $g(y)=-ay^2+2by$ with $a>0, b\in\mathbb{R}$, has its maximum given by $\max_y g(y)=g(b/a)=b^2/a$, so equivalently $F=\sum_x \max_{y_x}\{-\qxphitheta y_x^2+2\dotqxphitheta y_x\}$. Using a series \emph{ansatz} $y_x=\sum_{k=0}^K \alpha_k w(x)^k$ for some chosen function $w(x)$ with $K$ smaller than the cardinality of the probability distribution, we obtain a lower bound $\Fdown{K}$ on FI after maximizing now over the finite set $\{\alpha_k\}_{k=0}^K$. By construction, we have $\Fdown{0}\leq \Fdown{1}\leq \Fdown{2}\leq \cdots\leq\Fdown{K}\leq F$, and $\Fdown{K}$ can be computed straightforwardly as 
\begin{align}
\label{eq:AFKmoment}
\Fdown{K} = 
	&-\max_{\{\alpha\}}\Bigg[\sum_{k=0}^K\alpha_k\sum_{j=0}^K\alpha_j \sum_x \qxphitheta w(x)^{k+j}\nonumber\\
   &\phantom{-\max_{\{\alpha\}}\Bigg[}\;+\;2\sum_{k=0}^K\alpha_k\sum_x\dotqxphitheta w(x)^k\Bigg]\nonumber\\
	&=\max_{\boldsymbol{\alpha}}-\boldsymbol{\alpha}^T A \boldsymbol{\alpha}+2\boldsymbol{b}^T\boldsymbol{\alpha}=\boldsymbol{b}^T A^{-1}\boldsymbol{b}, 
\end{align}
where $\boldsymbol{\alpha}=\begin{pmatrix} \alpha_0,\,\alpha_1,\,\cdots\,,\alpha_K \end{pmatrix}^T$, and $A$ and $\boldsymbol{b}$ are as in \refmain{Eqs.~(26, 27) in the main text}, with the more general replacement $\mathbb{E}[x^j]\rightarrow \mathbb{E}[w(x)^j]$ and $\dot{\mathbb{E}}[x^j]\rightarrow \dot{\mathbb{E}}[w(x)^j]$. If we choose further $w(x)=x$, $\Fdown{K}$ becomes a lower bound on $F$ that takes into account up to the $2K$-th moment of the distribution $\qphitheta$. For $K=0$ and $K=1$, we have explicitly $\Fdown{0}=0$ and $\Fdown{1}=\dot{\mathbb{E}}[w(x)]^2/(\mathbb{E}[w(x)^2]-\mathbb{E}[w(x)]^2)$. For $K\geq2$, \eqnref{eq:AFKmoment} can be computed numerically and efficiently by standard matrix inversion techniques.

\subsection{Proof of Theorem 1 and convergence rate to the perfect QFI}
\label{sec:proof_of_thm1}
As the results and expressions derived in Lemma~1 are independent of the number of probes, it follows that Equations (\ref{eq:AF}-\ref{eq:Agamma}) hold still. 
Applying specifically to the multi-probe scenario discussed in the main text with the structure of measurement as $\cM^{\otimes N}$, i.e. each probe is still measured independently but in an imperfect manner, \eqnsref{eq:AF2}{eq:Agamma} read
\begin{align}\label{eq:AFN2}
F_N=&4\;\gamma(\Phivec,\psi^N(\theta))\;\langle \partial_\theta\psi_\perp^N(\theta)\ket{\partial_\theta\psi_\perp^N(\theta)}, 
\end{align}
with
\begin{align}\label{eq:AgammaN}
&\gamma(\Phivec,\psi^N(\theta)):=\nonumber\\
&\qquad\frac{1}{4}\sum_{\xvec}  \frac{[\bra{\psi_\perp^N(\theta)}\VPhi^\dagger M_{\xvec}\VPhi\ket{\psi^N(\theta)}+\cc]^2}{\bra{\psi^N(\theta)}\VPhi^\dagger M_{\xvec}\VPhi\ket{\psi^N(\theta)}},
\end{align}
where now $\xvec=(x_1,x_2,\cdots,x_N)$, with $\cM^{\otimes N}\sim\{M_{\xvec}\}$,
\begin{align}\label{eq:AMxvec}
M_{\xvec}=M_{x_1}\otimes M_{x_2}\otimes\cdots\otimes M_{x_N},
\end{align}	 
where $ M_{x_\ell}$ is the noisy measurement operator for outcome $x_\ell$ for the $\ell$-th probe.

A lower bound on $\gamma(\Phivec,\psi^N(\theta))$, and hence on $F_N$, for any given choice of $\VPhi$ can be constructed as follows. Define $\ket{\psi_{\pm}^N(\theta)}:=\frac{1}{\sqrt{2}}\Big(\ket{\psi^N(\theta)}\pm\ket{\psi_\perp^N(\theta)}\Big)$, and  $p_{\pm}(\xvec):=\bra{\psi_{\pm}^N(\theta)}\VPhi^\dagger M_{\xvec}\VPhi\ket{\psi_{\pm}^N(\theta)}$, where in order not to overload the notation, we have kept the $\theta$ and $\Phivec$ dependence implicit. Then, the numerator of $4\gamma(\Phivec,\psi^N(\theta))$ is
\begin{align}\label{eq:Agammanum}
\big(p_{+}(\xvec)-p_{-}(\xvec)\big)^2,
\end{align}
while the denominator is
\begin{align}\label{eq:Agammadenom}
&\frac{1}{2} \Big(p_{+}(\xvec)+p_{-}(\xvec)\nonumber\\
&+\bra{\psi_{-}^N}\VPhi^\dagger M_{\xvec}\VPhi\ket{\psi_{+}^N}+\bra{\psi_{+}^N}\VPhi^\dagger M_{\xvec}\VPhi\ket{\psi_{-}^N}\Big),
\end{align}
which by the Cauchy-Schwartz inequality is smaller or equal to 
\begin{align}\label{eq:Agammadenombound}
\frac{1}{2}\Big(\sqrt{p_{+}(\xvec)}+\sqrt{p_{-}(\xvec)}\Big)^2.
\end{align}
Hence, we have
\begin{align}\label{eq:Agammabound}
\gamma(\Phivec,\psi^N(\theta))&\geq \frac{1}{2}\sum_{\xvec} \Big(\sqrt{p_{+}(\xvec)}-\sqrt{p_{-}(\xvec)}\Big)^2\nonumber\\
&=\Big(1-\sum_{\xvec} \sqrt{p_{+}(\xvec)p_{-}(\xvec)}\Big).
\end{align}
Let us now denote:
\begin{align}\label{eq:AVPhichoice}
\VPhi\ket{\psi^N_{+}(\theta)}&=\ket{\zeta^N},\nonumber\\
\VPhi\ket{\psi^N_{-}(\theta)}&=\ket{\zeta^N_\perp},
\end{align}
and note that $p_{\pm}(\xvec)$ depends solely on our choice of $\ket{\zeta^N},\ket{\zeta^N_\perp}.$
To attain the lower bound in \refmain{Eq.~(17) of Thm.~1}, consider a specific
choice of $\ket{\zeta^N}, \ket{\zeta^N_\perp}$ (obtained by a suitable $\VPhi$) : 
\begin{align}\label{eq:AVPhichoice2}
\ket{\zeta^N}=\ket{\zeta}^{\otimes N},\nonumber\\
\ket{\zeta^N_\perp}=\ket{\zeta_\perp}^{\otimes N}
\end{align}
for some orthogonal single-qudit states $\{\ket{\zeta},\ket{\zeta_\perp}\}$. Then, with $p_{+}(x):=\bra{\zeta}M_x\ket{\zeta}, p_{-}(x):=\bra{\zeta_\perp}M_x\ket{\zeta_\perp}$, and $c:=\sum_{x}\sqrt{p_+(x)p_-(x)}$, we have $p_{\pm}(\xvec)=\prod_{j=1}^N p_{\pm}(x_j)$, and thus 
\begin{align}\label{eq:AcN}
\sum_{\xvec} \sqrt{p_{+}(\xvec)p_{-}(\xvec)} = c^N.
\end{align}
Evidently, we have $0\leq c\leq1$, and so we complete the proof of Thm~1. \hfill $\blacksquare$

A few remarks are in order:
Given a specific choice of $\{\ket{\zeta}^{\otimes N},\ket{\zeta_{\perp}}^{\otimes N} \}$ (equivalently $V_{\Phivec}$ and $p_{\pm}(\xvec)\}$) 
the FI converges to the perfect QFI exponentially fast with an approximate prefactor (see \eqnsref{eq:Agammabound}{eq:AcN}):
\begin{align}
1-c^N&=
1-\left(\sum_{x}\sqrt{p_{+}\left(x\right)p_{-}\left(x\right)}\right)^{N}\nonumber\\
&=1-\left(\underset{\xvec}{\sum}\sqrt{p_{+}\left(\xvec\right)p_{-}\left(\xvec\right)}\right).
\end{align}
Note that this prefactor is the Hellinger distance, $H\left(p_{+} \left(\xvec\right),p_{-}\left(\xvec\right)\right)$, between the distributions $p_{+}\left(\boldsymbol{x}\right)$ and $p_{-}\left(\boldsymbol{x}\right)$~\cite{Basu2019}, while the convergence rate $\chi$  ($c\equiv\upe^{-\chi}$) reads
\begin{align} \label{convergence_rate}
\chi&=-\log\left(\sum_{x}\sqrt{p_{+}\left(x\right)p_{-}\left(x\right)}\right)\nonumber\\
&=-\frac{1}{N}\log\bigg(1-H\big( p_{+}\left(\xvec\right) , p_{-}\left(\xvec\right) \big) \bigg).
\end{align}
For very close distributions it can be observed that the convergence rate is equal to the Fisher metric: $\sum_{x}\frac{\left(p_{+}\left(x\right)-p_{-}\left(x\right)\right)^{2}}{8p_{+}\left(x\right)}$.

As $N\rightarrow\infty$, generally we can use the central limit theorem to approximate $H\left(p_{+}\left(\xvec\right),p_{-}\left(\xvec\right)\right)$
as the Hellinger distance between two Gaussian distributions. This would then imply a convergence rate of: $\frac{1}{4}\frac{\left(\mu_{+}-\mu_{-}\right)^{2}}{\sigma_{+}^{2}+\sigma_{-}^{2}}$, where $\mu_{\pm},\sigma_{\pm}$ are the average and standard deviation  of $p_{\pm}\left( x \right)$
respectively, which of course depends on our choice of $\ket{\zeta}, \ket{\zeta_\perp}$.

We can thus apply this analysis to obtain the convergence rate for specific cases.  For Poissonian channel (with coefficients $\lambda_{\ket{0}}, \lambda_{\ket{1}}$), such as in NV centres, we find a convergence rate of: $\frac{1}{2}\left(\sqrt{\lambda_{\ket{0}}}-\sqrt{\lambda_{\ket{1}}}\right)^{2}$.
This convergence rate is achieved by taking $\ket{\zeta^N}, \ket{\zeta_\perp^N}$ to be $|0\rangle^{\otimes N},|1\rangle^{\otimes N}$ (similar convergence rate is obtained by taking them to be any superposition of $|0\rangle^{\otimes N},|1\rangle^{\otimes N}$).
This implies that for realistic experimental values of NV centres, $\lambda_{\ket{0}}=0.1, \lambda_{\ket{1}}=0.07$ (without the use of nuclear spins as memory), the number of probes that obtains $95 \%$ of the perfect QFI would be $ \sim 2000$.
For a binary asymmetric bit-flip channel with probabilities $\pp,\qq$,
we find a convergence rate of: $\frac{1}{4}\frac{\left(\pp+\qq-1\right)^{2}}{\pp\left(1-\pp\right)+\qq\left(1-\qq\right)}$, given a similar choice of $|\zeta\rangle=|0\rangle^{\otimes N},|\zeta_{\perp}\rangle=|1\rangle^{\otimes N}$.
As a side note, we may also compute $c$ directly in this case, and it can be verified that indeed  $c^N=\left( \sqrt{\pp(1-\qq)}+\sqrt{\qq(1-\pp)} \right)^N \approx \exp\left(-\chi N\right)$ with the said $\chi$.

A note on optimality: Let us justify our choice of unitary, namely the choice of $|\zeta^{N}\rangle=|0\rangle^{\otimes N},|\zeta_{\perp}^{N} \rangle=|1\rangle^{\otimes N}$. Given the approximated value of the convergence rate, \eqnref{convergence_rate}, we claim that this choice yields the optimal convergence rate.
That is, we would like to choose $|\zeta^{N}\rangle,|\zeta_{\perp}^{N}\rangle$ that maximize the Hellinger distance between $\left\{ p_{+}\left( \xvec \right)=\langle\zeta^{N}|M_{\xvec}|\zeta^{N}\rangle\right\} _{\xvec} , \left\{ p_{-}\left( \xvec \right)=\langle\zeta_{\perp}^{N}|M_{\xvec}|\zeta_{\perp}^{N}\rangle\right\} _{\xvec}$. As before, we focus on the classical noise channel, namely commuting $M_{x}$: $M_{x}=\sum_{i}p\!\left(x|i\right)\Pi_{i}$. Observe that:
\begin{itemize}
\item  The Hellinger distance is convex in the probability distributions: $H\left(\lambda p_{1}+\left(1-\lambda\right)p_{2},q\right)\leq\lambda H\left(p_{1},q\right)+\left(1-\lambda\right)H\left(p_{2},q\right)$.
\item  Let $\left\{ |j\rangle\right\} _{j}$ be the common eigenbasis of $\left\{ M_{x}\right\} _{x}$, then given $|\zeta\rangle=\sqrt{\lambda}|j_{1}\rangle+\sqrt{1-\lambda}|j_{2}\rangle$: $\langle\zeta|M_{x}|\zeta\rangle=\lambda\langle j_{1}|M_{x}|j_{1}\rangle+\left(1-\lambda\right)\langle j_{2}|M_{x}|j_{2}\rangle$. In words: taking superpositions of the eigenstates leads to convex combinations of the probabilities.
\end{itemize}
The above two observations imply that the maximal Hellinger distance is achieved by taking $|\zeta^{N}\rangle,|\zeta_{\perp}^{N}\rangle$ to be elements in the common eigenbasis (and not superpositions of them). Hence, we just need to find the two basis states that yield maximal Hellinger distance. For $N$ qudits, we thus need to find the two states $|j_{1}\rangle,|j_{2}\rangle$ with maximal Hellinger distance out of the $d$-dimensional eigenbasis, and then the optimal choice of $|\zeta^{N}\rangle,|\zeta_{\perp}^{N}\rangle$ would be $|j_{1}\rangle^{\otimes N},|j_{2}\rangle^{\otimes N}$. For the case of NV centres with the local projective measurement $\{\Pi_1=\proj{0}, \Pi_2=\proj{1}\}$ for each of the $N$ NV centres, this immediately implies that the optimal $|\zeta^{N} \rangle,|\zeta_{\perp}^{N}\rangle$ are $|0\rangle^{\otimes N},|1\rangle^{\otimes N}$.

In fact this intuition together with some numerical evidence leads us to conjecture
that for any classical noise channel, $M_{x}=\sum_{i}p\!\left(x|i\right)\Pi_{i}$ that is applied independently on each of the $N$ probes, the optimal $|\zeta^{N} \rangle,|\zeta_{\perp}^{N}\rangle$ take the form of ``cat states":
\begin{eqnarray*}
\begin{split}
|\zeta^{N} \rangle &=\cos\left(\theta\right)|j\rangle^{\otimes N}+\sin\left(\theta\right)|k\rangle^{\otimes N},\\
|\zeta_{\perp}^{N} \rangle &=-\sin\left(\theta\right)|j\rangle^{\otimes N}+\cos\left(\theta\right)|k\rangle^{\otimes N},
\end{split}
\end{eqnarray*}
 where $ \ket{j}, \ket{k}$ can be found numerically for just a single probe, and $\theta$ depends on $N$ and should be found numerically. In particular for $N$ qubits we connjecture that the optimal $|\zeta^{N} \rangle,|\zeta_{\perp}^{N}\rangle$ take the form of $\cos\left(\theta\right)|0\rangle^{\otimes N}+\sin\left(\theta\right)|1\rangle^{\otimes N},-\sin\left(\theta\right)|0\rangle^{\otimes N}+\cos\left(\theta\right)|1\rangle^{\otimes N}.$

\subsection*{Proof of Lemma 2}
With the encoded state $\rho_r^N(\theta)=r\,\psi^N(\theta)+(1-r)\id_{d^N}/{d^N}$, where $\psi^N(\theta)=\ket{\psi^N(\theta)}\bra{\psi^N(\theta)}$, similar to the proof of Thm.~1, it is straightforward to establish that the FI can be written as
\begin{align}\label{eq:AFNWerner}
F_N=&4\langle \partial_\theta\psi_\perp^N(\theta)\ket{\partial_\theta\psi_\perp^N(\theta)} \gamma_r(\VPhi,\ket{\psi^N(\theta)}), 
\end{align}
where
\begin{align}\label{eq:AgammaWerner}
&\gamma_r(\Phivec,\psi^N(\theta))\nonumber\\
=&\frac{1}{4}\sum_{\xvec}  \frac{r^2[\bra{\psi_\perp^N(\theta)}\VPhi^\dagger M_{\xvec}\VPhi\ket{\psi^N(\theta)}+\cc]^2}{r\bra{\psi^N(\theta)}\VPhi^\dagger M_{\xvec}\VPhi\ket{\psi^N(\theta)}+(1-r)p'(\xvec)},
\end{align}
with $p'(\xvec):=\Tr\{M_{\xvec}\}/d^N$. Note that $p'(\xvec)$ is a legit probability distribution, i.e., $p'(\xvec)\geq0\,\forall\xvec$, and $\sum_{\xvec}p'(\xvec)=1$. Moreover, evidently $\gamma_r(\Phivec,\psi^N(\theta))\leq r \gamma(\Phivec,\psi^N(\theta))\leq r$, and so we arrive at an upper bound $F_N\leq 4r\langle\partial_\theta\psi_\perp^N(\theta)\ket{\partial_\theta\psi_\perp^N(\theta)}=r\FQ[\psi^N(\theta)]$.

Now, using essentially the same observation and notation leading to Eqs.~(\ref{eq:Agammanum}-\ref{eq:Agammadenombound}), the following inequality holds:  
\begin{align}\label{eq:AgammaWernerbound1}
&\gamma_r(\Phivec,\psi^N(\theta))\nonumber\\
\geq& \frac{1}{4}\sum_{\xvec} \frac{r^2\big(p_{+}(\xvec)-p_{-}(\xvec)\big)^2}{\dfrac{r}{2}\Big(\sqrt{p_{+}(\xvec)}+\sqrt{p_{-}(\xvec)}\Big)^2+(1-r)p'(\xvec)}.
\end{align}

Consider then three different ways of grouping all the $\xvec$. First, consider two sets, $A_+$ and $B_+$, defined as 
\begin{align}\label{eq:AsetABp}
A_+:=\{\xvec\, |\, (1-r)p'(\xvec)\leq r p_{+}(\xvec) \},\nonumber\\
B_+:=\{\xvec\, |\, (1-r)p'(\xvec)> r p_{+}(\xvec) \}.
\end{align}
Similarly, define the sets 
\begin{align}\label{eq:AsetABm}
A_-:=\{\xvec\, |\, (1-r)p'(\xvec)\leq r p_{-}(\xvec) \},\nonumber\\
B_-:=\{\xvec\, |\, (1-r)p'(\xvec)> r p_{-}(\xvec) \},
\end{align}
as well as
\begin{align}\label{eq:AsetAB}
A:=\{\xvec\, |\, (1-r)p'(\xvec)\leq r\max\big(p_{+}(\xvec),p_{-}(\xvec)\big)\},\nonumber\\
B:=\{\xvec\, |\, (1-r)p'(\xvec)> r\max\big(p_{+}(\xvec),p_{-}(\xvec)\big)\}.
\end{align}
Evidently, $A_+\cup A_- = A$, and $B_+\cap B_- = B$. Morever, define 
\begin{align}\label{eq:Aepspm}
\epsilon_{\pm}:=\sum_{\xvec} \min\big(r p_{\pm}(\xvec), (1-r)p'(\xvec)\big),
\end{align}
which can be interpreted as the minimal error in discriminating the two probability distributions $\{p_{\pm}(\xvec)\}$ and $\{p'(\xvec)\}$ with prior $r$ and $1-r$, respectively. Thus, 
\begin{align}\label{eq:Aepspmbound}
\epsilon_{+}+\epsilon_{-}=&\Big(\sum_{\xvec\in A_+}+\sum_{\xvec\in A_-}\Big) (1-r)p'(\xvec)\nonumber\\
&+\sum_{\xvec\in B_+} rp_{+}(\xvec) +\sum_{\xvec\in B_-} rp_{-}(\xvec)\nonumber\\
\geq&\sum_{\xvec\in A}(1-r)p'(\xvec)+ \nonumber\\
&+\sum_{\xvec\in B} r \max\big(p_{+}(\xvec),p_{-}(\xvec)\big).
\end{align}

Then, for the r.h.s. of \eqnref{eq:AgammaWernerbound1}, we shall evaluate the sum over all $\xvec$ into $A$ and $B$ respectively. For $\xvec\in A$, using the identity $(1+x)^{-1}\geq 1-x$ for any $(1+x)\in\mathbb{R}_+$, we get
\begin{align}\label{eq:AsumA}
\sum_{\xvec\in A}(\cdots)\geq& \,2r \sum_{\xvec\in A} \Big(\sqrt{p_{+}(\xvec)}-\sqrt{p_{-}(\xvec)}\Big)^2\nonumber\\
&-4 \sum_{\xvec\in A} (1-r)p'(\xvec).
\end{align}
Meanwhile, for $\xvec$ in the set $B$, observe that 
\begin{align}\label{eq:AsumB}
&2r\Big(\sqrt{p_{+}(\xvec)}-\sqrt{p_{-}(\xvec)}\Big)^2\nonumber\\
\leq&\, 2r \big(p_{+}(\xvec)+p_{-}(\xvec)\big) \nonumber \\
\leq&\, 4r\max \big(p_{+}(\xvec), p_{-}(\xvec)\big).
\end{align}
Hence, we have
\begin{align}\label{eq:AsumB2}
\sum_{\xvec\in B}(\cdots)\geq& \sum_{\xvec\in B} 2r\Big(\sqrt{p_{+}(\xvec)}-\sqrt{p_{-}(\xvec)}\Big)^2\nonumber\\
&-4r \sum_{\xvec\in B}\max \big(p_{+}(\xvec), p_{-}(\xvec)\big).
\end{align}
Combining Eqs.~(\ref{eq:AgammaWernerbound1}, \ref{eq:Aepspmbound}, \ref{eq:AsumA}, \ref{eq:AsumB2}), we obtain
\begin{align}\label{eq:AgammaWernerbound2}
&\gamma_r(\Phivec,\psi^N(\theta))\nonumber\\
\geq&\, \frac{1}{2}r\sum_{\xvec}  \Big(\sqrt{p_{+}(\xvec)}-\sqrt{p_{-}(\xvec)}\Big)^2-\epsilon_+-\epsilon_- \nonumber\\
=&\,r\Big(1-\sum_{\xvec} \sqrt{p_{+}(\xvec)p_{-}(\xvec)}\Big)-\epsilon_+-\epsilon_-.
\end{align}

Finally, upon choosing $\VPhi$ as in \eqnref{eq:AVPhichoice2}, we have $F_N(\VPhi)\geq \FQ[\psi^N(\theta)]\big(r(1-c^N)-\epsilon_+-\epsilon_-\big)$. To complete the proof of Lemma 2, note that the choice of $\VPhi$ in \eqnref{eq:AVPhichoice2} gives us $p_{\pm}(\xvec)=\prod_{j=1}^N p_{\pm}(x_j)$, and since $p'(\xvec)=\prod_{j=1}^N p'(x_j)$ where $p'(x_j)=\Tr\{M_{x_j}\}/d$, $\epsilon_{\pm}$ is now the minimal error in discriminating two probability distributions $\{p_{\pm}(x_j)\}$ and $p'(x_j)$ over $N$ repetitions, which goes to zero in the $N\rightarrow0$ limit.  \hfill $\blacksquare$

\subsection*{Proof of Lemma 3}
A particular Kraus representation of the (single-probe) channel 
$\Lambda_{\theta,\phivec}=\Lambda_\cM\circ\cV_{\phivec}\circ\cU_{\theta}$, with the quantum-classical channel $\Lambda_\cM$ being defined in \eqnref{eq:AQCchannel} for a given imperfect measurement $\cM$, reads
\begin{align}\label{eq:AcanonicalKraus}
\Lambda_{\theta,\phivec}\sim\{K_{x,j}(\theta, \phivec)=\ket{x}\bra{j}\sqrt{M_{x}}V_{\phivec}U_\theta\}_{{x,j}}.
\end{align}
Note that as mentioned earlier, the set of orthogonal bra basis $\{\bra{j}\}_{j=1}^{d}$ can be chosen arbitrarily, corresponding to different choice of quantum-classical channel $\Lambda$. Then, as elaborated in Methods in the main text, the asymptotic CE bound $\FNCEas$ is defined when there exists some other Kraus representation $\{\tilde{K}_{x,j}(\theta,\phivec)\}$ of $\Lambda_{\theta,\phivec}$ such that $\beta_{\tilde{K}}=0$ for any $\phivec$. Moreover, following Refs.~\cite{Fujiwara2008, Demkowicz2012, Kolodynski2013}, it suffices to consider Kraus representations that have the following properties:
\begin{align} 
\tilde{K}_{x,j}(\theta,\phivec)&=K_{x,j}(\theta,\phivec), \label{eq:AKrausCE1} \\
\dot{\tilde{K}}_{x,j}(\theta,\phivec)&=\partial_\theta \tilde{K}_{x,j}^\dagger(\theta,\phivec) \nonumber\\
&=\dot{K}_{x,j}(\theta,\phivec)-\upi\sum_{x',j'} \gg_{x,j;x',j'} K_{x',j'}(\theta,\phivec) \label{eq:AKrausCE2}
\end{align}
where $\gg(\phivec)$ is an arbitrary Herimitian matrix satisfying $\gg_{x,j';x',j'}(\phivec)=\gg_{x',j';x,j}^*(\phivec)$, and the $\beta_{\tilde{K}}=0$ condition is equivalent to the existence of $\gg$ such that
\begin{align}
&\upi\sum_{x,j}\dot{K}_{x,j}^\dagger K_{x,j}-\sum_{x,j,x',j'}\gg_{x,j;x',j'}(\phivec)K_{x,j}^\dagger K_{x',j'} =0. \label{eq:Abeta=00}
\end{align}
Putting in \eqnref{eq:AcanonicalKraus} into \eqnref{eq:Abeta=00}, the  $\beta_{\tilde{K}}=0$ condition is then given by ($U_{\theta}=\upe^{{\upi h\theta}}$)
\begin{align}\label{eq:Abeta=0}
h=\sum_{x}V_{\phivec}^{\dagger} \sqrt{M_{x}}A_{x}(\phivec) \sqrt{M_{x}}V_{\phivec},
\end{align}
with $A_{x}(\phivec):=2\sum_{j,j'} \gg_{x,j;x,j'}(\phivec)\ketbra{j}{j'}$, which is \refmain{Eq.~(22)} in Lemma 3---here, without specifying explicitly which probe we are referring to. \hfill $\blacksquare$

\subsection*{Proof of Corollary 2}
We say that a detection channel $\cP$  acts \emph{non-trivially} on a given subset $\tilde{I}\subseteq I$ of `inaccessible' outcomes, if for any pair $i,i'\in \tilde{I}$ 
there exists at least one `observable' outcome $x$ such that the transition probabilities of $\cP$ satisfy $p(x|i)p(x|i')>0$. Moreover, if this $\tilde{I}$ contains the outcomes spanning the subspace of the encoding Hamiltonian, i.e.~$h=\sum_{i,i'\in \tilde{I}} \hh_{i,i'}\ket{i}\bra{i'}$ with $\det\{\hh\}\neq0$, we say that $\cP$ acts non-trivially on the encoding subspace. However, for our purposes we consider $\cP$ that act non-trivially on \emph{all} the outcomes in $I$, and refer to these as \emph{non-trivial}.

Suppose the imperfect measurement $\cM$ is composed of a perfect von Neumann measurement, i.e., $\Pi\sim\{\Pi_{i}=\proj{i}\}_{i}$ with $\Pi_{i}\Pi_{i'}=\delta_{i,i'}\Pi_{i}$, followed by a noisy detection channel $\cP\sim\{p(x|i)\}$, such that $M_{x}=\sum_{i}p(x|i)\Pi_{i}$. 
As a result, the condition \eref{eq:Abeta=0} is equivalent to $h=\sum_{i,i'\in I} V_{\phivec}^{\dagger} \ket{i} \sum_x \sqrt{p(x|i)}\bra{i}A_{x}(\phivec)\ket{i'}\sqrt{p(x|i')}\bra{i'}V_{\phivec}$, where the entries $\bra{i}A_{x}(\phivec)\ket{i'}=2\gg_{x,i;x,i'}(\phivec)$ can be chosen arbitrarily (of some Hermitian matrix). Now, given that $\cP$ is \emph{non-trivial}, so that for any pair $i,i'\in I$ there exists $x$ such that $\sqrt{p(x|i)p(x|i')}\neq0$, we can define a Hermitian matrix $\mathsf{C}$ whose \emph{all} entries, $\mathsf{C}_{i,i'}:=2\sum_x \sqrt{p(x|i)p(x|i')}\,\gg_{x,i;x,i'}(\phivec)$ for all $i,i'\in I$, can be freely chosen (in particular, non-zero) by varying $\gg$. Consequently, we may again rewrite \eqnref{eq:Abeta=0} as $\hh=\mathsf{V}^\dagger\mathsf{C}\mathsf{V}$, where $\mathsf{V}:=\sum_{i,i'\in I}\bra{i}V_{\phivec}\ket{i'}\ket{i}\!\bra{i'}$ is an (invertible) unitary matrix. Hence, we can always satisfy the condition \eref{eq:Abeta=0} by choosing $\gg$ such that $\mathsf{C}=\mathsf{V}\hh\mathsf{V}^\dagger$. In summary, whenever the stochastic map $\cP$ representing the detection noise is non-trivial,
 \eqnref{eq:Abeta=0} can always be fulfilled, which by the virtue of the asymptotic CE bound forces the MSE to asymptotically follow the SS. 
\hfill $\blacksquare$
\subsection*{Computing $\FNCE$ and $\FNCEas$ by an SDP}
First, in view of the symmetry in the problem, and as suggested by the expression of $\bar{F}_N^{\mathrm{(CE,as)}}$, let us focus on having the local unitary settings all being the same, i.e., $\phivec_{\ell}=\phivec$ for all $\ell$. In this case then, we have
\begin{align}\label{eq:ACEbound}
&\FNCE(\{\phivec_\ell\})\Rightarrow\FNCE(\phivec):=\nonumber\\
&4\min_{ \tilde{K}(\theta,\phivec) } \Big\{N ||\alpha_{\tilde{K}(\theta,\phivec)}||+N(N-1)||\beta_{\tilde{K}(\theta,\phivec)}||^2\Big\}, 
\end{align}
and 
\begin{align} \label{eq:ACEasbound}
\FNCEas(\{\phivec_\ell\})\Rightarrow \FNCEas(\phivec):= 4\,N\min_{\substack{\tilde{K}(\theta,\phivec)\\ \beta_{\tilde{K}(\theta,\phivec)}=0}} ||\alpha_{\tilde{K}(\theta,\phivec)}||.
\end{align}

Then, note that the calculations for $\FNCE(\phivec)$ or $\FNCEas(\phivec)$ can be made simpler using the fact that we are looking at estimation precision locally around some underlying true value of $\theta$, say $\theta_0$. Consequently, instead of considering the most general unitaries $\uu$ with arbitrary $\theta$ dependence, such that
$\tilde{K}_{x,j}(\theta,\phivec)=\sum_{x',j'} \uu_{x,j;x',j'}(\theta,\phivec)K_{x',j'}(\theta,\phivec)$, it suffices to consider all the Kraus representations $\tilde{K}(\theta,\phivec)$ that differ from the canonical one $K(\theta,\phivec)$ only by their first derivatives with respect to $\theta$. That is, we only need to consider unitaries $\uu=\upe^{\upi (\theta-\theta_0)\gg}$ for some Hermitian generator $\gg$, such that at $\theta=\theta_0$ eventually, the Kraus operators obey Eqs~(\ref{eq:AKrausCE1} and \ref{eq:AKrausCE2}). As result we can replace abstract minimization $\min_{\tilde{K}}$ in Eqs.~(\ref{eq:ACEbound}) and (\ref{eq:ACEasbound}) by $\min_{\gg}$, i.e.~minimization over all Hermitian matrices $\gg$ of dimension $d|X| \times d|X|$.

The bounds $\FNCE(\phivec)$ and $\FNCEas(\phivec)$ involve calculations of operator norms $\|\alpha_{\tilde{K}}\|$ and $\|\beta_{\tilde{K}}\|$, which can be cast as a SDP problem. We refer the readers again to Refs.~(\cite{Demkowicz2012,Kolodynski2013}) for its complete derivation, and for here we shall just outline the algorithm and result. In essence, upon defining $\lambda_a^2:=\|\alpha_{\tilde{K}}\|$ and $\lambda_b^2:=\|\beta_{\tilde{K}}\|^2$, and stacking up all the Kraus operators into a vector of matrices, such that \eqnref{eq:AKrausCE1} now reads $\tilde{\krausvec}=\krausvec:=[K_{x=0,j=1}(\theta,\phivec),K_{x=1,j=1}(\theta,\phivec),\cdots]^T$ and $\dot{\tilde{\krausvec}}=\dot{\krausvec}-\upi \gg\krausvec$, we can rewrite \eqnref{eq:ACEbound} as
\begin{align}\label{eq:AFNCEsdp}
\FNCE(\phivec)&=4N\min_{\gg} \{\lambda_a^2+(N-1)\lambda_b^2\}
\quad\text{with}\quad
\boldsymbol{A},\boldsymbol{B}\geq0, 
\end{align}
where 
\begin{align} \label{eq:ASDPmatAB}
\boldsymbol{A}&=\begin{bmatrix}
\sqrt{\lambda_a}\id_{d} & \dot{\tilde{\krausvec}}^\dagger \\
\dot{\tilde{\krausvec}} & \sqrt{\lambda_a}\id_{d_\mathrm{out}}
\end{bmatrix}, \quad
\boldsymbol{B}&=\begin{bmatrix}
\sqrt{\lambda_b}\id_{d} & (\upi\dot{\tilde{\krausvec}}^\dagger\krausvec)^\dagger \\
\upi\dot{\tilde{\krausvec}}^\dagger\krausvec & \sqrt{\lambda_b}\id_{d_\mathrm{out}}
\end{bmatrix},
\end{align}
and $d_\mathrm{out}=d(|X|^2+1)$. In order to evaluate $\FNCEas(\phivec)$ of \eqnref{eq:ACEasbound}, an additional constraint should just be added to \eqnref{eq:AFNCEsdp}, i.e.~$\upi\dot{\krausvec}^\dagger\krausvec=\krausvec^\dagger \gg\krausvec$ that is simply equivalent to the condition $\beta_{\tilde{K}}=0$ imposed in \eqnref{eq:ACEasbound}.

For particular simple examples analytical answers can be obtained. In particular, for the case of $N$ qubits each sensing the phase $\theta$ in the encoding $\cU_{\theta}\sim\{\upe^{\upi\theta\sigma_{z}/2}\}$, and are subject to measurement noise corresponding to binary asymmetric channel $\cP$ mixing each binary outcome of measuring $\{\Pi_i=\ket{\Pi_i}\bra{\Pi_i}\}$ with $\ket{\Pi_{1(2)}}=\ket{\pm}$, starting from the canonical Kraus representation, we find that by selecting $\gg=r(\pauliz\oplus\pauliz)$ in \eqnref{eq:AKrausCE2} with
\begin{align}\label{eq:Aqubitgmat}
r=\frac{\delta-\sqrt{\pp(1-\qq)}+\sqrt{\qq(1-\pp)}}{2\delta},
\end{align}
$\FNCEas(\phivec)$ is equal to 
\begin{align}
N\Big(\frac{\sqrt{\pp(1-\pp)}-\sqrt{\qq(1-\qq)}}{\pp-\qq}\Big)^{2}. \label{eq:AFNCEas}
\end{align}
Interestingly this result is independent of the choice of $\phivec$ with the local control unitaries taking the form $\Vphi=\upe^{{\upi \sigma_{z}\phi}}$, which is found to be optimal over all choices of $\Vphi$, and so we have actually obtained $\bar{F}_N^{\mathrm{(CE,as)}}=\max_{\phivec}\FNCEas(\phivec)$ in \eqref{eq:AFNCEas}, and hence, \refmain{Eq.~(24) in the main text}.

We also take note that the bound \eqref{eq:AFNCEas} can also be obtained from a conjecture, as a consequence of the $G$-covariance formalism (Observation 2 in main text), applied to the case with local control unitaries. That is, upon conjecturing that the optimal local control unitaries take the form of $\Vopt=\upe^{{\upi \sigma_{z}\phi}}$ (which as said is supported by our numerical findings), the generalisation of \eqnref{eq:AimcQFIbound4} to $N$ independent copies of channel leads to
\begin{align}
\FQimbar_{N}\leq \FQbar [(\Lambda\circ\cU_\theta)^{\otimes N}],
\end{align}
where $\Lambda$ is some conjugate-map decomposition $\Lambda$ satisfying the $G$-covariant condition \eqref{eq:ALambdachanneldef2}, e.g., the quantum-classical channel \eqref{eq:AQCchannel}. As $(\Lambda\circ\cU_\theta)^{\otimes N}$ is of the form of uncorrelated noisy encoding, then, we may apply the standard technique of CE formalism to upper bound $\FQbar [(\Lambda\circ\cU_\theta)^{\otimes N}]$ \cite{Kolodynski2013}, which, in this case, turns out to be given exactly by \eqref{eq:AFNCEas}.


\subsection*{Error-propagation formula with imperfect measurement}
The mean squared error of estimators obtained from measuring the mean of some observable $\hat{O}$ with large number of repetitions $\nu$, is well approximated by the so-called ``error-propagation formula"\cite{Wineland1992}
\begin{align}
\nu\Delta^2\tilde{\theta}_N=\frac{\Delta^2 \hat{O}}{\Big|\dfrac{\partial\langle\hat{O}\rangle}{\partial\theta}\Big|^2}, \label{eq:AEPF1}
\end{align}
where $\Delta^2{O}=\langle \hat{O}^2\rangle-\langle\hat{O}\rangle^2$, with $\langle A\rangle$ being the expectation value of the operator $A$ over the quantum state $\rho(\theta)$, i.e. $\langle A\rangle=\Tr\{\rho(\theta)A\}$. 

For our quantum-classical channel scenario with noisy detection channel represented by the stochastic map $\cP\sim\{p(x|i)\}$, while we have the freedom to choose the measurement basis $\Pi_{i,\phivec}$, we need to keep in mind that the only observable and effective measurement that we have is $\{\Mphi{x}=\sum_i p(x|i)\Pi_{i,\phivec}\}$, and it is not projective in general. The observable that we measure is thus $\hat{O}=\sum_x f_x\Mphi{x}$ for some $\{f_x\}$ defining the observable (which can be chosen quite arbitrarily). For $N$ independent quantum-classical channel, we can then construct the joint observable $\hat{O}=\sum_{j=1}^N \hat{O}^{(j)}=\sum_{j=1}^N\sum_x f_x\Mphi{x}^{(j)}$, where $j$ labels the different channels. Alternatively, we may also consider a second kind of joint observable, where instead of summing over the constituent single-particle operators, we perform \emph{product} over them: $\hat{O}=\prod_{j=1}^N \hat{O}^{(j)}=\prod_{j=1}^N\big(\sum_x f_x\Mphi{x}\big)^{(j)}$.

While it maybe tempting, we cannot however simply use $\hat{O}^2=\Big(\sum_{j=1}^N \hat{O}^{(j)}\Big)^2$ or $\hat{O}^2=\Big(\prod_{j=1}^N \hat{O}^{(j)}\Big)^2$ to compute $\Delta^2\hat{O}$ in \eqnref{eq:AEPF1}. The reason is, \eqnref{eq:AEPF1} uses the implicit assumption that the observable $\hat{O}$ is measured at its eigenbasis, and that is not the case here. The effective $\hat{O}^2$ that we should use in \eqnref{eq:AEPF1} is one which mimics the statistics that we would get as if we are measuring the eigenbasis, i.e., \emph{as if} $\{\Mphi{x}\}$ are projective. That is, we have, for the first kind of observables,
\begin{align}
\hat{O}^2=&\sum_j \Big(\sum_x f_x\Mphi{x}^{(j)}\Big)^2\nonumber\\
&+\sum_{j\neq k} \Big(\sum_x f_x\Mphi{x}^{(j)}\Big)\Big(\sum_x f_x\Mphi{x}^{(k)}\Big)\\
\longrightarrow \hat{O}'^2=&\sum_j \sum_x f_x^2\Mphi{x}^{(j)}\nonumber\\
&+\sum_{j\neq k} \Big(\sum_x f_x\Mphi{x}^{(j)}\Big)\Big(\sum_x f_x\Mphi{x}^{(k)}\Big),
\end{align}
and, for the second kind of observables,
\begin{align}
\hat{O}^2=&\prod_j \Big(\sum_x f_x\Mphi{x}^{(j)}\Big)^2\nonumber\\
\longrightarrow \hat{O}'^2=&\prod_j \Big(\sum_x f_x^2\Mphi{x}\Big)^{(j)}\nonumber\\
\end{align}
and then
\begin{align}
\nu\Delta^2\tilde{\theta}_N=\frac{\langle\hat{O}'^2\rangle-\langle\hat{O}\rangle^2}{\Big|\dfrac{\partial\langle\hat{O}\rangle}{\partial\theta}\Big|^2} \label{eq:AEPF2}
\end{align}
for our quantum-classical channel.

We apply \eqnref{eq:AEPF2} to the case of $N$ qubits, each of which undergoes a projective measurement $\Pi_{1,\phivec}=\ket{+}\bra{+}, \Pi_{2,\phivec}=\ket{-}\bra{-}$ where $\paulix\ket{\pm}=\pm\ket{\pm}$, and $|X|=2$. We consider a binary mixing channel $\cP$ that flips the measurement outcomes with $p(1|1)=\pp$, $p(2|2)=\qq$, so that the effective measurements corresponding to the observed outcomes read $\Mphi{1}=(1+\delta)\id/2+\eta\paulix/2$ and $\Mphi{2}=(1-\delta)\id/2-\eta\paulix/2$ with $\eta:= \pp+\qq-1, \delta:= \pp-\qq$. Then, for the first kind of observables, we have
\begin{align}
\hat{O}=&(f_1-f_2)\eta \Jx+\frac{N}{2}[f_1+f_2+(f_1-f_2)\delta],\\
\hat{O}'^2=&(f_1^2-f_2^2)\eta \Jx + \frac{N}{2}[f_1^2+f_2^2+(f_1^2-f_2^2)\delta] \nonumber\\
&+(\Jx^2-\frac{N}{4})\eta^2(f_1-f_2)^2 \nonumber\\
&+(N-1)[f_1+f_2+(f_1-f_2)\delta]\nonumber\\
&\phantom{+}\Big(\eta(f_1-f_2)\Jx+\frac{N}{4}[f_1+f_2+(f_1-f_2)\delta]\Big),
\end{align}
where $\hat{J}_\ell=\sum_{j=1}^N\frac{\sigma_\ell^{(j)}}{2}$ is the usual total angular momentum operator in the $\ell$-direction with $\ell=\{\mathrm{x,y,z}\}$. After some straightforward algebra, one obtains 
\begin{align}
\nu\Delta^2\tilde{\theta}_N=\frac{\Delta^2 \Jx}{|\partial_\theta\langle \Jx\rangle|^2}-\frac{\delta \langle \Jx\rangle}{\eta|\partial_\theta\langle \Jx\rangle|^2}+\frac{N}{4\eta^2}\frac{1-\eta^2-\delta^2}{|\partial_\theta\langle \Jx\rangle|^2}.
\end{align}
Given $U_\theta=\upe^{\upi\theta\pauliz/2}$ to be the unitary encoding the estimated parameter $\theta$ onto each probe, the state of all the probes just before the measurement reads $\rho^N(\theta):=\upe^{\upi\theta \Jz}\rho^N\upe^{-\upi\theta \Jz}$, where $\rho^N$ is the input $N$-qubit probe state. In such as case, we have that $\partial_\theta\langle \Jx\rangle=-\langle\sin\theta \Jx+\cos\theta \Jy\rangle_{\rho^N}=\Tr\{\rho^N (\sin\theta \Jx+\cos\theta \Jy)\}$.

For the second kind of observables, let us consider for example the (imperfect) parity operator, i.e., with $f_1=-f_2=1$. Then, we have 
\begin{align}
\hat{O}&=\hat{P}\equiv\prod_{j=1}^N \big(\Mphi{1}-\Mphi{2}\big)^{(j)}=\prod_j \Big(\delta\id+\eta\paulix\Big)^{(j)},\\
\hat{O}'^2&=\prod_{j=1}^N \big(\Mphi{1}+\Mphi{2}\big)^{(j)}=\id,
\end{align}
and finally 
\begin{align}
\nu\Delta^2\tilde{\theta}_N= \frac{1-\langle\hat{P}\rangle^2}{|\partial_\theta \langle\hat{P}\rangle|^2}.
\end{align}

Consider the measurement of the (imperfect) parity operator, $\hat{P}=\prod_{j=1}^N \big(\Mphi{1}^{(j)}-\Mphi{2}^{(j)}\big)$, with $\Mphi{1}=(1+\delta)\id/2+\eta\paulix/2$ and $\Mphi{2}=(1-\delta)\id/2-\eta\paulix/2$ as above. Then, by the error-propagation formula once more (see Supplement for details), we have
\begin{align}\label{eq:GHZparity1}
\nu\Delta^2\tilde{\theta}_N=&\frac{1-\langle\hat{P}\rangle^2}{|\partial_\theta \langle\hat{P}\rangle|^2},
\end{align}
where $\langle\hat{P}\rangle=\Tr\{\upe^{\upi\theta\Jz}\rho^N\upe^{-\upi\theta\Jz} \hat{P}\}$. Using the (rotated) GHZ input state, $\rho^N=\ket{\psi}\bra{\psi}$, $\ket{\psi}=\upe^{\upi\phi\Jz}\frac{1}{\sqrt{2}}\big(\ket{0\dots0}+\ket{1\dots1}\big)$, we thus get  
 \begin{align}\label{eq:GHZparity2}
\nu\Delta^2\tilde{\theta}_N=&\frac{1-\big(\delta^N+\eta^N \cos(N\varphi))^2}{N^2\eta^{2N}\sin^2(N\varphi)},
\end{align}
with $\varphi=\phi+\theta$ as before. While the parity measurement with GHZ state will perform poorly for large $N$ by virtue of the exponential factor $\eta^{-2N}$, it does however make a good candidate for small $N$ regime where the $1/N^2$ factor dominates. Indeed, upon optimising over $\phi$, we obtain the blue curve in \refmain{Fig.~6 of the main text}.

\bibliographystyle{myapsrev4-1}
\bibliography{QMwIM-bib}

\begin{thebibliography}{96}
\bibitem [{foo({\natexlab{a}})}]{footnote1}%
  \BibitemOpen
  \href@noop {} {}\bibinfo {note} {We focus on unitary encodings for clarity,
  but we formulate the theorems and lemmas in their most general form, often
  applicable beyond the unitary setting.}\BibitemShut {Stop}%
\bibitem [{foo({\natexlab{b}})}]{footnote5}%
  \BibitemOpen
  \href@noop {} {}\bibinfo {note} {In the picture of ``standard'' noisy
  metrology protocols, this corresponds to correlated decoherence and, hence,
  the no-go theorems precluding the HS no longer apply.}\BibitemShut {Stop}%
\bibitem [{foo({\natexlab{c}})}]{footnote2}%
  \BibitemOpen
  \href@noop {} {}\bibinfo {note} {Note that the structure of the derivation in
  Methods suggests \eqnref{eq:CEasboundmax} to hold also for adaptive
  protocols, in which the $\ell$th control unitary $\cV_{\phivec_\ell}$ may be
  adjusted based on the outcomes of local measurements performed on the
  previous probes $(1,2,\dots,\ell-1)$.}\BibitemShut {Stop}%
\bibitem [{foo({\natexlab{e}})}]{footnote4}%
  \BibitemOpen
  \href@noop {} {}\bibinfo {note} {This is due to the fact that for every
  $\boldsymbol{x}$ either $p_{+}(\boldsymbol{x})p_{-}(\boldsymbol{x})=0$ or
  $p_{+}(\boldsymbol{x})=p_{-}(\boldsymbol{x})$.}\BibitemShut {Stop}%
\end{thebibliography}

\end{document}